\documentstyle[epsf,12pt,eqsecnum,aps,amssymb]{revtex}

\draft
\begin{document}

\title{Polymer depletion effects near mesoscopic particles}
\author{A. Hanke$^{1}$, E. Eisenriegler$^{2}$, and S. Dietrich$^{1}$}
\address{$^{1}$Fachbereich Physik, Bergische Universit\"at Wuppertal,\\
D-42097 Wuppertal, Federal Republic of Germany}
\address{$^{2}$Institut f\"ur Festk\"orperforschung, 
Forschungszentrum J\"ulich,\\
D-52425 J\"ulich, Federal Republic of Germany}
\date{\today}
\maketitle
\bigskip
\begin{abstract}
The behavior of mesoscopic particles dissolved in a dilute 
solution of long, flexible, and nonadsorbing polymer chains 
is studied by field-theoretic methods. For spherical 
and cylindrical particles the solvation free energy for 
immersing a single particle in the solution is calculated
explicitly. Important features are qualitatively different 
for self-avoiding polymer chains as compared with ideal 
chains. The results corroborate the validity of the
Helfrich-type curvature expansion for general particle
shapes and allow for quantitative experimental tests.
For the effective interactions between a small sphere and a wall,
between a thin rod and a wall, and between two small spheres 
quantitative results are presented.
A systematic approach for studying 
effective many-body interactions is provided. The  
common Asakura-Oosawa approximation modelling 
the polymer coils as hard spheres turns out to
fail completely for small particles and still fails
by about $10\,\%$ for large particles. 
\end{abstract}
\bigskip

\narrowtext

\newpage

\section{Introduction}
\label{secI}

In colloidal suspensions the depletion interaction 
between mesoscopic dissolved particles and nonadsorbing 
free polymer chains represents one of the basic 
and tunable effective interactions
(see, e.g., Ref.\,\cite{colloid} for a review).
For example, adding free polymer chains to the 
solvent of a colloidal solution leads to an 
effective attraction between the particles 
which may lead to flocculation \cite{sperry}.
For two individual colloidal particles
or for a single particle near a planar wall
this effective interaction can be measured 
even directly \cite{verma,ohshima}.
In view of its importance it is surprising that 
for a long time the interaction between polymers 
and colloidal particles has been modelled only rather
crudely by approximating the polymer chains by 
nondeformable hard spheres 
\cite{asakura,colloid,verma,ohshima}.

Chain flexibility has been taken into account only more
recently. Mainly the following two cases have been considered:
(a) strongly overlapping chains (semidilute solution) which 
are described within a self-consistent field theory or 
within the framework of a phenomenological scaling theory 
\cite{joanny1,gennes1,odijk,ie}; (b) nonoverlapping
chains (dilute solution) which to a certain extent
can be modelled by
random walks without self-avoidance (ideal chains)
\cite{joanny2,jansons,meijer,lipowsky,ehd}. In three 
dimensions this latter situation is closely realized in a 
theta solvent \cite{gennes2}.

Besides presenting some new results for ideal chains the main
emphasis of the present contribution is on the generic case 
of a {\em good\/} solvent and we investigate systematically the 
consequences of the ensuing excluded volume interaction 
(EV interaction) \cite{confusion}
on depletion effects in a dilute and monodisperse
polymer solution. The interaction of {\em long\/} flexible 
chains with mesoscopic particles leads to {\em universal} 
results which are independent of most microscopic details 
\cite{gennes2,cloizeaux,schafer,eisen} and depend only on 
a few gross properties such as the shape of the particles. 
By focusing on such systems we obtain results 
which are free of nonuniversal model parameters. 
Due to the universality of the corresponding properties
it is sufficient to choose a simple model for calculating 
these results. For example, in a lattice model the interaction 
between a particle and a nonadsorbing chain can be implemented 
as the purely geometrical restriction that the chain must not
intersect the particle \cite{meijer}.
For our investigations we use an Edwards-type model 
\cite{gennes2,cloizeaux,schafer} for the polymer chain 
which allows for an expansion in terms of the EV interaction
and which is amenable to a field-theoretical
treatment. The basic elements in this expansion are 
partition functions $Z_{[0]}({\bf r}, {\bf r}\,')$ for chain
segments without EV interaction (as indicated by the subscript $[0]$)
and with the two ends of the segment fixed at 
${\bf r}$ and ${\bf r}\,'$. In this coarse grained description 
the interaction of the nonadsorbing polymer with the particle 
is implemented by the boundary condition that 
the segment partition function vanishes \cite{walk}
as ${\bf r}$ or ${\bf r}\,'$ approaches the surface $S$ of
the particle \cite{gennes2,eisen}, i.e.,
\begin{equation} \label{eq117}
Z_{[0]}({\bf r}, {\bf r}\,') \, \to \, 0 \, \, ,
\quad {\bf r} \, \to \, S \, \, .
\end{equation}

The only relevant property which characterizes one of 
the interacting polymer chains is its mean square end-to-end 
distance ${\cal R}_{E}^{\,2}$ in the absence of particles and 
other chains. Within the perturbative treatment of the EV 
interaction it will be necessary to generalize the
three-dimensional space to a space of $D$ spatial dimensions. 
In this respect it is convenient \cite{cloizeaux} to introduce 
\begin{equation} \label{I5}
{\cal R}_{x}^{\,2} \, = \, {\cal R}_{E}^{\,2} \, / D \, \, ,
\end{equation}
the mean square of the projection of the end-to-end distance
vector onto a particular direction, say, the $x$-axis, in the 
$D$-dimensional space. For industrially produced 
polymers such as polystyrene values of ${\cal R}_x$ up to the 
order of $\mu\text{m}$ are easily accessible.

The simplest particle shapes relevant for applications are spheres
and rods \cite{colloid} but the particles can also have more
complex structures such as those of closed bilayer membranes in the case
of vesicles \cite{vesicle}. We note that the radius $R$ of
spherical particles can be quite small as compared to accessible
values of ${\cal R}_x$, e.g., $R \approx 0.012\,\mu\text{m}$ in the 
case of Ludox silica particles \cite{kaler}. Rodlike objects are 
provided, e.g., by fibers or colloidal rods \cite{buining}, 
semiflexible polymers with a large persistence length 
$\ell_p$ such as actin for which $\ell_p \approx 17 \mu\text{m}$ 
\cite{gittes}, and microtubuli \cite{gittes}. 
The ratio of the length $l$ and the radius $R$ of rodlike
particles may be of the order of $40$ or larger, in conjunction with
a quite small radius such as $R \approx 0.007\,\mu\text{m}$ in the case 
of colloidal boehmite rods \cite{buining}. As the interaction
between rodlike particles and polymers is concerned we 
consider {\em long\/} rods, i.e., $R, {\cal R}_x \ll l$, and
neglect effects which may arise due to their finite length $l$. 
In order to be able to treat spheres and cylinders in a unified 
way and in general dimensionality,
we are thus led to consider a {\em generalized cylinder\/} $K$ 
with an infinitely extended `axis' of dimension $\delta$. Such a 
generalized cylinder has been introduced in Ref.\,\cite{ehd}, in 
the following denoted as I. The `axis' can be the axis of an 
ordinary infinitely elongated cylinder ($\delta = 1$), or the
midplane of a slab ($\delta = D - 1$),
or the center of a sphere ($\delta = 0$). For general integer 
$D$ and $\delta$ the explicit form of $K$ is
\begin{equation} \label{I10} 
K \, = \, \Big\{ {\bf r} = ({\bf r}_{\perp},
{\bf r}_{\parallel} \,) \in {\mathbb R}^{D - \delta}
\times {\mathbb R}^{\delta}; \, \,  | {\bf r}_{\perp} | \le R \Big\}
\end{equation}
with ${\bf r}_{\perp}$ and ${\bf r}_{\parallel}$ perpendicular and
parallel to the axis, respectively. Note that ${\bf r}_{\perp}$
is a $d$-dimensional vector with 
\begin{equation} \label{I15} 
d = D - \delta \, \, .
\end{equation}
The radius $R$ of the generalized cylinder $K$ is the radius in 
the cases of an ordinary cylinder or a sphere and it is half of
the thickness in the case of a slab. For the slab the geometry 
reduces to the much studied case of (two decoupled) half spaces 
\cite{eisen}. We stress that the generalization of $D$ to values 
different from three is introduced only for technical reasons 
because $D_{uc} = 4$ marks the upper critical dimension for the 
relevance of the EV interaction in the bulk 
\cite{gennes2,cloizeaux,schafer}. Eventually we are 
interested in $-$ and will obtain results for $-$ the 
experimentally relevant case $D = 3$. These results concern 
the solvation free energy for a single particle and the 
depletion interaction between particles.


\subsection{Solvation free energy of a particle}
\label{secIA}

We consider the increase in configurational free energy of a
dilute solution of long flexible polymers with number density 
$n_p$ upon immersing a single particle. For $\delta > 0$ we 
actually consider a generalized cylinder with a large but {\em finite\/} 
axis length $l^\delta$ (i.e., an ordinary cylinder with axis
length $l$ or a slab with cross section area $l^{D-1}$) and
study the increase $n_{p\,} f_K^{\,(1)}$ in free energy per 
$k_B T$ and per $l^{\delta}$ in the limit $l \to \infty$, for 
which $l$ drops out \cite{superscript}. For a sphere 
$n_{p\,} f_K^{\,(1)}$ is simply the free energy increase 
per $k_B T$. The additional increase in free energy upon 
immersing the particle in the polymer free (i.e., $n_p = 0$) 
solvent is regarded as a background term
which, in an experiment, can be determined separately.
In the asymptotic regime where both ${\cal R}_{x}$ 
and $R$ are large on the microscopic scale (such as the 
monomer length or the diameter of the solvent molecules)
it turns out that $f_K^{\,(1)}$ takes the scaling form
\begin{equation} \label{I20} 
f_K^{\,(1)} \, = \, R^{\,d} \, \, Y_{d,D\,}(x) \, \, ,
\end{equation}
where $Y_{d,D}$ is a universal scaling function of the scaling 
variable
\begin{equation} \label{I23} 
x \, = \, {\cal R}_{x} \, / R \, \, .
\end{equation}
For ideal chains (no EV interaction) and $d$ fixed 
the function $Y_{d,D} = Y_d^{\,(id)}$ is independent of $D$ 
(compare I where $f_K^{\,(1)}$ was denoted as $\delta f_K$). 
Results for $Y_d^{\,(id)}$ for $d=3$ (sphere) and $d=2$ 
(cylinder) have been given in Ref.\,\cite{jansons} 
and in I. Here we calculate the scaling function 
$Y_{d,D}(x)$ for chains with EV interaction
perturbatively in terms of $\varepsilon = 4 - D$ with the 
upper critical dimension $D_{uc} = 4$. In particular we 
investigate the following features of $f_K^{\,(1)}$:

(a) For {\em short\/} chains, i.e., $x \ll 1$, 
we assume that $Y_{d,D}(x)$ is analytic so that it 
can be expanded into a Taylor series around $x = 0$. This is 
plausible since for short chains the thickness $\sim {\cal R}_{x}$ 
of the polymer depletion layer is much smaller than the
particle radius $R$ so that a small curvature expansion 
is applicable to $f_K^{\,(1)}$ in which a volume term 
$\sim R^{\,d}$ is followed by a surface term $\sim R^{\, d - 1}$ 
and by successive terms $\sim R^{\, d - 2}$, $R^{\, d - 3}$, etc.,  
generated by the surface curvature. We note, however, that it can be
rather difficult to actually prove this assumption.

The first Taylor coefficients of the expansion of $Y_{d,D}(x)$ 
around $x = 0$
also determine the curvature energies of a particle ${\cal K}$ 
of {\em more general shape\/} provided its surface $S$ is smooth 
and all principal radii of curvature are much larger than the 
polymer size ${\cal R}_{x}$ (compare Ref.\,\cite{membran} and I). 
Consider the increase $F_{\cal K}$ in configurational free energy
upon immersing a particle ${\cal K}$ with finite volume $v_{\cal K}$
into the dilute polymer solution 
with bulk pressure $n_{p\,} k_B T$. Due to general arguments 
\cite{david} in three dimensions one expects an expansion of the 
Helfrich-type \cite{helfrich}
\begin{mathletters}
\label{I25}
\begin{equation} \label{I25a}
F_{\cal K} \, - \, n_p \, k_B T \, v_{\cal K} \, = \, 
\int\limits_{S} dS \, \Big\{ \, \Delta \sigma \, + \, 
\Delta \kappa_1 K_m \, + \, \Delta \kappa_2 K_m^{\,2} \, + \,
\Delta \kappa_G K_G \, + \, \ldots \, \Big\}
\end{equation}
with the local mean curvature
\begin{equation} \label{I25b}
K_m \, = \frac{1}{2}
\, \left( \, \frac{1}{R_1} + \frac{1}{R_2} \, \right)
\end{equation}
and the local Gaussian curvature
\begin{equation} \label{I25c}
K_G \, = \, 1 \, / \, (R_1 R_2) \, \, ,
\end{equation}
where $R_1$ and $R_2$ are the two principal local radii of curvature.
We use the convention that $R_1, R_2 > 0$ means that the boundary
surface is bent {\em away\/} from the polymer solution located in 
the exterior of ${\cal K}$. 
Provided that the expansion (\ref{I25a}) is valid
the surface tension $\Delta \sigma$ and the curvature energies 
$\Delta \kappa_1$, $\Delta \kappa_2$, and $\Delta \kappa_G$ are 
determined uniquely by the special cases that ${\cal K}$ is a 
sphere and a cylinder, respectively. Our explicit results for 
$Y_{d,D}(x)$ provide a strong indication that the Helfrich-type 
expansion (\ref{I25}) is indeed valid and, moreover, 
does yield quantitative 
estimates of the surface tension and of the curvature energies
for the polymer depletion problem in the presence of EV interaction.
These values are the extra contributions 
(as indicated by the $\Delta$'s) to the solvation free 
%
%
\unitlength1cm
\begin{figure}[t]
\begin{picture}(16,6)
\put(-0.5,0.5){
\setlength{\epsfysize}{5.5cm}
\epsfbox{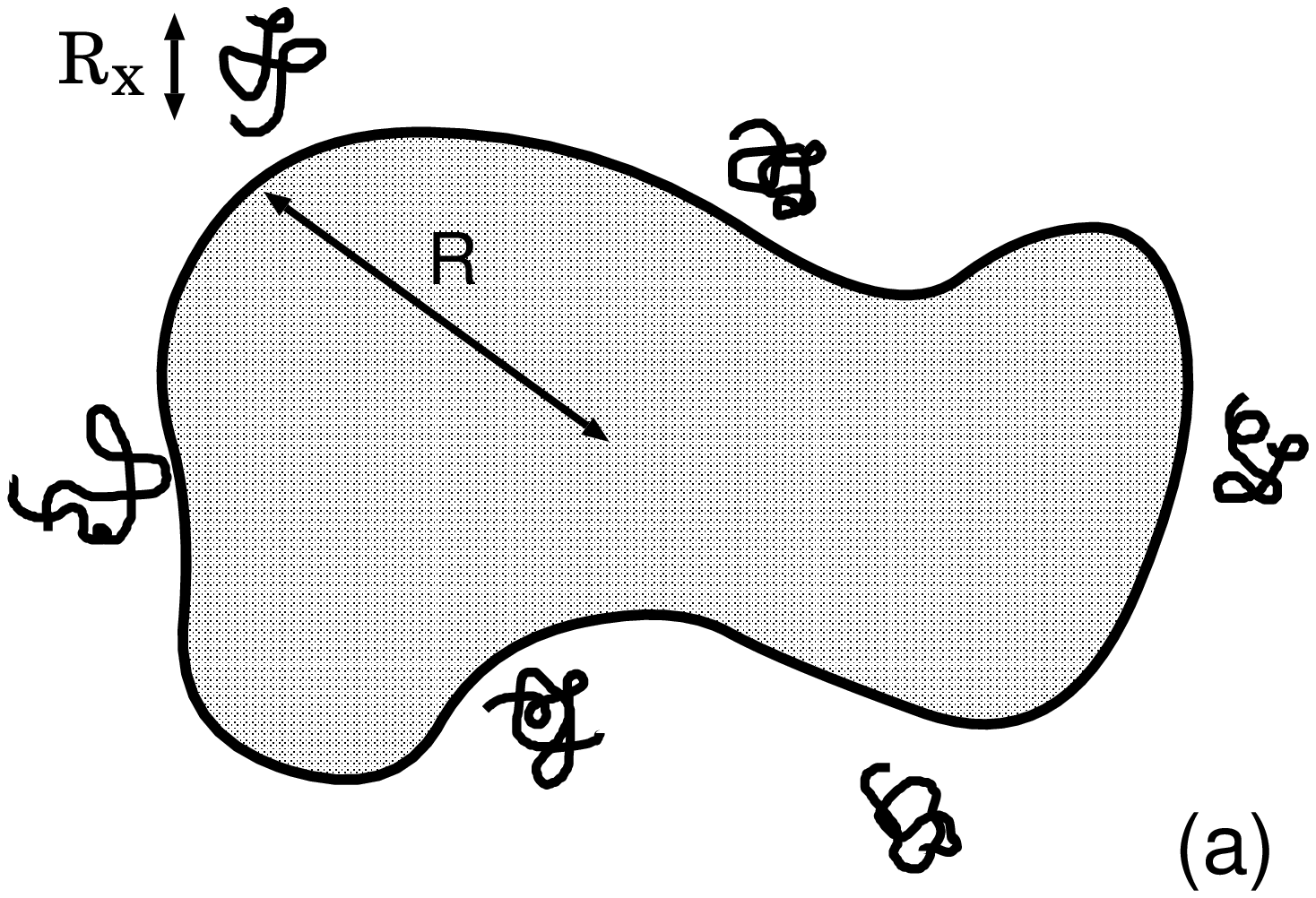}}
\put(8.5,0.5){
\setlength{\epsfysize}{5.5cm}
\epsfbox{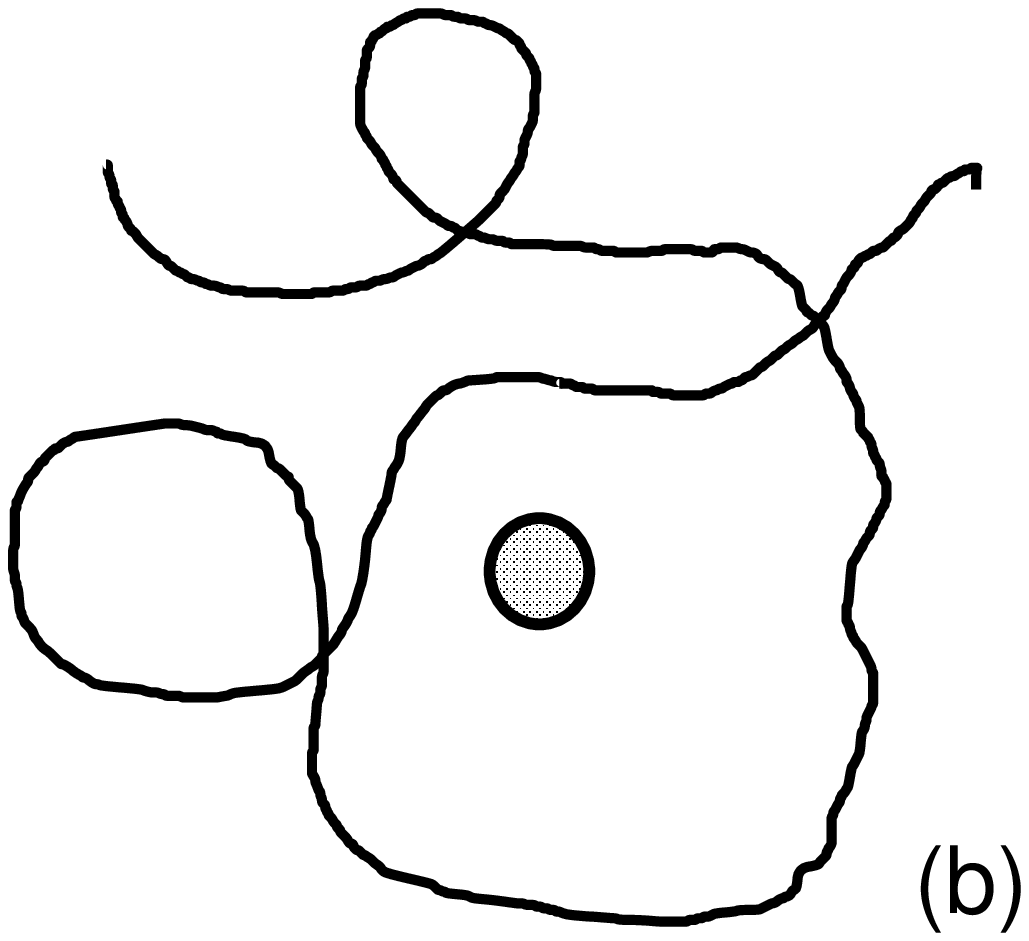}}
\end{picture}
\caption{Situations of short and long chains
in which the limiting behavior of the 
scaling function $Y_{d,D}({\cal R}_{x} / R)$ can be applied: (a) 
For ${\cal R}_{x} \ll R$ the function $Y_{d,D}$ determines the 
change of the surface tension $\Delta \sigma$ and curvature energies    
$\Delta \kappa_1$, $\Delta \kappa_2$, and $\Delta \kappa_G$ in the Helfrich-type 
expansion (\ref{I25}) of a membrane upon exposing one side of 
it to a dilute polymer solution. 
(b) For ${\cal R}_{x} \gg R$ the polymer can deform 
in order to avoid the space occupied by the particle
and coil around a spherical (or rodlike) particle and the  
function $Y_{d,D}$ exhibits the power law (\ref{I30}) 
with the Flory exponent $\nu$.}
\label{fig1}
\end{figure}
%
%
\noindent
energy 
of a particle in addition to its background value for the polymer
free solvent (i.e., $n_p = 0$), not included in Eq.\,(\ref{I25a}).
To the best of our knowledge this is the first check of the 
expansion (\ref{I25}) for a nontrivial interacting system that 
can be realized in nature. 

For other types of systems the expansion (\ref{I25}) can be violated. 
For example, as pointed out by Yaman et al. \cite{pincus}, a somewhat 
counter-intuitive behavior arises for the case in which a surface is 
exposed on one side to a dilute solution of thin 
{\em rigid rods} (needles): even for arbitrarily
small surface curvature the free energy in this case cannot be
expanded in the analytical and local form of the Helfrich-type 
expansion (\ref{I25}). However, for flexible ideal chains instead 
of needles the expansion is known to apply (see I). In particular 
the asymmetry in the curvature contribution $\sim R^{-2}$ between the 
in- and outside of a spherical or cylindrical surface as reported by 
the above authors for needles does not occur for flexible ideal chains
\cite{ideal}.

We note that the curvature energies are experimentally accessible.
For example the expansion (\ref{I25}) determines the change in
surface tension and in the first and second order curvature
energies of a {\em flexible\/} surface such as a {\em membrane\/} 
upon exposing one side of it to a solution of polymers which
are depleted near the membrane (see Fig.\,\ref{fig1}\,(a)).
Thus the addition of polymers to a solution of closed membranes,
i.e. vesicles, should influence the phase diagram of vesicle shapes
in a quantitatively controllable way (see, e.g., Ref.\,\cite{dobereiner}). 
An additional experimental access to the solvation free energy 
will be discussed at the end of this subsection.

\end{mathletters}

(b) For {\em long\/} chains, i.e., $x \gg 1$, a single chain 
can deform in order to avoid the space occupied by the 
particle and coil around a spherical or rodlike
particle (see Fig.\,\ref{fig1}\,(b)). In this case
it turns out that $Y_{d,D}(x)$ exhibits a power law
\begin{equation} \label{I30} 
Y_{d,D\,}(x \to \infty) \, \to \, A_{d,D} \, x^{\, 1/\nu}
\end{equation}
with a dimensionless and universal amplitude $A_{d,D}$, provided
\begin{equation} \label{I35} 
d \, > \, 1 / \nu
\end{equation}
so that $f_K^{\,(1)}$ vanishes for $R \to 0$ (see Eq.\,(\ref{I20})).
Here $\nu$ is the Flory exponent characterizing
the power law dependence ${\cal R}_{x} \sim N^{\nu}$ 
of ${\cal R}_{x}$ on the number $N$ of monomers per
chain if $N$ is large. The properties described by 
Eqs.\,(\ref{I30}) and (\ref{I35}) follow from a 
{\em s}mall {\em r}adius operator
{\em e}xpansion (SRE)
(see Sec.\,\ref{secIB} below).

Finally, we emphasize that $f_K^{\,(1)}$ is experimentally 
accessible by monitoring the dependence 
of the number density $n_c$ of the {\em c}olloidal particles
on the number density $n_p$ of the {\em p}olymers  
in a sufficiently dilute solute of immersed particles
which is in thermal equilibrium with a surrounding ideal 
gas phase with given partial pressure $p_c^{\,(0)}$ of 
the particles \cite{medium}. Accordingly $n_c$ is 
determined by a Henry-type law
\begin{mathletters}
\label{I50}
\begin{equation} \label{I50a} 
n_c \, = \, \frac{p_c^{\,(0)}}{k_B T} \, \, \Lambda^{-1}
\end{equation}
where $\Lambda$ measures the change
of the solubility of the colloidal particles 
due to the presence of the polymers and is given by
\begin{equation} \label{I50b} 
\Lambda \, = \, \exp( n_p f_K^{\,(1)\,} l^{\delta} ) \, \, .
\end{equation}
For the dilute immersed particles the reduced free energy 
increase $n_p f_K^{\,(1)\,} \l^{\delta}$ constitutes a reduced 
one-particle potential or, equivalently, an increase in 
chemical potential, so that Eq.\,(\ref{I50}) follows upon 
equating the chemical potentials of the particles in the 
ideal gas phase and in the solution phase.

\end{mathletters}


\subsection{Colloidal particles with small radii}
\label{secIB}

We consider the case in which a polymer chain interacts
with a spherical or cylindrical particle whose radius $R$ 
$-$ albeit being large on the microscopic scale $-$ is much 
smaller than the size ${\cal R}_{x}$ of the chain and other 
characteristic lengths \cite{lengths}. In this limiting case 
the effect of the spherical particle upon the 
configurations of the chain can be 
represented by a $\delta$-function potential located at the center 
of the particle which repels the monomers of the chain.
For a generalized cylinder $K$ with a small radius $R$ this
$\delta$-function potential is smeared out over its axis. Thus
the Boltzmann weight $W_K\{ {\bf y}_i \}$ for the chain \cite{neu}  
arising from the presence of $K$ (whose axis includes the origin)
is replaced by
\begin{equation} \label{I60}
W_K\{ {\bf y}_i \} \, \to \,
\, 1 \, - \, A_{d,D} \, R^{\,d - 1/\nu} \, w_K
\end{equation}
with
\begin{equation} \label{I61}
w_K \, = \, \left\{ \begin{array}{l@{\quad}l}
\int\limits_{{\mathbb R}^{\delta}}
d^{\, \delta} r_{\parallel} \, \,
\rho({\bf r}_{\perp}=0, {\bf r}_{\parallel})
\, \, ,    & d < D \, \, , \\[2mm]
\, \, \rho(0) \, \, \, ,   & d = D \, \, ,
\end{array} \right.
\end{equation}
provided $d > 1 / \nu$. The positions 
$\{ {\bf y}_{i\,}; i = 1, \ldots, N \}$
of the $N$ chain monomers which define the chain configuration 
appear in Eq.\,(\ref{I61}) in terms of the modified monomer density
\begin{equation} \label{I70}
\rho({\bf r}) \, = \, \frac{ {\cal R}_{x}^{\, 1/\nu} }{N} \,
\sum_{i=1}^{N} \, \delta^{\,(D)\,}({\bf y}_i - {\bf r} ) \, \, .
\end{equation}
The sum of $\delta$-functions in Eq.\,(\ref{I70}) is the usual 
monomer number density at a point ${\bf r}$. We have chosen its 
prefactor such that $\rho({\bf r})$ is less dependent on the 
microscopic monomer structure (i.e., on what is considered as 
a monomer) than the sum itself. In particular 
$\int d^D r \, \rho({\bf r}) = {\cal R}_{x}^{\, 1/\nu}$
is independent of these details while $N$ is not. 
The scaling dimension $D - 1 / \nu$ of $\rho({\bf r})$
equals its naive inverse length dimension so that the
exponent of $R$ in Eq.\,(\ref{I60}) follows by comparing naive
dimensions. The amplitude $A_{d,D}$ is dimensionless and 
universal \cite{needle}.

The monomer positions $\{ {\bf y}_i \}$ are statistical variables so that 
Eq.\,(\ref{I60}) is a relation between fluctuating quantities which is
to be used inside polymer conformation averages such as the ratio
of polymer partition functions with and without the presence of $K$.
One can use Eq.\,(\ref{I60}) for a variety of different
situations. If $K$ is the only particle within reach of the polymer
chain, Eq.\,(\ref{I60}) leads to the free energy change 
given by Eq.\,(\ref{I20}) in the limit discussed in Eq.\,(\ref{I30}).
This is the reason why the same amplitude $A_{d,D}$ appears
in Eqs.\,(\ref{I30}) and (\ref{I60}).
If there are in addition other particles or walls $K'$, 
Eq.\,(\ref{I60}) can be used to calculate the polymer mediated 
free energy of interaction (potential of mean force) between 
$K'$ and $K$ (compare I and Sec.\,\ref{secIC} below). 
Equation (\ref{I60}) simplifies the theoretical treatment of
these problems significantly
because $K$ is replaced by the monomer density $\rho({\bf r})$. 
While the remaining, simpler averages depend on the particular 
problem under consideration, the universal amplitude $A_{d,D}$ 
is always the same.

In this work we study the small radius expansion (\ref{I60})
for the generalized cylinder $K$ for the case of polymers in
a good solvent. Our main objective is to present quantitative 
estimates for the universal amplitudes $A_{3,3}$ and $A_{2,3}$ 
corresponding to a sphere and to an infinitely elongated cylinder 
in three dimensions. The cylinder (i.e., $d=2$) is particularly
interesting since in this case the EV interaction changes the 
behavior {\em qualitatively\/}: while for ideal chains a thin 
cylinder is a marginal perturbation which can lead to a logarithmic 
behavior \cite{joanny2} and for which Eq.\,(\ref{I60}) does 
{\em not\/} apply, for chains with EV interaction the power law 
exponent $d - 1 / \nu \approx 0.30$ is {\em positive\/} and
Eq.\,(\ref{I60}) holds. This peculiarity for $d=2$ is reflected
in the $\varepsilon$-expansion of $A_{d,D}$ 
for $D = 4 - \varepsilon$.


\subsection{Interactions between particles}
\label{secIC}

Polymer mediated interactions between particles are in general 
{\em not pairwise additive\/}, i.e., they cannot be written as 
a superposition of pair interactions \cite{colloid,meijer}. 
For a dilute polymer solution with polymer density $n_p$
we consider the total increase in reduced configurational
free energy $n_{p\,} f_{\text{tot}}^{\,(M)}$ upon immersing
spherical particles $K_1, \, \ldots \, , \, K_{M}$ centered
at ${\bf r}_1, \, \ldots \, , \, {\bf r}_{M}$. The quantity 
$f_{\text{tot}}^{\,(M)}$ has the form
\begin{eqnarray}
f_{\text{tot}}^{\,(M)}({\bf r}_1, \, \ldots \, , {\bf r}_{M})  \, & = & \,
\sum_{i=1}^{M} \, f_{K_i}^{\,(1)} \, + \,
\sum_{pairs \atop i<j}^{M} \, f_{K_i,K_j}^{\,\,(2)\,}({\bf r}_i,{\bf r}_j)
\label{I80}\\
& & \, + \, \ldots \, + \, 
f_{K_1, \, \ldots \, , K_{M}}^{\,\,(M)\,}
({\bf r}_1, \, \ldots \, , {\bf r}_{M}) \, \, .
\nonumber
\end{eqnarray}
The $m$-body contributions $f^{(m)}$ for $2 \le m \le M$
on the rhs of Eq.\,(\ref{I80}) are defined inductively by 
considering first two particles in order to define $f^{(2)}$ 
via Eq.\,(\ref{I80}), then three, and so on.
For spherical particles
the dimension of $f^{(m)}$ is that of a volume, i.e., of 
$(\text{length})^D$. The existence of polymer mediated 
nonpairwise interactions has first been noticed within 
the PHS approximation, which consists in replacing the 
{\em p}olymer by a {\em h}ard {\em s}phere 
\cite{asakura}. Here we consider 
the limit for which the polymer is flexible
and much longer than the particle radii, i.e.,
${\cal R}_{x} \gg R$, and where the small-radius expansion 
(\ref{I60}) gives a simple and {\em quantitative\/} description.  
We find that the polymer mediated interaction for particles 
with small $R$ is {\em drastically\/} different from the 
depletion interaction for large $R$ in which case the PHS 
approximation is reasonable and has been widely used.
This confirms the generally accepted belief that for the 
applicability of the PHS approximation a large size ratio
$R / {\cal R}_x$ is crucial and refutes an opposite claim 
in Ref.\,[9(a)]. 

As illustration we consider three spherical particles 
$A$, $B$, $C$ with radii $R_A$, $R_B$, $R_C$ much smaller 
than their mutual distances and than ${\cal R}_{x}$. It 
is easy to see that $f_{\text{tot}}^{\,(3)}$ is
determined by the Boltzmann weights of the particles
introduced in the text preceding Eq.\,(\ref{I60}) 
in the form 
\begin{equation} \label{I90}
f_{\text{tot}}^{\,(3)}({\bf r}_A, {\bf r}_B, {\bf r}_C)
\, = \, \int\limits_{{\mathbb R}^{D}}
d^{\, D} y \, \, \left\{ 1 \, - \, W_A \, W_B \, W_C \right\}_{{\bf y}}
\end{equation}
where $\{\,\,\}_{\bf y}$ denotes the average over all conformations
of a single chain in {\em free\/} space (i.e., no particles) under 
the constraint that one end of the chain is fixed at the point 
${\bf y}$. In the limit of small radii $R$ one finds by using 
Eq.\,(\ref{I60}) that in addition to the one-body contributions
$f_{A}^{\,(1)}$, $f_{B}^{\,(1)}$, and $f_{C}^{\,(1)}$, each 
exhibiting the scaling form described by Eq.\,(\ref{I20}) in the 
limit given by Eq.\,(\ref{I30}), there arise two-body contributions
\begin{mathletters}
\label{I100}
\begin{equation} \label{I100a}
f_{A,B}^{\,\,(2)} \, \to \,
- \, (A_{D,D})^{\,2} \, (R_A R_B)^{\,D - 1/\nu} \, \,
C_2({\bf r}_A, {\bf r}_B) \, \, ,
\end{equation}
$f_{A,C}^{\,\,(2)}$, and $f_{B,C}^{\,\,(2)}$,
and a three-body contribution 
\begin{equation} \label{I100b}
f_{A,B,C}^{\,\,(3)} \, \to \,
(A_{D,D})^{\,3} \, (R_A R_B R_C)^{\,D - 1/\nu} \, \,
C_3({\bf r}_A, {\bf r}_B,  {\bf r}_C) \, \, . 
\end{equation}
The arrows in the above relations indicate the leading behavior
for small radii. Here $C_2$ and $C_3$ are pair and triple
correlation functions corresponding to 
\end{mathletters}
\begin{equation} \label{I110}
C_m({\bf r}_1, {\bf r}_2, \, \ldots \, , {\bf r}_m) \, = \,
\int\limits_{{\mathbb R}^{D}} d^{\, D} y \, \,
\{\rho({\bf r}_1) \rho({\bf r}_2) \, \ldots \,
\rho({\bf r}_m)\}_{{\bf y}}
\end{equation}
of the (modified) monomer density $\rho({\bf r})$ 
defined in Eq.\,(\ref{I70}) for a single polymer chain
in free space. Since ${\cal R}_{x}$ and the relative distances 
$r_{AB} = |{\bf r}_A - {\bf r}_B|$ are large on the microscopic
scale these correlation functions exhibit the scaling forms
\begin{mathletters}
\label{I120}
\begin{equation} \label{I120a}
C_2({\bf r}_A, {\bf r}_B) \, = \, {\cal R}_{x}^{\,2/\nu\, - D} \, \,
g(z_{AB}) \, \, ,
\end{equation}
with $z_{AB} = r_{AB} / {\cal R}_{x}$, and 
\begin{equation} \label{I120b}
C_3({\bf r}_A, {\bf r}_B,  {\bf r}_C) \, = \,
{\cal R}_{x}^{\,3/\nu\, - 2 D} \, \,
h(z_{AB}, z_{AC}, z_{BC}) \, \, ,
\end{equation}
which follow from the scaling dimension $D - 1 / \nu$ of $\rho({\bf r})$.
Thus for three spherical particles with equal radii $R$ and with
center to center distances $r_{AB}$, $r_{AC}$, $r_{BC}$ which are of
the order of ${\cal R}_{x}$ but much larger than $R$, the three-body
interaction is smaller than the two-body interaction
by a factor $\sim (R / {\cal R}_{x})^{D - 1/\nu}$.

\end{mathletters}

Similar fluctuation induced, not pairwise additive 
interactions arise between 
particles which are immersed in a near-critical fluid mixture
\cite{burkhardt}. In this case one encounters order parameter 
correlation functions instead of the present monomer density 
correlation functions.

The small radius expressions (\ref{I100}) cease to apply $-$ even 
if the equal radii $R$ are much smaller than ${\cal R}_x$ $-$
if some of the relative distances between the spheres 
become comparable with $R$. However, there are other types 
of short distance expansions which are capable to
describe these latter situations. In particular we shall discuss a 
`small dumb-bell' expansion for a pair of spheres $A$, $B$ for which 
both $R$ and $r_{AB}$ are much smaller than the other
lengths. The structure of this expansion is similar to 
Eq.\,(\ref{I60}) in conjunction with the lower part of 
Eq.\,(\ref{I61}), but the amplitude corresponding to $A_{D,D}$
now depends on the ratio $r_{AB} / R$. 
We calculate this new amplitude function
for the case of ideal chains.    

In Sec.\,\ref{secII} we discuss in detail the 
solvation free energy for a single particle.
In Sec.\,\ref{secIII} we consider 
the depletion interaction between particles.   
Section \ref{summary} contains our conclusions.  
In Appendix \ref{neuA} we derive the asymptotic 
expansions for a small and large size ratio ${\cal R}_x / R$ 
required for Sec.\,\ref{secII}. In Appendix \ref{appB} we   
discuss the perturbative treatment of the small radius operator 
expansion. Finally, in Appendix\,\ref{appC}, we derive a 
short-distance amplitude which characterizes the behavior of 
monomer density correlation functions in free space as needed
in Sec.\,\ref{secIII}.
 

\newpage

\section{Solvation free energy of a particle}
\label{secII}

The free energy for immersing a particle in a dilute solution 
of freely floating chains with or without self-avoidance can 
be expressed in terms of the
density profile of chain ends in the presence of the particle
(compare, e.g., Eq.\,(3.7) in I). For the scaling function introduced
in Eq.\,(\ref{I20}) this implies
\begin{equation} \label{II20}
Y_{d,D}(x) \, = \, \frac{\Omega_{d}}{d} \, + \, 
\Omega_{d} \, Q_{d,D}(\eta) \, \, , \quad \eta = x^2 / 2 \, \, ,
\end{equation}
with $\Omega_{d} = 2 \pi^{d/2}/\,\Gamma(d/2)$ the surface area of
the $d$-dimensional unit sphere and
\begin{equation} \label{A240}
Q_{d,D\,}(\eta) \, = \, \int\limits_{1}^{\infty} d \rho \, \rho^{\,d-1} \,
\left[1 - M_E(\rho, \eta) \right] \, \, .
\end{equation}
In Eq.\,(\ref{A240}) the scaling function 
$M_E(r_{\perp}/R, \eta)$ is the bulk normalized density 
profile of chain ends at a distance $r_{\perp} - R$ from
the particle surface.
In Sec.\,\ref{secIIvoran} we derive the explicit form of 
$Q_{d,D}(\eta)$ in the presence of EV interaction to lowest 
nontrivial order in $\varepsilon = 4 - D$. 
In Secs.\,\ref{secIIA} and \ref{secIIB}
we discuss the resulting behavior of $Y_{d,D}(x)$ in the limit 
of short and long chains, respectively. Finally, we obtain 
in Sec.\,\ref{secIIC}
approximations for the full scaling function $Y_{d,D}(x)$ 
corresponding to a sphere and a cylinder in $D = 3$.


\subsection{Density of chain ends and polymer magnet
analogy} \label{secIIvoran}

We employ the polymer magnet analogy (PMA) in order to calculate
the density profile ${\cal M}_E$ of chain ends in a dilute 
solution of chains with EV interaction which 
arises in the presence of the nonadsorbing 
generalized cylinder $K$ introduced in Eq.\,(\ref{I10}).
As in I we define ${\cal M}_E$ as bulk normalized so that it 
approaches one far from the particle. It is given by
\begin{eqnarray}
& & {\cal M}_E(r_{\perp}; L_0, R, u_0) \label{E10}\\[1mm]
& & = \, \int\limits_{V} d^D r' \,
Z({\bf r}, {\bf r}\,' ; L_0, R, u_0) \, \Big/ \,
\int\limits_{V} d^D r' \, 
Z_b({\bf r}, {\bf r}\,' ; L_0, u_0) \, \, . \nonumber
\end{eqnarray}
Here $Z$ and $Z_b$ are partition functions of a single chain
with the two ends fixed at ${\bf r}$, ${\bf r}\,'$ in the
presence and absence, respectively, of the generalized cylinder $K$
(the subscript $b$ stands for `bulk'). The volume $V$ available for 
the chain is the outer space $V = {\mathbb R}^D \setminus K$ of $K$.
The parameter $u_0$ 
characterizes the strength of the EV interaction and $L_0$
determines the monomer content or `length' of the chain 
such that $2 L_0$ equals the mean square
${\cal R}_{x}^{\,2}$ of the projected end-to-end distance
of the chain in the absence of $K$ and of the EV interaction,
i.e., for $u_0 = 0$. The usual arguments of the PMA
\cite{gennes2,cloizeaux,schafer,eisen} carry over to the
present case and imply the correspondence
\begin{equation} \label{E20}
Z({\bf r}, {\bf r}\,' ; L_0, R, u_0) \, = \,
{\cal L}_{t_0 \to L_0} \, 
\langle \Phi_1({\bf r}) \Phi_1({\bf r}\,') \rangle
\Big|_{{\cal N} = 0}
\end{equation}
between $Z$ and the two-point correlation function 
$\langle \Phi_1({\bf r}) \Phi_1({\bf r}\,') \rangle$ 
in a ${\cal O}({\cal N})$ symmetric field theory for an 
${\cal N}$-component order parameter field 
${\bf \Phi} = (\Phi_1, \, ... \,, \Phi_{\cal N})$ 
in the restricted volume $V = {\mathbb R}^D \setminus K$.
In Eq.\,(\ref{E20}) the operation
\begin{equation} \label{E30}
{\cal L}_{t_0 \to L_0} \, = \,
\frac{1}{2 \pi i} \int\limits_{\cal C} d t_0 \, e^{L_0 t_0}
\end{equation}
acting on the correlation function is an inverse Laplace transform
with ${\cal C}$ a path in the complex $t_{0\,}$-plane to the right
of all singularities of the integrand.
The Laplace-conjugate $t_0$ of $L_0$ and
the excluded volume strength $u_0$ appear, respectively, as the
`temperature' parameter and the prefactor of the 
$({\bf \Phi}^2)^2$-term in the Ginzburg-Landau Hamiltonian 
\begin{mathletters} \label{A10}
\begin{equation} \label{A10a}
{\cal H}_K \{ {\bf \Phi} \}
= \int\limits_V d^D r \, \left\{ \frac{1}{2} (\nabla {\bf \Phi})^2
+ \frac{t_0}{2} \, {\bf \Phi}^2 
+ \frac{u_0}{24} \, ({\bf \Phi}^2)^2 \right\}
\end{equation}
which provides the statistical weight 
$\exp(- {\cal H}_K \{ {\bf \Phi} \})$ for the field theory.
The position vector ${\bf r}$ covers the volume $V$ and its
boundary, which is the surface of $K$. In order to be consistent
with Eq.\,(\ref{eq117}) we have to impose the Dirichlet condition
\begin{equation} \label{A10b}
{\bf \Phi}({\bf r}) \, = \, 0
\, \, , \, \,  \text{if} \, \, \, 
|{\bf r}_{\perp}| = R \, \, ,
\end{equation}
on the boundary. This corresponds to the fixed point boundary 
condition of the so-called ordinary transition \cite{binder,diehl}
for the field theory. For our renormalization group improved
perturbative investigations we use a dimensionally regularized 
continuum version of the field theory which we shall 
renormalize by minimal subtraction
of poles in $\varepsilon = 4 - D$ \cite{amit} (this is related
via Eq.\,(\ref{E20}) to a corresponding procedure in the 
Edwards model \cite{cloizeaux,schafer,eisen}). The basic element
of the perturbation expansion is the 
Gaussian two-point correlation function (or propagator) 
$\langle \Phi_{i}({\bf r}) \, \Phi_{j}({\bf r}\,' ) \rangle_{[0]}$
where the subscript $[0]$ denotes $u_0 = 0$. It is given by
\end{mathletters}
\begin{mathletters}
\label{A20}
\widetext
\begin{eqnarray}
& & \langle \Phi_{i}({\bf r}) \, \Phi_{j}({\bf r}\,' ) \rangle_{[0]} 
\, = \, \delta_{ij} \, G({\bf r}, {\bf r}\,'; t_{0\,}, R) \, = \,
\delta_{ij} \, \widehat{G}(r_{\perp}, r_{\perp}^{\,\,'}, \vartheta, 
| {\bf r}_{\parallel} - {\bf r}_{\parallel}^{\,\,'} | \, ; t_{0\,}, R) 
\label{A20a}\\
& & = \, \delta_{ij} \, \times \, \left\{ \begin{array}{l@{\,\,\,}l}
{\displaystyle \sum_{n=0}^{\infty} }
\, \, W_{n}^{(\alpha)}(\vartheta) \,
{\displaystyle \int\limits_{{\mathbb R}^{\delta}}
\frac{d^{\, \delta}P}{(2 \pi)^{\delta}} } \, \,
\exp [i \, {\bf P} \,
({\bf r}_{\parallel} - {\bf r}_{\parallel}^{\,\,' }) ] \, \,
\widetilde{G}_{n\,}(r_{\perp},r_{\perp}^{\,\,'}; {\cal S}, R)
\, \, , \quad & d < D \, , \\
{\displaystyle \sum_{n=0}^{\infty} }
\, \, W_{n}^{(\alpha)}(\vartheta) \, \,
\widetilde{G}_{n\,}(r_{\perp},r_{\perp}^{\,\,'}; t_{0\,}, R)
\, \, , \quad & d = D \, ,\end{array} \right. \nonumber
\end{eqnarray}
\narrowtext
where $\alpha = (d-2)/2$, ${\cal S} = P^2 + t_0$, 
$r_{\perp} = |{\bf r}_{\perp}|$, and 
$\vartheta$ is the angle between 
${\bf r}_{\perp}$ and ${\bf r}_{\perp}^{\,\,'}$
(compare Fig.\,1 in I). 
Note that for $d = D$ (last line in Eq.\,(\ref{A20a})) 
there is no parallel component 
${\bf r}_{\parallel} - {\bf r}_{\parallel}^{\,\,'}$ and
hence no Fourier variable ${\bf P}$.
The functions $W_{n}^{(\alpha)}(\vartheta)$ are given by
\begin{equation} \label{A20b}
W_{n}^{(\alpha)}(\vartheta) = \left\{ \begin{array}{l@{\,\,\,}l}
(2 \pi^{d/2})^{-1} \, \Gamma(\alpha)
\, (n + \alpha) \, \, C_{n}^{\alpha}( \cos
\vartheta) \, \, , & d \not= 2 \, ,\\ \\
(2 \pi)^{-1} \, (2 - \delta_{n, 0}) \,
\cos (n \vartheta) \, \, , & d = 2 \, ,
\end{array} \right.
\end{equation}
where $\Gamma$ is the gamma function, $C_{n}^{\alpha}$ are
Gegenbauer polynomials \cite{abr}, and $\delta_{n, 0} = 1$
for $n = 0$ and zero otherwise.
The functions $W_{n}^{(\alpha)}$ are normalized so that
$\int d\Omega_d \, W_{n}^{\, (\alpha)} = \delta_{n,0\,}$.
The propagator $\widetilde{G}_{n}$ has the form
\begin{eqnarray}
& & \widetilde{G}_{n}
(r_{\perp}, r_{\perp}^{\,\,'}; {\cal S}, R)  \, = \,
(r_{\perp}^{(<)} \, r_{\perp}^{(>)})^{-\alpha}
\, K_{\alpha + n}( \sqrt{\cal S} \, r_{\perp}^{(>)} )
\label{A20c}\\[2mm]
& & \, \, \times \,
\left[ \, I_{\alpha + n}( \sqrt{\cal S} \, r_{\perp}^{(<)} )
\, - \, \frac{I_{\alpha + n}(\sqrt{\cal S} \, R)}
{K_{\alpha + n}(\sqrt{\cal S} \, R)} \,
K_{\alpha + n}( \sqrt{\cal S} \, r_{\perp}^{(<)} ) \, \right] \, \, ,
\nonumber 
\end{eqnarray}
where $r_{\perp}^{(<)} = \min(r_{\perp}, r_{\perp}^{\,\,'\,})$ and 
$r_{\perp}^{(>)} = \max(r_{\perp}, r_{\perp}^{\,\,'\,})$. For $d = D$ the
variable ${\cal S}$ is replaced by $t_0$.
$I_{\alpha}$ and $K_{\alpha}$ denote modified 
Bessel functions \cite{abr}.

\end{mathletters}

The numerator in the density profile ${\cal M}_E$ in Eq.\,(\ref{E10})
can be obtained from the integrated two-point correlation function,
i.e., the local susceptibility $\chi$, for $t_0 > 0$.
Due to rotational invariance around and translational invariance
along the axis of $K$ the local susceptibility $\chi$ only
depends on the radial component $r_{\perp}$
of the point ${\bf r} = ({\bf r}_{\perp}, {\bf r}_{\parallel\,})$.
The loop expansion of $\chi$ reads
\begin{equation} \label{A30}
\chi(r_{\perp} ; t_{0\,}, R, u_0) \, = \,
\chi^{[0]}(r_{\perp} ; t_{0\,}, R) \, + \,
u_0 \, \chi^{[1]}(r_{\perp} ; t_{0\,}, R) \, \, + \, 
{\cal O}( u_0^2 )
\end{equation}
where the zero-loop contribution $\chi^{[0]}$
is given by the integrated propagator
\begin{equation} \label{A40}
\chi^{[0]}(r_{\perp} ; t_{0\,}, R)
= \, \int\limits_V \, d^D r\,' \, \,
G({\bf r}, {\bf r}\,' ; t_{0\,}, R)
\, = \, \frac{R^2}{\tau_0} \left[ \, 1 - \, 
\frac{ \rho^{-\alpha} \, K_{\alpha}(\, \rho \, \sqrt{\tau_{0}} \,) }
{ K_{\alpha}(\sqrt{\tau_{0}} \,) } \, \right] \, \, .
\end{equation}
The greek symbols on the rhs denote dimensionless variables
expressed in terms of
the radius $R$ of $K$: 
\begin{equation} \label{A50}
\tau_0 \, = \, t_{0\,} R^2 \, \, , \quad 
\rho \, = \, r_{\perp\,} / R \, \, .
\end{equation}
According to standard perturbation theory
the one-loop contribution is given by
\begin{eqnarray}
& & u_0 \, \chi^{[1]}(r_{\perp}; t_{0\,}, R) \label{A60}\\[2mm]
& & \, = \, - \, \frac{{\cal N} + 2}{3} \, \frac{u_0}{2} \, 
\int\limits_V \, d^D y \, \, 
G({\bf r}, {\bf y} ; t_{0\,}, R) \, \,
G({\bf y}, {\bf y} ; t_{0\,}, R) \, \,
\chi^{[0]}(y_{\perp} ; t_{0\,}, R) \nonumber\\
& & \, =  - \, \frac{{\cal N} + 2}{3} \, \frac{u_0}{2} \,
\, R^{\,2+\varepsilon} \,
\int\limits_{1}^{\infty} \, d\psi \, \psi^{\, \varepsilon-1} \, \,
{\cal G}(\rho, \psi, \tau_0) \,
g(\psi, \tau_{0}, \varepsilon) \, {\cal X}^{[0]}(\psi, \tau_0)  
\nonumber
\end{eqnarray}
where $\psi = y_{\perp\,} / R$ (compare Eq.\,(\ref{A50})).
The functions in the integrand of the last line 
in Eq.\,(\ref{A60}) are dimensionless and defined by
\begin{equation} \label{A62}
{\cal X}^{[0]}(\psi, \tau_0) = R^{- 2} \,
\chi^{[0]}(y_{\perp}; t_{0\,}, R) \, \, , 
\end{equation}
\begin{equation} \label{A64}
{\cal G}(\rho, \psi, \tau_0) \, = \, R^{2 \alpha} \,
\widetilde{G}_{n=0}
(r_{\perp}, y_{\perp}; {\cal S} = t_0, R) \, \, ,
\end{equation}
\begin{equation} \label{A66}
y_{\perp\,}^{\, \varepsilon} \, g(\psi, \tau_0, \varepsilon) \, = \, 
y_{\perp\,}^{\,d} \, R^{- 2 \alpha} \, G({\bf y}, {\bf y} ; t_{0\,}, R) \, \, .
\end{equation}
The function $g$ can be split into $g = g_b + g_s$ where
\begin{mathletters}
\label{A80}
\begin{equation} \label{A80b}
g_{b}(\psi, \tau_0, \varepsilon) \, = \, \tau_{0}^{\,1 - \varepsilon / 2} \,
\psi^{d - \varepsilon} \, \,
\frac{\Gamma(\varepsilon/2 - 1)}{(4 \pi)^{D/2}}
\end{equation}
stems from the bulk contribution of
$G({\bf y}, {\bf y} ; t_{0\,}, R)$ and
\begin{eqnarray}
& & g_{s}(\psi, \tau_0, \varepsilon) \, = \, - \, \psi^{2 - \varepsilon} \,
\sum_{n=0}^{\infty} \, W_{n}^{(\alpha)}(\vartheta = 0)
\label{A80c}\\
& & \qquad \times \, \left\{ \begin{array}{l@{\,\,\,}l}
{\displaystyle \frac{\Omega_{\delta}}{(2 \pi)^{\delta}} } \,
{\displaystyle \int\limits_{0}^{\infty} d q \, \, q^{\delta-1} } \,
{\displaystyle
\frac{I_{\alpha+n\,}(\sqrt{q^2 + \tau_{0}}\,) }
     {K_{\alpha+n\,}(\sqrt{q^2 + \tau_{0}}\,) } } \, \,
\left[ 
K_{\alpha+n\,}(\psi \, \sqrt{q^2 + \tau_{0}}\,) \right]^{2} \, \, , &
d < D \, , \\ \\
{\displaystyle
\frac{ I_{\alpha+n\,}(\sqrt{\tau_{0}}\,) }
     { K_{\alpha+n\,}(\sqrt{\tau_{0}}\,) } } \, \,
\left[K_{\alpha+n\,}(\psi \, \sqrt{\tau_{0}}\,)\right]^{2} \, \, , &
d = D \, . \end{array} \right. \nonumber
\end{eqnarray}
Note that $\delta = D - d = 4 - \varepsilon - d$.
In the case $d < D$ we shall consider $d$ (and $\alpha = (d-2)/2$) 
as a variable which is {\em independent\/} of $D = 4 - \varepsilon$
whereas in the case $d = D$ the variable
$\alpha = 1 - \varepsilon / 2\,$ depends of course on $\varepsilon$.
One can check that in the case $d = 1$, for which 
$W_{n}^{(\alpha)}(\vartheta = 0)$ 
with $\alpha = - 1/2$ contributes
only for $n=0$ and $1$, the upper part of Eq.\,(\ref{A80c})
leads indeed to the half-space result
\begin{eqnarray}
& & G({\bf y},{\bf y}; t_0, R) \, - \, 
G_b({\bf y},{\bf y}; t_0) \label{A13}\\[2mm]
& & = \, - \, \frac{\Omega_{D-1}}{(2 \pi)^{D-1}}
\int\limits_0^{\infty} dP \, \frac{P^{D-2}}{2 \sqrt{P^2 + t_0}} \,
\exp[- 2 \sqrt{P^2 + t_0} \, \, (y_{\perp}-R)] \, \, , 
\quad d = 1 \, \, ,
\nonumber
\end{eqnarray}
where the integral can be expressed in terms of a modified Bessel
function. We add the following two remarks about the behavior of 
$g_s$ for $d > 1$ if $R \to \infty$ or $R \to 0$.

(i) It is instructive to see how the behavior for the half-space 
arises by taking the limit $R \to \infty$ with $t_0$ and $y_{\perp}-R$
fixed. Consider, e.g., the case $d = D = 4$ corresponding to the
sphere in four dimensions.
Since upon approaching the above limit the arguments of the Bessel
functions in the lower Eq.\,(\ref{A80c}) become large and
since many terms contribute in the sum over $n$ one has to use the
uniform asymptotic expansion of the Bessel functions 
(compare, e.g., sections 9.7.7 and 9.7.8 in Ref.\,[40(a)])
and may replace the sum by an integral. This yields that 
$G({\bf y},{\bf y}; t_0, R) - G_b({\bf y},{\bf y}; t_0)$ for 
$d = D = 4$ does indeed tend to the half-space expression on the rhs
of Eq.\,(\ref{A13}) with $D = 4$, where the role of the length $P$ 
of the wavevector ${\bf P}$ is taken by the ratio $n/R$.

(ii) For $d > 2$ and fixed nonvanishing 
lengths $y_{\perp}$ and $t_0^{\, -1/2}$ the
quantity $g_s(\psi, \tau_0, \varepsilon)$ has a finite
limit for $R \to 0$, i.e., 
\begin{eqnarray}
& & g_s^{\,(\text{as})}(\psi \sqrt{\tau_0}\,, \varepsilon) \, \equiv \,
\lim_{R \to 0} \,  g_s(\psi, \tau_0, \varepsilon) \label{A82}\\
& & = \, - \, \frac{2^{2 - d}}{\pi^{d/2\,} \Gamma(\alpha)} \, 
\left\{ \begin{array}{l@{\,\,\,}l}
\displaystyle{ \frac{\Omega_{\delta}}{(2 \pi)^{\delta}}\,
\int\limits_{0}^{\infty} d k \, \, k^{\delta-1} \,
\left( k^2 + \psi^2 \tau_0 \right)^{\alpha} \, 
\left[K_{\alpha} ( \sqrt{k^2 + \psi^2 \tau_{0}}\,) \right]^{2} } 
\, \, , & d < D \, , \\ \\
\displaystyle{\left( \psi^2 \tau_0 \right)^{\alpha} \, 
\left[K_{\alpha} ( \psi \sqrt{\tau_{0}}) \right]^{2} }  
\, \, , & d = D \, , \end{array} \right. \nonumber
\end{eqnarray}
which depends only on the $R$-independent product 
$\psi \sqrt{\tau_0} = y_{\perp} \sqrt{t_0}$ and describes 
the behavior of $g_s$ for $R \ll y_{\perp}$, $t_0^{\,-1/2}$. 
This is consistent with the operator expansion for small
radius $R$ of the Boltzmann weight representing $K$ 
when applied to a Gaussian field theory (compare I). While 
$g_s^{\,(\text{as})}$ decays exponentially for 
$\psi \sqrt{\tau_0} \to \infty$ it approaches a finite constant for 
$\psi \sqrt{\tau_0} \to 0$,
which equals $- \alpha / (4 \pi^2)$ for $\varepsilon = 0$
and characterizes the behavior of $g_s$
for $R \ll y_{\perp} \ll t_0^{\,-1/2}$. 
This should be compared with the behavior 
$g_s \sim - (\psi - 1)^{2 - D}$ which applies 
{\em close to the surface\/} of $K$, i.e., for
$0 < y_{\perp} - R \ll R$, $t_0^{\,-1/2}$.

\end{mathletters}

The reparametrizations \cite{amit}
\begin{mathletters} \label{A54}
\begin{equation} \label{A54c}
u_0 \, = \, 16 \, \pi^2 \, f(\epsilon) \, \mu^{\varepsilon} 
\, Z_{u\,} u \, \, , \qquad Z_u = 1 + {\cal O}(u) \, \, ,
\end{equation}
and 
\begin{equation} \label{A54a}
t_0 \, = \, \mu^2 \, Z_t \, t \, = \,
\mu^2 \, \Big( 1 + \frac{{\cal N} + 2}{3} \,
\frac{u}{\varepsilon}  \, + \, {\cal O}(u^2) \Big) \, t
\end{equation}
of the bare bulk parameters $u_0$ and $t_0$ in terms of 
their renormalized and dimensionless counterparts $u$ and $t$
are not affected by the presence of the surface \cite{symanzik,diehl}.
Here $\mu$ is the inverse length scale which determines the
renormalization group flow and
$f(\varepsilon) = 1 + \varepsilon \, f_1 + {\cal O}(\varepsilon^2)$.
The coefficient $f_1$ drops out from universal quantities 
and therefore can be chosen arbitrarily.
Equation (\ref{A54a}) implies the renormalized counterpart
\begin{equation} \label{A54b}
\tau \, = \, (\mu R)^2 \, t
\end{equation}
of $\tau_0$. The renormalized, i.e., pole-free, local susceptibility
$\chi_{ren}$ is related to $\chi$ by \cite{diehl,amit} 
\begin{eqnarray}
\chi_{ren\,}(r_{\perp} ; t, R, u) \, & = & \, 
\chi(r_{\perp} ; \, t_{0\,}, R, u_0) \, / \, Z_{\Phi}(u) \label{A57}\\[1mm]
& = & \, \chi(r_{\perp} ; \, t_{0\,}, R, u_0)  \, \, + \, 
{\cal O}( u^2 ) \nonumber
\end{eqnarray}
\end{mathletters}
\noindent
with the renormalization factor $Z_{\Phi}$ of the field ${\bf \Phi}$
which deviates from one only in second order in $u$. 
The only pole in $\chi^{[1]}$ is due to the bulk contribution $g_{b}$
in Eq.\,({\ref{A80b}).
When the results for $\chi^{[0]}$ and
$u_0 \, \chi^{[1]}$ in Eqs.\,(\ref{A40}) and (\ref{A60})
are substituted into Eq.\,(\ref{A30}) and when the bare parameters
$\tau_0$ and $u_0$ are
expressed in terms of their renormalized counterparts
$\tau$ and $u$ according to 
Eqs.\,(\ref{A54}), the poles in $\varepsilon$
cancel indeed \cite{symanzik}. 
This cancellation can be traced back to the relation 
\begin{equation} \label{A88}
\int\limits_{1}^{\infty} \, d\psi \, \psi^{d-1} \, \,
{\cal G}(\rho, \psi, \tau) \, {\cal X}^{[0]}(\psi, \tau) \, = \,
- \frac{\partial}{\partial \tau} {\cal X}^{[0]}(\rho, \tau) \, \, .
\end{equation}
The resulting renormalized and scaled
local susceptibility ${\cal X}_{ren} = R^{-2} \, \chi_{ren}$
up to one loop order reads
\begin{eqnarray}
& &{\cal X}_{ren\,}(\rho, \tau, \mu R, u) \, = \, 
{\cal X}^{[0]}(\rho, \tau) \label{A90}\\[2mm]
& & \quad + \, \, \frac{{\cal N} + 2}{3} \, u \, \Big[
\Big( \frac{\ln \tau}{2} - \ln(\mu R) + B \Big) \,  
\tau \, \frac{\partial}{\partial {\tau}} \, 
{\cal X}^{[0]}(\rho, \tau) 
\, + \, {\cal E}_d(\rho, \tau) \Big] \, \, + \, 
{\cal O}(u^2) \nonumber
\end{eqnarray}
with the nonuniversal constant
\begin{equation} \label{A100}
B \, = \, \frac{C_E}{2} \, - \frac{1}{2} \, - \, f_1 \, - \,
\frac{\ln(4 \pi)}{2} \, \, ,
\end{equation}
where $C_E$ is Euler's constant, and the function
\begin{equation} \label{A110}
{\cal E}_d(\rho, \tau) \, = \, - \, 8 \pi^2 \,
\int\limits_{1}^{\infty} \, d\psi \, \psi^{\, -1} \, \,
{\cal G}(\rho, \psi, \tau) \,
g_s(\psi, \tau, \varepsilon = 0) \, {\cal X}^{[0]}(\psi, \tau) \, \, .
\end{equation}
Since ${\cal E}_d$ belongs to the one loop contribution
and because we assume that 
in the last line of Eq.\,(\ref{A60}) the order of the
$\psi$-integration and the limit $\varepsilon \to 0$
can be interchanged we set $\varepsilon = 0$ in the integrand
on the rhs of Eq.\,(\ref{A110}). This implies that in the case 
$d = D$ only ${\cal E}_4$ enters into 
Eq.\,(\ref{A90}) (compare the remark below Eq.\,(\ref{A80c})).
The integral on the rhs of Eq.\,(\ref{A110})
is well-defined since the divergence of
$g_s(\psi, \tau, \varepsilon = 0)$ for $\psi \searrow 1$ becomes integrable
due to the Dirichlet behavior of ${\cal G}$ and  ${\cal X}^{[0]}$ 
as implied by Eq.\,(\ref{A10b}).
We also need the bulk value (far away from $K$)
of the renormalized local susceptibility up to one loop order,
which reads
\begin{equation} \label{A120}
{\cal X}_{ren, \,b}(\tau, \mu R, u) \, = \, \frac{1}{\tau} \,
- \, \, \frac{{\cal N} + 2}{3} \, \frac{u}{\tau} \,
\Big( \frac{\ln \tau}{2} - \ln(\mu R) + B \Big)  \, \, + \, 
{\cal O}(u^2) \, \, .
\end{equation}

The perturbative result (\ref{A90}) can be improved using standard
renormalization group arguments \cite{diehl}. Although we
need only the results (\ref{A90}) and (\ref{A120})
for the discussion of the polymer depletion problem, 
we note that in the asymptotic limit for which 
$r_{\perp}$, $R$, and the bulk
correlation length $\xi_{+}$ for $t > 0$ are large
compared with microscopic lengths the ratio
\begin{equation} \label{A130}
{\cal X}_{ren\,}(\rho, \tau, \mu R, u) \, /
\, {{\cal X}_{ren, \, b}(\tau, \mu R, u)}
\, \to \, \Xi_{\cal N}(\rho, \gamma)
\end{equation}
yields a scaling form expressed in terms of
the universal scaling function 
$\Xi_{\cal N}(\rho, \gamma)$ with the scaling variables
$\rho = r_{\perp} / R$ and $\gamma = R^2 / \xi_{+}^{\,2\,}$.
The function $\Xi_{\cal N}$ depends on the
number ${\cal N}$ of components of ${\bf \Phi}$,
on the parameter $d$ which characterizes the shape of $K$,
and on the space dimension $D$.
While the amplitude $\xi_0^{\,+}$ in the bulk relation
$\xi_+ = \xi_0^{\,+} \, t^{- \nu({\cal N})}$ is nonuniversal,
the exponent $\nu({\cal N})$ is universal and
depends only on ${\cal N}$ and $D$.
The asymptotic scaling behavior is governed by the infrared
(long-distance) stable fixed point for which
\begin{equation} \label{A140}
u \, = \, u^* \, = \, \frac{3 \, \varepsilon}{{\cal N} + 8} \, + \, \,
{\cal O}(\varepsilon^2)
\end{equation}
and
\begin{equation} \label{A150} 
\nu({\cal N}) \, = \, \frac{1}{2} \, + \, \frac{1}{4} \, \,
\frac{{\cal N} + 2}{{\cal N} + 8} \, \, \varepsilon
\, + \, \, {\cal O}(\varepsilon^2) \, \, .
\end{equation}
The bulk correlation length $\xi_+$ can be defined 
in various ways. For definiteness we assume that
$\xi_{+}^{\,2}$ is defined \cite{definition} as the 
second moment of the two-point correlation function 
divided by $2 D$, which implies
\begin{equation} \label{A160}
(\xi_0^{\,+})^2 \, = \, ( D_t(u) )^{- 2 \nu({\cal N})} \,
\Big\{\mu^{-2} \, \Big[ 1 \, - \, 
\frac{{\cal N} + 2}{{\cal N} + 8} \, \, \varepsilon \, B
 \, + \, {\cal O}(\varepsilon^2) \Big] \Big\}
\end{equation}
with the nonuniversal constant $B$ defined in Eq.\,(\ref{A100}). 
Here the curly bracket equals $\xi_{+}^{\,2}$ for $t = 1$ and 
$u = u^{*}$ and the dependence of $(\xi_0^{\,+})^2$ on $u$
is contained in the dimensionless amplitude $D_t$ which can be
expressed in terms of Wilson functions corresponding 
to the renormalization group flow of $t$ and $u$ 
\cite{amit,cloizeaux,schafer,eisen}.
When Eqs.\,(\ref{A140}) - (\ref{A160})
are combined with Eqs.\,(\ref{A90}) and (\ref{A120}) one
finds that ${\cal X}_{ren} / {\cal X}_{ren, \, b}$ at the 
fixed point is indeed consistent with Eq.\,(\ref{A130})
and that the scaling function $\Xi_{\cal N}$ is given by
\begin{equation} \label{A170}
\Xi_{\cal N}(\rho, \gamma) \, = \, \gamma \, 
{\cal X}^{[0]}(\rho, \gamma) \, + \, 
\frac{{\cal N} + 2}{{\cal N} + 8} \, \, \varepsilon
\, \, \gamma \, {\cal E}_d(\rho, \gamma)
\, + \, \, {\cal O}(\varepsilon^2) \, \, .
\end{equation}
Equation (\ref{A170}) provides the general result
for the bulk normalized local susceptibility of the magnetic 
analogue in the presence of $K$.

The density ${\cal M}_E$ of chain ends as defined in Eq.\,(\ref{E10})
can be related to ${\cal X}_{ren} = R^{-2} \chi_{ren}$, 
with $\chi_{ren}$ from Eq.\,(\ref{A57}), by
means of Eqs.\,(\ref{E20}) and (\ref{A54}). The result is
\begin{mathletters} \label{xi}
\begin{equation} \label{A200}
{\cal M}_E(r_{\perp}; L_0, R, u_0) \, = \, 
{\cal Z}_{ren\,}(\rho, \lambda, \mu R, u) \, /
\, {\cal Z}_{ren, \, b}(\lambda, \mu R, u)
\end{equation}
where
\begin{equation} \label{A180}
{\cal Z}_{ren\,}(\rho, \lambda, \mu R, u) \, = \,
{\cal L}_{\tau \to \lambda} 
\left\{{\cal X}_{ren\,}(\rho, \tau, \mu R, u) \right\}
\Big|_{{\cal N}=0}
\end{equation}
is the renormalized and scaled version of the integrated chain
partition function in the numerator of the rhs of Eq.\,(\ref{E10}).
Here ${\cal L}$ is the operation in Eq.\,(\ref{E30}) and 
\begin{equation} \label{A190}
\lambda \, = \, L \, / \, (\mu R)^2 \, = \, Z_{t\,} L_0 / R^2 
\end{equation}
is the scaled counterpart of the renormalized and dimensionless
chain `length' $L$ \cite{cloizeaux,schafer,eisen} so that
$\lambda \tau = L t = L_{0\,} t_0$.
For large $r_{\perp}$, $L_0$, $R$ the end density exhibits the
scaling behavior
\begin{equation} \label{d}
{\cal M}_E(r_{\perp}; L_0, R, u_0) \, \to \,
M_E(\rho, \eta) \, \, ,
\end{equation}
where $M_E$ is a universal scaling function of $\rho = r_{\perp} / R$
and the scaling variable
\begin{equation} \label{A204}
\eta \, = \, \frac{{\cal R}_{x}^{\,2}}{2 R^2} \, \, .
\end{equation}
\end{mathletters}
\noindent
According to our definition in Eq.\,(\ref{I5}) of
${\cal R}_{E}^{\,2} = D {\cal R}_{x}^{\,2}$ as
the second moment of the bulk partition function
$Z_b({\bf r}, {\bf r}\,')$ 
the nonuniversal prefactor $r_0^{\,2}$ in the asymptotic behavior 
${\cal R}_{E}^{\,2} / (2 D) = r_0^{\,2} \, L^{2 \nu}$
with $\nu = \nu({\cal N} = 0)$ has the form
\begin{equation} \label{A220}
r_0^{\,2} \, = (D_L(u))^{2 \nu} \, \Big\{ \mu^{-2} \, \Big[ 1 \, - \, 
\frac{\varepsilon}{4} \, 
\Big(B + 1 - \frac{C_E}{2} \Big) \, + 
\, {\cal O}(\varepsilon^2) \Big] \Big\}
\end{equation}
with $B$ from Eq.\,(\ref{A100}).
The curly bracket equals ${\cal R}_{E}^{\,2} / (2 D)$ for $L = 1$
and $u = u^{*}$ and the dependence of $r_0^{\,2}$ on $u$
is contained in the amplitude $D_L = 1 / D_t$ with $D_t$ from
Eq.\,(\ref{A160}) (compare, e.g., Ref.\,\cite{eisen}).
Obviously $\eta$ plays a similar role as the inverse
of the scaling variable $\gamma = R^2 / \xi_{+}^{\,2}$
in Eq.\,(\ref{A170}) in the magnetic analogue.
By using Eqs.\,(\ref{A200}) and (\ref{A180}) and by
carrying out the same steps which lead
to the scaling function $\Xi_{\cal N}$ in Eq.\,(\ref{A170})
of the magnetic analogue one arrives at
\begin{mathletters}
\label{A230}
\begin{equation} \label{A230a}
M_E(\rho, \eta) \, = \, M_E^{[0]}(\rho, \eta)
\, + \, \frac{\varepsilon}{4} \, M_E^{[1]}(\rho, \eta) \, +
\, \, {\cal O}(\varepsilon^2)
\end{equation}
where
\begin{equation} \label{A230b}
M_E^{[0]}(\rho, \eta) \, = \, {\cal L}_{\tau \to \eta} 
\left\{ {\cal X}^{[0]}(\rho, \tau) \right\}
\end{equation}
is the zero loop, i.e., Gaussian contribution and \cite{laplace}
\begin{eqnarray}
& & M_E^{[1]}(\rho, \eta) \, = \, 
{\cal L}_{\tau \to \eta} \left\{ \, {\cal E}_d(\rho, \tau) \, \right\}
\, + \,
{\cal L}_{\tau \to \eta}
\Big\{ \frac{\ln \tau}{2} \,
\Big[ \tau \, \frac{\partial}{\partial {\tau}} \, 
{\cal X}^{[0]}(\rho, \tau) 
+ \frac{1}{\tau} \Big] \Big\} \label{A230c}\\
& & \quad + \, \frac{1}{2} \Big[ 1 - M_E^{[0]}(\rho, \eta) - 
\eta \, \frac{\partial}{\partial {\eta}} \, 
M_E^{[0]}(\rho, \eta) \Big] 
\Big[ \ln \eta + C_E \Big] \, + \,
\eta \, \frac{\partial}{\partial {\eta}} \, 
M_E^{[0]}(\rho, \eta) \nonumber \, \, .
\end{eqnarray}

\end{mathletters}

Equation (\ref{A230}) provides the general result for the bulk
normalized density of chain ends $M_E$ in a dilute polymer 
solution in the presence of $K$. 
According to Eq.\,(\ref{A240}) for the scaling function
$Y_{d,D}$ we only need the integrated form.
The terms in Eq.\,(\ref{A230c}) have been arranged such 
that the $\rho$-integration in Eq.\,(\ref{A240})
can be carried out in each bracket separately.
This leads to
\begin{mathletters}
\label{A250}
\begin{equation} \label{A250a}
Q_{d,D\,}(\eta) \, = P_{d}^{\,[0]}(\eta)
\, + \, \frac{\varepsilon}{4} \, P_{d}^{\,[1]}(\eta) \, +
\, \, {\cal O}(\varepsilon^2) \, \, ,
\end{equation}
where
\begin{equation} \label{A250b}
P_{d}^{\,[0]}(\eta) \, = \, {\cal L}_{\tau \to \eta} \left\{  
\frac{K_{\alpha+1}(\sqrt{\tau})}
{\tau^{3/2} \, K_{\alpha}(\sqrt{\tau})} \right\}
\end{equation}
is the zero loop, i.e., Gaussian contribution and
\begin{eqnarray}
& & P_{d}^{\,[1]}(\eta) \, = \, - \, C_{d}(\eta) \, + \, 
{\cal L}_{\tau \to \eta}
\left\{ \frac{\tau \ln \tau}{2} \, \, \frac{\partial}{\partial \tau}
\left[ \frac{K_{\alpha+1}(\sqrt{\tau})}
{\tau^{3/2} \, K_{\alpha}(\sqrt{\tau})}\right] \right\} \label{A250c}\\
& & \quad - \, \frac{1}{2} \Big[ P_{d}^{\,[0]}(\eta) +
\eta \, \frac{\partial}{\partial {\eta}} \, 
P_{d}^{\,[0]}(\eta) \Big] 
\Big[ \ln \eta + C_E \Big] \, + \,
\eta \, \frac{\partial}{\partial {\eta}} \, 
P_{d}^{\,[0]}(\eta) \, \, . \nonumber
\end{eqnarray}
In Eq.\,(\ref{A250c}) we have introduced the function
\end{mathletters}
\begin{mathletters}
\label{A260}
\begin{equation} \label{A260a}
C_{d}(\eta) \, = \, {\cal L}_{\tau \to \eta} 
\left\{ {\cal C}_d(\tau) \right\}
\end{equation}
with
\begin{eqnarray}
{\cal C}_d(\tau) \, & = & \, \int\limits_{1}^{\infty} 
d \rho \, \rho^{\,d-1} \,
{\cal E}_d(\rho, \tau) \label{A260b}\\
& = & \, - \, 8 \pi^2 \, 
\int\limits_{1}^{\infty} \, d\psi \, \psi^{\, -1} \, \,
g_s(\psi, \tau, \varepsilon = 0) \, 
\left[{\cal X}^{[0]}(\psi, \tau) \right]^{2} \, \, , \nonumber
\end{eqnarray}
where Eq.\,(\ref{A110}) has been used.
The functions in the integrand of the last line
in Eq.\,(\ref{A260b}) are given by Eqs.\,(\ref{A80c}) and
(\ref{A62}) in conjunction with (\ref{A40}). 
In the case $d = D$ we have to consider 
${\cal C}_4(\tau)$ only (compare the remark 
below Eq.\,(\ref{A110})).

\end{mathletters}


\subsection{Short chains: $Y_{d,D}(x)$ for $x \to 0$} \label{secIIA}

The aim of this subsection is to determine
the surface tension $\Delta \sigma$ and the curvature energies 
$\Delta \kappa_1$, $\Delta \kappa_2$, and $\Delta \kappa_G$ in the 
expansion (\ref{I25}) to first order in 
$\varepsilon = 4 - D$ by considering  
the special cases that the particle ${\cal K}$ is a 
generalized cylinder $K$ with $d = D$, $3$, and $2$.

In accordance with the discussion in Sec.\,\ref{secIA}
we assume that the function
$Y_{d,D}(x)$ in Eq.\,(\ref{II20}) is analytic 
at $x = 0$ which implies that $Q_{d,D\,}(\eta)$ 
can be expanded into a Taylor series in 
$\sqrt{\eta} = x / \sqrt{2}$.
In the following we determine the first three terms 
of this expansion. 
The expansion is consistent with the behavior 
\begin{equation} \label{II40}
{\cal C}_d(\tau) \, = \, {\cal C}_{0} \, \tau^{- 3/2} \, + \,
{\cal C}_{1}^{\,(d)} \, \tau^{-2} \, + \, 
{\cal C}_{2}^{\,(d)} \, \tau^{-5/2} \, +
\, \, {\cal O}(\tau^{-3})
\end{equation}
for large $\tau = (\mu R)^2 \, t$ of the function ${\cal C}_d(\tau)$
in Eq.\,(\ref{A260b}) which we verify in Appendix \ref{neuA}.
Its form \cite{laplace}
\begin{eqnarray} 
& & Q_{d,D\,}(\eta) \, = \, \frac{2 \, \eta^{1/2}}{\sqrt{\pi}} \, 
\Big\{ 1 - \frac{\varepsilon}{4} 
\Big[ 1 - \frac{3 \ln 2}{2} + {\cal C}_{0} \Big] \Big\}
\, + \, \eta \, \left\{ \frac{d-1}{2} - \frac{\varepsilon}{4} \,
{\cal C}_{1}^{\,(d)} \right\} \label{II150}\\[2mm]
& & \qquad + \, \, \frac{4 \, \eta^{3/2}}{3 \sqrt{\pi}} \, \Big\{
\frac{(d-1)(d-3)}{8} \, 
\Big[ 1 - \frac{\varepsilon}{4} \, 
\Big( \frac{11}{6} - \frac{5 \ln 2}{2} \Big) \Big] \, - \,
\frac{\varepsilon}{4} \, {\cal C}_{2}^{\,(d)} \Big\}  \, +
\, \, {\cal O}(\eta^2, \varepsilon^2) \nonumber
\end{eqnarray}
follows from Eqs.\,(\ref{A250}) and (\ref{A260}) by inserting
Eq.\,(\ref{II40}) and the large $\tau$ behavior
\begin{equation} \label{II30}
\frac{K_{\alpha+1}(\sqrt{\tau})}
{\tau^{3/2} \, K_{\alpha}(\sqrt{\tau})} \, = \, 
\tau^{- 3/2} \, + \, \frac{d-1}{2} \, \tau^{-2} \, + \,
\frac{(d-1)(d-3)}{8} \, \tau^{-5/2} \, +
\, \, {\cal O}( \tau^{-3} ) \, \, .
\end{equation}
Since ${\cal C}_{0}$ is related to the surface tension $\Delta \sigma$ 
it should not depend on the shape of $K$, i.e., on the value of $d$.
Using, e.g., Eqs.\,(\ref{A260b}), (\ref{A66}), 
and (\ref{A13}) corresponding to a planar wall 
(i.e., $d = 1$) one finds
\begin{equation} \label{II45}
{\cal C}_{0} \, = \, - \, \frac{\pi}{2} \, + \, 
\frac{\pi}{\sqrt{3}} \, \, .
\end{equation}
The evaluation of the coefficients 
${\cal C}_{1}^{\,(d)}$ and ${\cal C}_{2}^{\,(d)}$
in Eq.\,(\ref{II40}) for $d = D$, $3$, and $2$
is carried out in Appendix \ref{neuA} by extending
the method explained after Eq.\,(\ref{A13}) to the
next-to-leading terms. 
For $d = D$ we have to consider ${\cal C}_4(\tau)$ only
and find
\begin{mathletters}
\label{II100}
\begin{equation} \label{II100b}
{\cal C}_{1}^{\,(4)} \, = \, - \, \frac{17}{6} \, + \, 
\frac{15 \, \pi}{4} \, - \, \frac{3 \sqrt{3} \, \pi}{2} \, \, ,
\end{equation}
\begin{equation} \label{II100c}
{\cal C}_{2}^{\,(4)} \, = \, - \, 66 \, + \, \frac{8011 \, \pi}{128}
\, - \, \frac{191 \sqrt{3} \, \pi}{8} \, \, ;
\end{equation}
for $d = 3$ we find
\end{mathletters}
\begin{mathletters}
\label{II120}
\begin{equation} \label{II120a}
{\cal C}_{1}^{\,(3)} \, = \, - \, \frac{17}{9} \, + \, 
\frac{5 \, \pi}{2} \, - \, \sqrt{3} \, \pi \, \, ,
\end{equation}
\begin{equation} \label{II120b}
{\cal C}_{2}^{\,(3)} \, = \, - \, \frac{551}{15}
\, + \, \frac{1673 \, \pi}{48}
\, - \, \frac{40 \, \pi}{\sqrt{3}} \, \, ;
\end{equation}
and for $d = 2$
\end{mathletters}
\begin{mathletters}
\label{II140}
\begin{equation} \label{II140a}
{\cal C}_{1}^{\,(2)} \, = \, - \, \frac{17}{18} \, + \, 
\frac{5 \, \pi}{4} \, - \, \frac{\sqrt{3} \, \pi}{2} \, \, ,
\end{equation}
\begin{equation} \label{II140b}
{\cal C}_{2}^{\,(2)} \, = \, - \, \frac{221}{15}
\, + \, \frac{1791 \, \pi}{128}
\, - \, \frac{43 \, \sqrt{3} \, \pi}{8} \, \, .
\end{equation}
 
\end{mathletters}


We now determine the surface tension $\Delta \sigma$ and the
curvature energies $\Delta \kappa_1$, $\Delta \kappa_2$, and $\Delta \kappa_G$ in
the expansion (\ref{I25}).
To this end we need to generalize this expansion
to be applicable to ($D - 1$)-dimensional surfaces of 
general shape with values of $D$ different from three. 
According to differential geometry for integer $D \ge 3$ 
the expansion has again the form (\ref{I25a}) and the 
corresponding curvatures are given by 
\cite{geometry}
\begin{mathletters}
\label{II160}
\begin{equation} \label{II160a}
K_m \, = \, \frac{1}{2} \, \sum_{i=1}^{D-1} \, \frac{1}{R_i}
\, = \, \, \frac{d-1}{2} \, \frac{1}{R}
\end{equation}
and 
\begin{equation} \label{II160b}
K_G \, = \, \sum_{pairs \atop i<j}^{D-1} \, \frac{1}{R_i R_j} \, = \, \,
\frac{(d-1)(d-2)}{2} \, \frac{1}{R^2}
\end{equation}
where $R_i$ are the $D - 1$ principal local radii of curvature.
The last expressions on the rhs of Eq.\,(\ref{II160}) apply 
to the surface of a generalized cylinder $K$ with integer 
$d \le D$. These expressions hold because the surface of $K$ has
$d-1$ finite local radii of curvature
$R_i = R$ which allow for $(d-1)(d-2)/2$ different pairings.
Note that for $D = 3$ Eq.\,(\ref{II160}) reduce to 
Eqs.\,(\ref{I25b}) and (\ref{I25c}).
Applying Eqs.\,(\ref{I25a}) and (\ref{II160}) to 
generalized cylinders $K$ in $D$ dimensions one infers from 
the definition (\ref{I20}) of $Y_{d,D}$ and Eq.\,(\ref{II20})
the general form
\end{mathletters}
\begin{eqnarray}
& & n_p \, k_B T \, R \, Q_{d,D\,}(\eta) \, = \, 
\Delta \sigma \, + \, \Delta \kappa_1 \, \frac{d-1}{2} \, \frac{1}{R} 
\label{II167}\\[1mm]
& & \quad + \, \Big[ \Delta \kappa_2 \, \frac{(d-1)^2}{4} \, + \, 
\Delta \kappa_G \, \frac{(d-1)(d-2)}{2} \Big] \frac{1}{R^2}
\, \, + \, {\cal O}( R^{-3} ) \nonumber
\end{eqnarray}
of $Q$ for small $\eta$. Explicit results for $\Delta \sigma$, $\Delta \kappa_1$,
$\Delta \kappa_2$, and $\Delta \kappa_G$ follow from the results
(\ref{II45}) - (\ref{II140})
for the coefficients ${\cal C}_{i}^{\,(d)}$ by comparing 
Eq.\,(\ref{II167}) with Eq.\,(\ref{II150}). 
Using $\eta = {\cal R}_{x}^{\,2} / (2 R^2)$ we find for the surface 
tension to first order in $\varepsilon = 4 - D$
\begin{eqnarray}
\Delta \sigma \, & = & \, n_p \, k_B T \, 
{\cal R}_{x} \, \sqrt{ \frac{2}{\pi} } \,
\Big\{ 1 - \frac{\varepsilon}{4} \, 
\Big[ 1 - \frac{3 \ln 2}{2} + {\cal C}_{0} \Big] \Big\}
\, \, + \, {\cal O}(\varepsilon^2) \label{II170}\\[1mm]
& \approx & \, n_p \, k_B T \, {\cal R}_{x} \, \,
0.798 \, (1 - 0.0508 \, \varepsilon) 
\, \, + \, {\cal O}(\varepsilon^2) \, \, . \nonumber
\end{eqnarray}
Here and in the rest of this subsection by taking  
$\varepsilon = 1$ one obtains the corresponding estimate 
for the physical dimension $D = 3$. By setting 
$d = 2$, $3$, and $D$ in Eq.\,(\ref{II167}), in which 
the generalization of $d$ to noninteger values is obvious,
we find for the curvature energies 
\begin{eqnarray}
\Delta \kappa_1 \, & = & \, n_p \, k_B T \, \frac{{\cal R}_{x}^{\,2}}{2} \, 
\Big\{ 1 - \frac{\varepsilon}{2} \, {\cal C}_{1}^{\,(2)} \Big\}
\, \, + \, {\cal O}(\varepsilon^2) \label{II180}\\[1mm]
& \approx & \, n_p \, k_B T \, {\cal R}_{x}^{\,2} \, \, 
0.5 \, (1 - 0.131 \, \varepsilon) 
\, \, + \, {\cal O}(\varepsilon^2) \, \, , \nonumber
\end{eqnarray}
\begin{eqnarray}
\Delta \kappa_2 \, & = & \, - \, n_p \, k_B T \, 
\frac{ {\cal R}_{x}^{\,3}}{3 \, \sqrt{2 \pi}} \,
\Big\{ 1 - \frac{\varepsilon}{4} \Big[
\Big( \frac{11}{6} - \frac{5 \ln 2}{2} \Big)
- 8 \, {\cal C}_{2}^{\,(2)} \Big] \Big\} 
\, \, + \, {\cal O}(\varepsilon^2) \label{II190}\\[1mm]
& \approx & \, - \, n_p \, k_B T \, {\cal R}_{x}^{\,3} \, \,
0.133 \, \left(1 - 0.0713 \, \varepsilon \right) 
\, \, + \, {\cal O}(\varepsilon^2) \nonumber \, \, ,
\end{eqnarray}
and finally 
\begin{eqnarray}
\Delta \kappa_G \, & = & \, - \, \Delta \kappa_2 \, - \, \varepsilon
\, n_p \, k_B T \, 
\frac{ {\cal R}_{x}^{\,3}}{3 \, \sqrt{2 \pi}} \,
\frac{{\cal C}_{2}^{\,(3)}}{2} 
\, \, + \, {\cal O}(\varepsilon^2) \label{II200}\\[1mm]
& \approx & \, n_p \, k_B T \, {\cal R}_{x}^{\,3} \, \,
0.133 \, (1 - 0.177 \, \varepsilon) 
\, \, + \, {\cal O}(\varepsilon^2) \nonumber \, \, .
\end{eqnarray}
Note that $\Delta \kappa_1$ is fixed by considering only 
one of the cases $d = 2$, $3$, and $D$ 
(we chose $d = 2$ in Eq.\,(\ref{II180})).
However, since $\Delta \kappa_1$ must not depend 
on the value of $d$ one derives the two conditions
\begin{mathletters}
\label{II210}
\begin{equation} \label{II210a} 
{\cal C}_{1}^{\,(2)} \, = \, 
\frac{{\cal C}_{1}^{\,(3)}}{2} \, = \,
\frac{{\cal C}_{1}^{\,(4)}}{3} 
\end{equation}
which must be fulfilled if the expansion (\ref{I25}) is
consistent up to one loop order in the EV interaction
of the polymer chains. 
Similarly, $\Delta \kappa_2$ and $\Delta \kappa_G$ are fixed by 
considering only two of the cases $d = 2$, $3$, and $D$ 
(we chose $d = 2$ and $3$ 
in Eqs.\,(\ref{II190}) and (\ref{II200})).
Thus one derives the third condition
\begin{equation} \label{II210b}
{\cal C}_{2}^{\,(4)} \, = \, 
3 \Big[ {\cal C}_{2}^{\,(3)} - {\cal C}_{2}^{\,(2)} \Big] \, .
\end{equation}
By using the values 
of ${\cal C}_{1}^{\,(d)}$ and ${\cal C}_{2}^{\,(d)}$ as 
derived in the cases (a), (b), (c) above one finds that all three 
conditions (\ref{II210}) are indeed fulfilled. This confirms 
to first order in $\varepsilon$ the assumption preceding 
Eq.\,(\ref{II40}) that the scaling function 
$Y_{d,D\,}(x)$ is analytic at $x = 0$ and that the 
Helfrich-type expansion (\ref{I25}) is applicable
to the present polymer depletion problem for chains with EV 
interaction. Considering the involved 
analytical means which were necessary to derive the
coefficients ${\cal C}_{i}^{\,(d)}$ (see Appendix \ref{neuA})
we regard this as a 
very valuable and important check of our calculation 
and in addition as a 
strong evidence that the above statements
for $Y_{d,D\,}(x)$ are general properties in $D = 3$ which 
hold beyond the present perturbative treatment. 

\end{mathletters}

Note that the EV interaction 
of the polymer chains reduces the absolute values of the
surface tension $\Delta \sigma$ and of the curvature energies 
$\Delta \kappa_1$, $\Delta \kappa_2$, and $\Delta \kappa_G$ as compared to 
ideal chains. This trend can be anticipated because
the EV interaction of the chain monomers effectively
{\em reduces\/} the depletion effect of the particle surface 
(compare, e.g., Ref.\,\cite{eisen}). 
However, the corresponding corrections are relatively small
so that the overall behavior is changed only quantitatively. 
Thus exposing one side of a flexible membrane to a solution of
polymers which are depleted by the membrane
favors a bending of the membrane surface towards 
the solution \cite{nanoparticles}
and leads to a weakening of its surface rigidity. 
The sign of the Gaussian curvature energy $\Delta \kappa_G$ will 
generally favor surfaces with higher genuses
(see the Introduction and I). If the resolution of 
an experimental setup is high enough to observe these effects
quantitatively, the corrections due to the presence of 
the EV interaction of the polymer chains as compared to the 
behavior for ideal chains should be detectable.
Specificly we consider the experiments for
vesicles reported by D\"obereiner et al.\,\cite{dobereiner}.
The intrinsic spontaneous curvature energy $\kappa_1$
of the bilayer membrane is to be identified with their
quantity $-2 \kappa \bar{c}_0 / R_A$ 
(compare Eq.\,(9) in Ref.\,\cite{dobereiner}).
The difference $\Delta \kappa_1$ (see Eq.\,(\ref{II180}))
should be added in the presence of polymers in the solution.
The length $R_A$ is of the order of the size of the vesicle.
Upon inserting the values 
$\kappa \approx 10^{-19}\,\mbox{J}$ and $\bar{c}_0 \approx 10$ 
(compare Fig.\,9 in Ref.\,\cite{dobereiner}) one infers
$\kappa_{1} R_A \approx - 2 \times 10^{-18}\,\mbox{J}$. 
On the other hand, 
for $T = 300 \, {\mbox K}$ and $n_p {\cal R}_x^{\,3}$ of 
order unity, which means that the polymer solution is
still in the dilute regime so that the result 
(\ref{II180}) is valid, one has
$\Delta \kappa_{1} {\cal R}_x \approx 2 \times 10^{-21}\,\mbox{J}$. 
The size ratio ${\cal R}_x / R_A$ is of the order of 
$1 / 100 \ll 1$ for realistic values 
$R_A \approx 10 \, \mu{\mbox m}$ and 
${\cal R}_x \approx 0.1 \, \mu{\mbox m}$. 
We conclude that $\Delta \kappa_1$ can reach a value
up to about $10\,\%$ of $\kappa_1$ in a 
quantitatively controllable way. This can be expected
to lead to observable effects near a shape transition of the 
vesicle.

Ideal chains lead to the behavior that all contributions in 
curly brackets on the rhs of the expansion (\ref{I25a})
of second and higher order in the curvature {\em vanish\/}
for the case of a generalized cylinder with $d = 3$ and 
$D \ge 3$ arbitrary (compare, e.g., Eqs.\,(3.9) and (3.11) in I). 
This encompasses, in particular, the three-dimensional sphere 
for which $d = D = 3$. For the contribution of second order
in the curvature the reason is a combination of the general
property $K_G = K_{m}^{\,2}$ for $d = 3$ (compare Eq.\,(\ref{II160}))
with the property $\Delta \kappa_G = - \Delta \kappa_2$ valid for any
dimension $D$ if the chains are ideal. However, the last property is
rather special and is violated for polymers with EV interaction 
in $D$ slightly below $4$ since Eq.\,(\ref{II200}) implies 
\begin{eqnarray} 
\Delta \kappa_2 \, + \, \Delta \kappa_G \, & = & \,
- \, \varepsilon \, n_p \, k_B T \, 
\frac{{\cal R}_{x}^{\,3}}{3 \, \sqrt{2 \pi}} \,
\frac{{\cal C}_{2}^{\,(3)}}{2} 
\, \, + \, {\cal O}(\varepsilon^2) \label{II215}\\[1mm]
& \approx & \,
- \, \varepsilon \, n_p \, k_B T \, {\cal R}_{x}^{\,3}
\, \, 0.0141 \, + \, {\cal O}(\varepsilon^2) \, \, . \nonumber
\end{eqnarray}
There is no reason to believe that this violation 
is removed in $D=3$. Rather the crossover to a behavior 
$Q_{3,\,3} \sim ({\cal R}_x / R)^{1 / \nu}$ for 
${\cal R}_x / R \to \infty$ 
with the Flory exponent $\nu \approx 0.588$ 
(see Eqs.\,(\ref{I30}), (\ref{II20}), and Sec.\,\ref{secIIB}) 
implies infinitely many nonvanishing terms in the 
small curvature expansion (\ref{I25a}) in $D=3$. 
Thus in the physically important 
case of the three-dimensional sphere the appearance of the 
EV interaction does lead to a {\em qualitative\/} change. 

As an illustration, consider a spherical membrane in the 
dilute polymer solution with {\em both\/} sides of the membrane 
exposed to the polymers. In this case the contributions 
to $\Delta \kappa_1 K_m$ in the expansion (\ref{I25a}) from each side 
cancel and Eq.\,(\ref{II215}) implies that for chains with EV 
interaction the free energy cost for immersing the spherical 
membrane is {\em smaller\/} as compared to a flat membrane with
the same area. This is different from the behavior for ideal 
chains for which the solvation free energies for a spherical 
and a flat membrane with same area are equal in this case.


\subsection{Long chains: $Y_{d,D}(x)$ for $x \to \infty$}
\label{secIIB}

Figure \ref{fig2} shows in the $(d,D)$-plane the dashed
line $d = 1 / \nu(D)$ \cite{shape}.
It separates generalized cylinders $K$ 
which are relevant perturbations for long polymer chains with 
EV interaction (such as the strip in $D = 2$ or the plate in 
$D = 3$) from those which are irrelevant and for which 
Eq.\,(\ref{I60}) applies. 
The latter are located in the shaded region above the line 
and comprise the disc in $D = 2$ 
and the sphere and the cylinder in $D = 3$ and are of main 
interest here. For the sphere and the cylinder in $D = 3$ 
we show within an expansion in $\varepsilon = 4 - D$ 
that the first order result for $Y_{d,D}(x)$ given
by Eqs.\,(\ref{A250}) and (\ref{II20}) is consistent with the
expected power law (\ref{I30}) and we determine the corresponding 
universal amplitude $A_{d,D}$ to first order in $\varepsilon$ for 
$d = D$, $3$, and $2$.
These results in conjunction with the known value 
for $A_{2,2}$ in $D = 2$  are used in order to derive improved 
estimates for $A_{3,3}$ and $A_{2,3}$ corresponding to a sphere 
and a cylinder, respectively, in $D = 3$.

The line $d = 1 / \nu(D)$ itself corresponds to marginal 
perturbations leading to a behavior which in general is
different \cite{log} 
from Eq.\,(\ref{I30}). We shall discuss neither this nor 
the crossover from marginal to power law behavior which may 
arise for points close above the line. Instead, in the case 
$d = 2$ and $D < 4$ we shall obtain the 
$\varepsilon$-expansion of $A_{2,D}$ by analytic continuation 
in $d$ from the corresponding value for $d > 2$. 

In the following we set $d$ to an arbitrary value with 
$2 < d \le D$. By inserting $Q_{d,D}(\eta)$ from Eq.\,(\ref{A250}) 
in Eq.\,(\ref{II20}) one finds
\begin{equation} \label{II300}
Y_{d,D}(x \to \infty) \, \to \, \Omega_d \left[ 2 \alpha \eta \, - \,
\frac{\varepsilon}{4} \, 
C_{d}(\eta) \right] \, \, + \, {\cal O}(\varepsilon^2) \, \, , 
\quad \eta = x^2 / 2 \, \, ,
\end{equation}

%
\unitlength1cm
\begin{figure}[t]
\begin{picture}(16,10)
\put(0,0.5){
\setlength{\epsfysize}{9.5cm}
\epsfbox{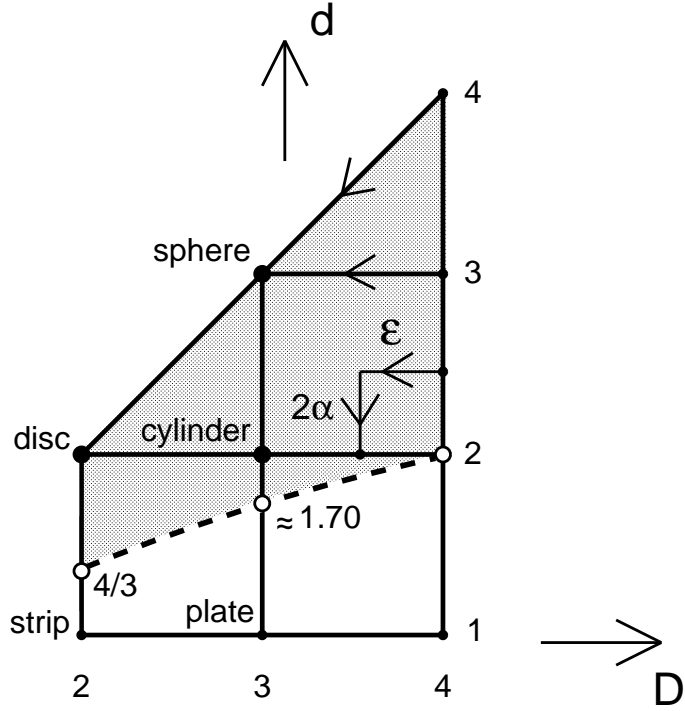}}
\end{picture}
\caption{Diagram of generalized cylinders $K$ which behave 
$-$ in the renormalization group sense $-$
as relevant or irrelevant perturbations 
for nonadsorbing polymers.
The parameter $d \le D$ characterizes the shape of $K$
and $D$ is the spatial dimension (see Eq.\,(\ref{I10})).
The point $(d,D) = (2,2)$ corresponds to a disc in $D=2$ and 
the points $(3,3)$ and $(2,3)$ to a sphere and an 
infinitely elongated cylinder in $D=3$, respectively. 
The line with $D = D_{uc} = 4$ and arbitrary $d$ represents
the upper critical dimension where the polymers behave like 
ideal chains and from which the perturbative expansion in
$\varepsilon = 4 - D$ starts in order to study the effects of 
the EV interaction. The open circles indicate points $(d,D)$
for which $d - 1 / \nu(D) = 0$. These points are
connected by the dashed line so that within the shaded region
{\em above\/} it the power law (\ref{I30}) applies
and $K$ represents an irrelevant perturbation. The paths 
indicated by the arrows are discussed in the main text.}
\label{fig2}
\end{figure}
%
%
\noindent
where $\alpha = (d - 2) / 2 > 0$.
The first term in square brackets stems from $P_{d}^{\,[0]}(\eta)$
in Eq.\,(\ref{A250b}) and $C_{d}(\eta)$ is given by Eq.\,(\ref{A260}).
Both the term $\Omega_d / d$ on the rhs of Eq.\,(\ref{II20})
and the sum of the terms following $- C_{d}(\eta)$ on the rhs of
Eq.\,(\ref{A250c}) are subdominant to the leading behavior
in Eq.\,(\ref{II300}). According to Appendix \ref{neuA}
this leads to
\begin{eqnarray}
& & Y_{d,D}(x \to \infty) \, \to \, \Omega_d \, 2 \alpha \eta 
\Big[ 1 - \frac{\varepsilon}{4} 
\Big( E_d +  \frac{\ln \eta}{2} \Big) \Big] 
\, \, + \, {\cal O}(\varepsilon^2) \label{II440}\\[2mm]
& & = \, \Omega_d \, 2 \alpha
\Big[ 1 - \frac{\varepsilon}{4} \, E_d \Big] \, 
\eta \Big[ 1 - \frac{\varepsilon}{4} \, \frac{\ln \eta}{2} \Big]
\, \, + \, {\cal O}(\varepsilon^2) \, \, ,
\quad \eta = x^2 / 2 \, \, . \nonumber
\end{eqnarray}
The constant $E_d$ is given by
\begin{equation} \label{II450}
E_d \, = \, - \, \frac{4 \pi^2}{\alpha} \, {\cal B}_d \, - \,
\frac{3}{2} \, + \, \ln 2 \, + \, \frac{\Psi(d/2)}{2} \, \, ,
\end{equation}
where for $d = D$ we have to consider $E_4$ only. 
The corresponding numbers ${\cal B}_d$ are:
\begin{equation} \label{II411}
{\cal B}_4 \, = \, 0 \, \, , \quad
{\cal B}_3 \, \approx \, 0.01047 \, \, , \quad 
{\cal B}_2 \, = \, \frac{1}{8 \pi^2} \, \, .
\end{equation}
The result in 
the second line of Eq.\,(\ref{II440}) for the behavior of
$Y_{d,D}(x \to \infty)$ is consistent with the power law 
(\ref{I30}) since 
\begin{equation} \label{II460}
\eta^{1/(2 \nu)} \, = \, 2^{\, - 1/(2 \nu)} \, x^{1/\nu} \, = \,
\eta \Big[ 1 - \frac{\varepsilon}{4} \, \frac{\ln \eta}{2} \Big]
\, \, + \, {\cal O}(\varepsilon^2)
\end{equation}
(see Eq.\,(\ref{A150}) for ${\cal N} = 0$).
The universal amplitude $A_{d,D}$
is determined by Eqs.\,(\ref{II440}) and (\ref{II450})
to first order in $\varepsilon = 4 - D$ with the results
\begin{mathletters}
\label{II471} 
\begin{eqnarray}
A_{D,D} \, & = & \, 2 \pi^2 \left\{
1 + \frac{\varepsilon}{4} 
\left[ 1 - 2 \ln \pi - \frac{\ln 2}{2} - \frac{3 \, C_E}{2} \right]
\right\} \, \, + \, {\cal O}(\varepsilon^2) \label{II471a}\\[2mm]
& \approx & 19.739 \, \left(1 - 0.625 \, \varepsilon \right)
\, \, + \, {\cal O}(\varepsilon^2) \, \, , \nonumber
\end{eqnarray}
\begin{eqnarray}
A_{3,D} \, & = & \, 2 \pi \left\{
1 + \frac{\varepsilon}{4} 
\left[8 \pi^2 {\cal B}_{3} +  
\frac{1}{2} + \frac{\ln 2}{2} + \frac{C_E}{2} \right]
\right\} \, \, + \, {\cal O}(\varepsilon^2) \label{II471b}\\[2mm]
& \approx & 6.283 \, \left(1 + \, 0.490 \, \varepsilon \right)
\, \, + \, {\cal O}(\varepsilon^2) \, \, , \nonumber
\end{eqnarray}
\begin{equation} \label{II471c}
A_{2,D} \, = \, \varepsilon \, 2 \pi^3 {\cal B}_{2}
\, + \, {\cal O}(\varepsilon^2) \, \approx \,
0.785 \, \varepsilon \, + \, {\cal O}(\varepsilon^2) \, \, ,
\end{equation}
where Eq.\,(\ref{II411}) has been used.

\end{mathletters}

From Eq.\,(\ref{II471c}) it is evident that 
$A_{2,D}$ {\em vanishes\/} in the limit
$D \nearrow 4$ which reflects the fact that 
for ideal chains, for which $1 / \nu = 2$ and 
the condition (\ref{I35}) is violated, the power law (\ref{I30})
does not apply \cite{log}. 
However, we succeeded in calculating
the amplitude $A_{2,D}$ for $D < 4$ to first
order in $\varepsilon = 4 - D$ by following a path in the
$(d,D)$-plane which circumvents the point $(2,4)$ 
as indicated by arrows in Fig.\,\ref{fig2} and along which
the power law (\ref{I30}) {\em does\/} apply with a positive 
amplitude $A_{d,D}$.
Accordingly, first one has to exponentiate
Eq.\,(\ref{II440}) with respect to $\varepsilon$ 
for $\alpha > 0$ fixed
in order to obtain the power law (\ref{I30}), 
and then one has to perform 
the limit $d - 2 = 2 \alpha \searrow 0$ for the resulting 
amplitude $A_{d,D}$ for $D = 4 - \varepsilon$ fixed.

We note that the values for $A_{3,3}$ and $A_{2,3}$
which follow from Eqs.\,(\ref{II471}) by setting 
$\varepsilon = 1$ are estimates which depend on the
path taken. For $\varepsilon = 1$, e.g., 
Eqs.\,(\ref{II471a}) and (\ref{II471b}) lead 
to the different estimates
$7.39$ and $9.36$, respectively, for the same quantity $A_{3,3}$
(the corresponding paths in the $(d,D)$-plane are indicated by the 
two upper arrows in Fig.\,\ref{fig2}). This discrepancy is caused
by the present perturbative calculation of $A_{d,D}$.

This unpleasant feature can be cured. 
As mentioned in Sec.\,\ref{secIB} the power law (\ref{I30}) is 
a special consequence of the small radius expansion (SRE)
in Eq.\,(\ref{I60}). Via the polymer magnet analogy this operator
expansion is related to a corresponding SRE in a field theory.
This allows one to understand not only the mechanism behind the SRE in
terms of perturbative field theoretic methods for $D$ 
slightly below $4$ (as demonstrated in Appendix \ref{appB}) but also
to use nonperturbative methods for $D = 2$ \cite{methods}
which incorporate the result $A_{2,2} = 3.81$
(see the end of Appendix \ref{appB}).
Improved estimates for the amplitudes $A_{3,3}$ and $A_{2,3}$
can be deduced
by combining the $\varepsilon$-expansion of $A_{d,D}$
in Eq.\,(\ref{II471}) with the above value for $A_{2,2}$. 
To this end we assume that $A_{d,D}$ is a smooth function 
of $d$ and $D$.
We consider the following interpolation 
schemes \cite{floter} for the functions 
$f(\varepsilon = 4 - D) = A_{D,D}$, $A_{2,D}$, $A_{D-1,D}$, and
$A_{6-D,D}$, which appear as curves in the $A_{d,D\,}$-surface
shown in Fig.\,\ref{fig_shell}:

(a) pure polynomial:

\begin{mathletters} \label{III10}

\begin{equation} \label{III10a}
f(\varepsilon) \, = \, f_a(\varepsilon) \, \equiv \,
f(0) \, + \, a_1 \, \varepsilon 
\, + \, a_2 \, \varepsilon^2 \, \, ,
\end{equation} 

(b) (1,1) - Pad\'{e} form: 

\begin{equation} \label{III10b}
f(\varepsilon) \, = \, f_b(\varepsilon) \, \equiv \, f(0) \, + \, 
\frac{b_1 \, \varepsilon}{1 + b_2 \, \varepsilon} \, \, \, .
\end{equation} 

\end{mathletters}

For $A_{D,D}$ and $A_{2,D}$ the coefficients on the rhs
of Eqs.\,(\ref{III10}) are fixed by Eqs.\,(\ref{II471a}) 
and (\ref{II471c}), respectively, in conjunction with 
$f(2) = A_{2,2}$. Note that the corresponding paths in 
the $(d,D)$-plane are straight lines, i.e., in particular
{\em smooth\/} paths, so that $A_{d,D}$ behaves smoothly
as function of $\varepsilon$ along these paths 
(see Fig.\,\ref{fig_shell}). We obtain estimates 
for $A_{3,3}$ and $A_{2,3}$ by the corresponding mean values 
$f_m(\varepsilon) = (f_a(\varepsilon) + f_b(\varepsilon)) / 2$
for $\varepsilon = 1$ and use the difference between the two
values $f_a(1)$ and $f_b(1)$ as an estimate for the error. 
For the sphere this leads to 
\begin{equation} \label{III20}
A_{3,3} \, = \, 9.82 \pm 0.3
\end{equation}
and for the cylinder to
\begin{equation} \label{III30}
A_{2,3} \, = \, 1.23 \pm 0.2 \, \, .
\end{equation}

\newpage

%
\unitlength1cm
\begin{figure}[t]
\begin{picture}(11,13)
\put(0,0.5){
\setlength{\epsfysize}{12.5cm}
\epsfbox{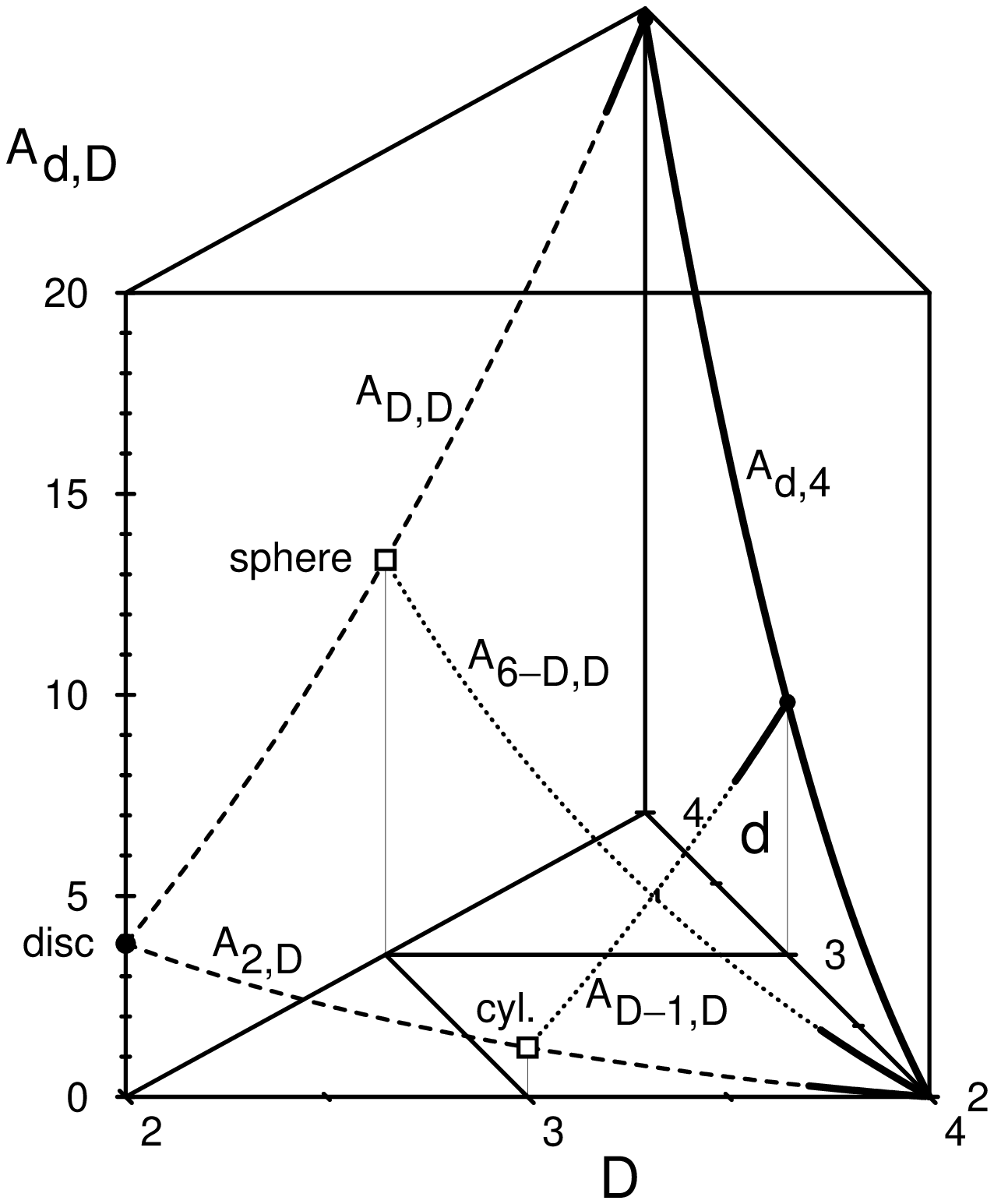}}
\end{picture}
\bigskip
\medskip
\caption{The universal amplitude $A_{d,D\,}$ corresponds to a 
two dimensional surface over the base plane $(d,D)$
(compare Fig\,\ref{fig2}).
The full dot corresponding to $A_{2,2}$  
and the thick solid lines represent the 
known parts of this surface.
The solid parts of the dashed and of the dotted lines indicate
the slopes of these lines at their end points
$D = 4$ according to Eq.\,(\ref{II471}).
The dashed lines themselves including
the desired estimates for $A_{3,3}$ and $A_{2,3}$ 
(open squares) display the corresponding mean values 
$f_m(\varepsilon) = (f_a(\varepsilon) + f_b(\varepsilon)) / 2$
of the two interpolation schemes described in
Eq.\,(\ref{III10}). The same holds for the dotted lines 
and for the two values of $A_{d,D}$ for
$(d,D) = (2.5, 3.5)$, which have been calculated for
a self-consistency check. These two values are connected by the 
short full line in order to indicate the deviation
caused by the fact that the two dotted lines miss each other 
slightly (for the exact surface $A_{d,D}$, of course, 
the two dotted lines do intersect at this point). The 
smallness of the deviation underscores the reliability 
of the interpolation scheme.}
\label{fig_shell}
\end{figure}
%
%
\newpage
\noindent
So far the $\varepsilon$-expansion of $A_{3,D}$ in Eq.\,(\ref{II471b})
has not been used. Now it can serve as a check for the reliability of the 
interpolation method leading to Eqs.\,(\ref{III20}) and (\ref{III30}).
In combination with the known curve
$A_{d,\,4} = 2 \pi^{d/2} / \Gamma((d-2)/2)$ 
Eq.\,(\ref{II471b}) determines
the plane tangent to the $A_{d,D\,}$-surface at $(d,D) = (3,4)$
which leads together with the value for $A_{2,3}$ in 
Eq.\,(\ref{III30}) to approximations of the form
(\ref{III10}) for the curve $A_{D-1,D}$. Corresponding
approximations for the curve $A_{6-D,D}$ follow from the 
known tangent plane at $(d,D) = (2,4)$ and the value for
$A_{3,3}$ in Eq.\,(\ref{III20}). The resulting mean values 
$f_{m}(\varepsilon)$ are shown as dotted 
lines in Fig.\,\ref{fig_shell}. A satisfactory 
self-consistency check for the accuracy is provided by the
observation that at the particular point $(d,D) = (2.5, 3.5)$ 
at which the two exact dotted lines should cross the approximate
ones in Fig.\,\ref{fig_shell} are only slightly off by the small 
amount of $0.3$.


\subsection{The complete scaling function $Y_{d,D}(x)$}
\label{secIIC}

The full scaling function $Y_{d,D}(x)$ 
describes the crossover between 
its analytic behavior for $x = {\cal R}_x / R \to 0$ and 
the power law (\ref{I30}) for $x \to \infty$ which have been 
discussed in 
Secs.\,{\ref{secIIA} and {\ref{secIIB}, respectively.
Here we present estimates for the complete functions 
$Y_{3,3}(x)$ and $Y_{2,3}(x)$ corresponding to a sphere
and a cylinder, respectively, in $D = 3$. The global 
behavior of $Y_{d,D}(x)$ is conveniently characterized 
in terms of the function
\begin{equation} \label{II500}
\Theta_{d,D}(x) \, = \, \frac{1}{x} \, 
\Big[ Y_{d,D}(x) - \frac{\Omega_d}{d} \Big] \, = \,
\Omega_d \, \frac{Q_{d,D}(\eta)}{x} \, \, , 
\quad \eta = x^2 / 2 \, \, ,
\end{equation}
where $Q_{d,D}(\eta)$ is defined in Eq.\,(\ref{A240}).
According to Eq.\,(\ref{II167}) the value $\Theta_{d,D}(0)$ is 
related to the surface tension $\Delta \sigma$ in the Helfrich-type 
expansion (\ref{I25}) and the first and second derivatives 
of $\Theta_{d,D}(x)$ at $x = 0$ are related to the corresponding first 
and second order curvature contributions, respectively 
(compare Sec.\,\ref{secIIA}). In the opposite limit 
$x \to \infty$ the function $\Theta_{d,D}(x)$ exhibits 
the power law
\begin{equation} \label{II520}
\Theta_{d,D}(x \to \infty) \, \to \, 
A_{d,D} \, x^{1/\nu \, - \, 1}
\end{equation}
as implied by Eqs.\,(\ref{II500}) and (\ref{I30}).

We derive estimates for $\Theta_{d,3}(x)$ for
$d = 3$ and $2$ by a combination of (a) an appropriate
exponentiation of the one loop order result for 
$\Theta_{d,D}(x)$ for large $x$ with (b) an interpolation to
the regular behavior of $\Theta_{d,D}(x)$ for small $x$.
The exponentiation is necessary because 
the actual forms of Eqs.\,(\ref{A250}) and (\ref{II500})
do not exhibit the 
power law (\ref{II520}) (compare Sec.\,\ref{secIIB}). 
By construction our estimates incorporate the improved
estimates for $A_{3,3}$ and $A_{2,3}$ as given by 
Eqs.\,(\ref{III20}) and (\ref{III30}) in conjunction with
the best available value $\nu(D=3) \approx 0.588$ [39(b)]
for the power law (\ref{II520}).
For $x \to 0$ they reproduce the regular behavior
as implied by Eqs.\,(\ref{II167}) - (\ref{II200}) with 
$\varepsilon = 4 - D$ set to one. 

(a) {\em Exponentiation\/}: There are numerous possibilities 
to add on the rhs of Eq.\,(\ref{A250a}) higher order terms 
in $\varepsilon$ such that the power law (\ref{II520}) is 
reproduced. For the sphere we choose to consider the 
path $d = D$ (compare the derivation of Eq.\,(\ref{III20}))
and define
\begin{equation} \label{II550}
\Theta_{D,D}^{\,(\infty)}(x) \, = \,
(1 + 0.179 \, \varepsilon^2) \, \Omega_D \, 
\frac{W(\eta, \varepsilon)}{x} \, 
\exp\Big[ \Big( 2 - \frac{1}{\nu} \Big) \,
\frac{P_{4}^{\,[1]}(\eta)}{P_{4}^{\,[0]}(\eta)} \, \Big]
\end{equation}
with $P_{4}^{\,[0]}$ and $P_{4}^{\,[1]}$ from Eq.\,(\ref{A250}).
The superscript $\infty$ refers to the behavior for $x \to \infty$.
Here 
\begin{equation} \label{II560}
W(\eta, \varepsilon) \, = \,
P_{4}^{\,[0]}(\eta) + \varepsilon \,
\Big[ - \eta + 
{\cal L}_{\tau \to \eta} \Big\{ \frac{1}{2 \tau^2} \,
\frac{[K_{0}(\sqrt{\tau})]^2}
{[K_{1}(\sqrt{\tau})]^2} \Big\} \Big] 
\end{equation}
is the expansion of $P_{D}^{\,[0]}(\eta)$
to first order in $\varepsilon$. The rhs of 
Eq.\,(\ref{II550}) is consistent with Eq.\,(\ref{A250}) 
in conjunction with Eq.\,(\ref{II500})
and leads to the power law (\ref{II520}). The constant 
$0.179 \, \varepsilon^2$ has been introduced in order to 
incorporate for $\varepsilon = 1$ the improved estimate 
$A_{3,3} \approx 9.82$ given by Eq.\,(\ref{III20}).
For the cylinder a proper exponentiation is more 
involved than for the sphere (compare Sec.\,\ref{secIIB}).
In this case we consider
\begin{eqnarray}
& & \Theta_{2,D}^{\,(\infty)}(x) \, = \, \frac{\Omega_2}{x} \,
\Bigg[
(1 + 0.412 \, \varepsilon) \, \frac{\varepsilon}{4} \,
\eta \, \exp\Big[ \Big( 2 - \frac{1}{\nu} \Big) 
E_{2+\varepsilon} \, \Big] \label{II566}\\
& & \quad + \, 
{\cal L}_{\tau \to \eta} \Big\{\frac{K_{1}(\sqrt{\tau})}
{\tau^{3/2} \, K_{\varepsilon/2}(\sqrt{\tau})} \Big\}
\Bigg] \,
\exp\Big[ \Big( 2 - \frac{1}{\nu} \Big) \,
\frac{P_{2+\varepsilon}^{\,[1]}(\eta)}
{P_{2+\varepsilon}^{\,[0]}(\eta)} \Big] \nonumber \, \, .
\end{eqnarray}
The constant $E_{d}$ is given by Eq.\,(\ref{II450}). 
The rhs of Eq.\,(\ref{II566}) is consistent with 
Eq.\,(\ref{A250}) in conjunction with Eq.\,(\ref{II500})
and leads to the power law (\ref{II520}).
The constant $0.412 \, \varepsilon$ has been 
introduced in order to incorporate the improved estimate 
$A_{2,3} \approx 1.23$ given by Eq.\,(\ref{III30}).
In order to carry out the exponentiation procedure 
described by Eq.\,(\ref{II440})
we use the properties \cite{abr}
$s K_{\alpha + 1}(s) = 2 \alpha K_{\alpha}(s) + s K_{\alpha - 1}(s)$
and $K_{\varepsilon} = K_0 + {\cal O}(\varepsilon^2)$
of the modified Bessel functions $K_{\alpha}$ and $K_{\alpha+1}$
which appear on the rhs of Eq.\,(\ref{A250b})
(compare the discussion after Eq.\,(\ref{II471c})).
For $\varepsilon = 1$ the quantities 
$E_{2+\varepsilon}$, $P_{2+\varepsilon}$,
and $K_{\varepsilon/2}$ in Eq.\,(\ref{II566})
correspond to $d = 3$.
This should be compared with Eq.\,(\ref{II550}) for the
sphere in which $P_4$ corresponds to $d = 4$.
Thus in a certain sense
the exponentiation procedure implied by Eq.\,(\ref{II566})
partially circumvents the point
$(d,D) = (2,4)$ in the $(d,D)$-plane
by connecting $(3,4)$ with $(2,3)$ on 
a straight line in the same way as Eq.\,(\ref{II550}) 
connects $(4,4)$ with $(3,3)$ (see Fig.\,\ref{fig2}).

(b) {\em Interpolation\/}: Since $\Theta_{d,D}(x \to 0)$ behaves 
regularly no exponentiation is needed in this limit. 
We are thus led to introduce a function 
$\Theta_{d,D}^{\,(0)}(x)$ by the first two 
contributions on the rhs of Eq.\,(\ref{A250a}) in 
conjunction with Eqs.\,(\ref{II500}).
For consistency with the discussion in 
Sec.\,\ref{secIIA} (compare Eq.\,(\ref{II167}))
for the sphere and the cylinder we consider
the function $\Theta_{d,D}$ for {\em fixed\/} 
$d=3$ and $d=2$, respectively. So far for both the sphere
and the cylinder we have constructed two functions: the one with
superscript $\infty$ describes well the limit $x \to \infty$ 
whereas the one with superscript $0$ describes well the limit
$x \to 0$. We smoothly interpolate between these
two functions using the switch function
\begin{equation} \label{II570}  
s(x) \, = \, \frac{1}{2} 
\Big[ \tanh\Big( \upsilon_1 x - 
\frac{\upsilon_2}{x} \Big) + 1 \Big]
\end{equation}
with $\upsilon_1, \upsilon_2 > 0$
so that $s(x \to 0) \to 0$ and $s(x \to \infty) \to 1$.
By using this switch function we constitute the estimate 
\begin{equation} \label{II580} 
\Theta_{d,3}(x) \, = \,
[1 - s(x)] \, \Theta_{d,3}^{\,(0)}(x) \, + \, 
s(x) \, \Theta_{d,3}^{\,(\infty)}(x)
\end{equation}
for $d = 3$ and $2$ in $D = 3$. The parameters 
$\upsilon_1, \upsilon_2$ in Eq.\,(\ref{II570}) 
should be adjusted such that $\Theta_{d,3}(x)$
behaves as smoothly as possible in the whole 
range of $x$. It turns out that within the 
corresponding region of $\upsilon_1, \upsilon_2$
the relative changes of 
$\Theta_{d,3}(x)$ are quite small. We shall use 
the values $\upsilon_1 = 0.1$ and $\upsilon_2 = 1.5$.
Note that the rhs of Eq.\,(\ref{II580}) reflects 
both the behavior $\Theta_{d,3}^{\,(0)}(x \to 0)$ 
and the behavior
$\Theta_{d,3}^{\,(\infty)}(x \to \infty)$ since $s(x)$
exhibits an essential singularity in both limits. 

The resulting functions $\Theta_{d,3}(x)$ for 
$d = 3$ and $2$ are shown in Fig.\,\ref{plot}. 
The functions which enter $\Theta_{d,3}^{\,(0)}(x)$
and $\Theta_{d,3}^{\,(\infty)}(x)$ in Eq.\,(\ref{II580})
have been derived in Sec.\,\ref{secIIvoran} 
and we have carried out the inverse Laplace
transforms in Eqs.\,(\ref{A250}) and (\ref{A260}) 
numerically (see also Table \ref{table} in Appendix \ref{neuA}). 
Figure \ref{plot} shows the behavior both for chains with 
EV interaction and for ideal chains.

It is evident that
the power law (\ref{II520}) does not only determine 
the asymptotic behavior of the scaling function 
for $x \to \infty$ but it also influences
the behavior down to values of $x$ 
of order unity. This implies that for a quantitative analysis 
it is indispensable to take the behavior (\ref{II520}) into 
account, in particular accurate values of the
amplitudes $A_{3,3}$ and $A_{2,3}$. 
For the cylinder and chains with EV interaction 
the approach towards the power law (\ref{II520}) is rather slow
which is consistent with the fact that in this case
the exponent $d - 1 / \nu \approx 0.30$ in Eq.\,(\ref{I60})
is positive but small.
Note that the functions $\Theta_{d,3}(x)$ exhibit
%
%
\unitlength1cm
\begin{figure}[t]
\begin{picture}(16,12)
\put(-0.5,0.5){
\setlength{\epsfysize}{11.5cm}
\epsfbox{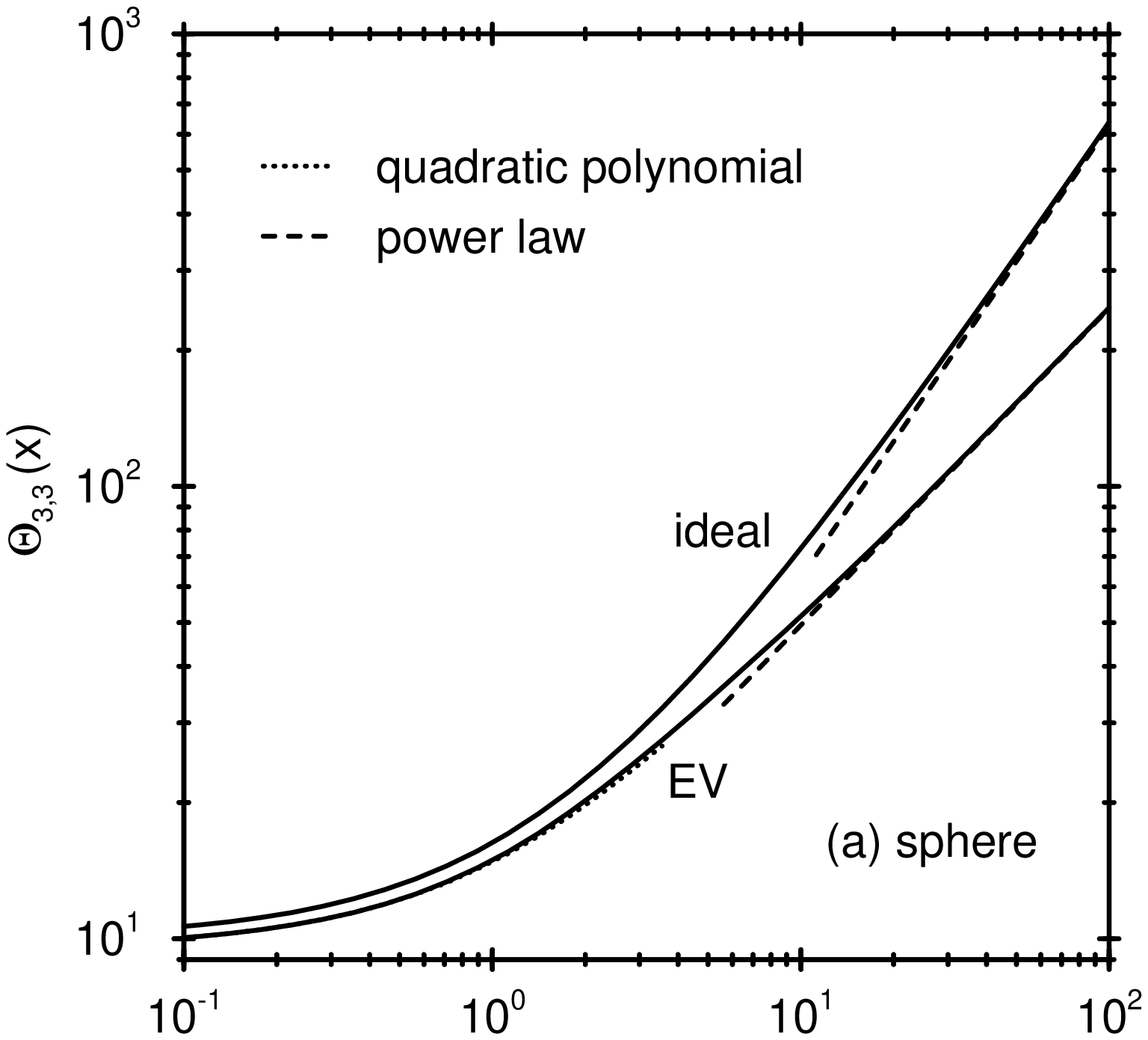}}
\put(-0.5,-11){
\setlength{\epsfysize}{11.5cm}
\epsfbox{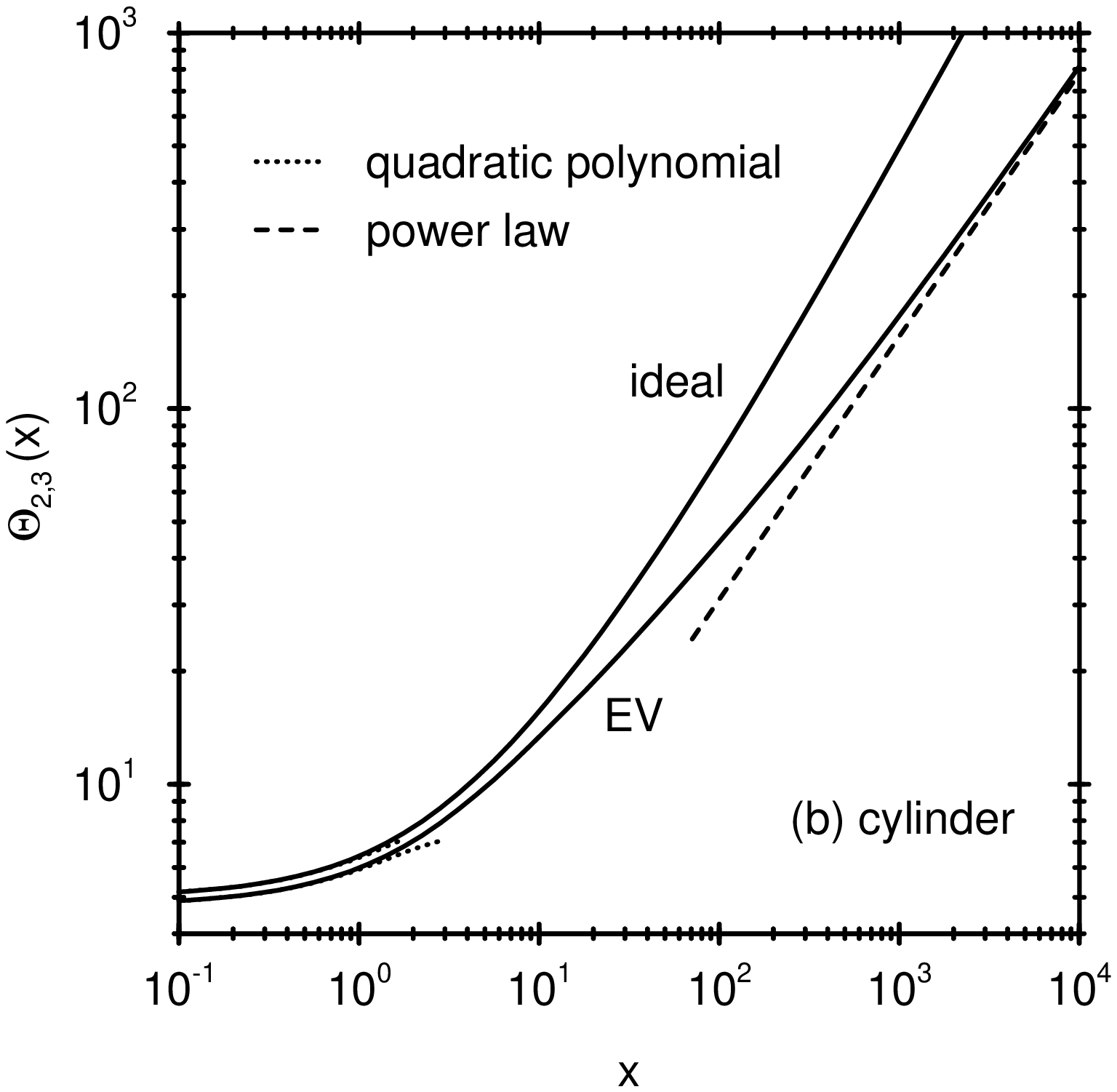}}
\end{picture}

\newpage

\caption{Scaling function $\Theta_{d,3}(x)$ (solid lines)
for (a) $d = 3$ and (b) $d = 2$ corresponding to a
sphere and a cylinder in $D = 3$, respectively
(see Eqs.\,(\ref{I20}) and (\ref{II500}), where
$\Omega_3 = 4 \pi$ and $\Omega_2 = 2 \pi$).
The lines labelled `EV' correspond to chains with
EV interaction and the lines labelled `ideal' to ideal chains.
The dotted lines display the quadratic polynomial in $x$
which characterizes the behavior $\Theta_{d,3}(x \to 0)$
(see Eqs.\,(\ref{II167}) and (\ref{II500})). 
For a sphere and ideal chains $\Theta_{3,3}(x)$ 
is simply a linear function of $x$
(compare Eqs.\,(3.9) and (3.11) in I).
The dashed lines display the power law (\ref{II520}).
For the curves corresponding to
chains with EV interaction the amplitudes
$A_{d,3}$ from Eqs.\,(\ref{III20}) and (\ref{III30}) 
and the value $\nu = 0.588$ have been incorporated.
For a cylinder exposed to ideal chains the scaling function 
$\Theta_{2,3}(x \to \infty)$ diverges as $x / \ln x$
instead of a pure power law.}
\label{plot}
\end{figure}
%

\medskip

\noindent
smaller values for 
chains with EV interaction than for ideal chains. This is 
consistent with the exponent $1 / \nu - 1 \approx 0.70$
for chains with EV interaction being smaller than the 
exponent $1 / \nu - 1 = 1$ for ideal chains. 
This difference in behavior is in accordance with the general
observation that the 
EV interaction effectively reduces the depletion effect of 
the immersed particle (compare the related discussion 
at the end of Sec.\,\ref{secIIA}).


\section{Depletion interaction between particles}
\label{secIII}

First, we consider the effective 
interaction between a thin rod and a planar wall confining 
the polymer solution. This is another example which 
demonstrates the importance of the qualitative difference between the 
behavior for ideal chains and chains with EV interaction 
which we have discussed in Sec.\,\ref{secIB}.
Then, we consider the effective interaction between two or
three small spherical particles in the unbounded solution.
When $R$ is small compared with ${\cal R}_x$ and the
distances between the particles, the small radius expansion 
(\ref{I60}) applies. On the other hand if both $R$ and some 
of these distances are small compared to ${\cal R}_x$
and the remaining distances, operator expansions slightly more 
complicated than Eq.\,(\ref{I60}) are expected to hold. In 
particular we shall consider a `small dumb-bell' expansion 
for two spheres. Finally, we compare our results with those
of the PHS model \cite{asakura}.

\subsection{Interaction of a thin rod with a planar wall}
\label{secIIIA}

In view of the depletion driven adsorption of colloidal rods
onto a hard wall \cite{sear} it is of interest to consider
a cylinder with radius $R$ and length $l$ immersed parallel to
and at a distance ${\cal D}$ of closest 
approach surface-to-surface from a planar wall $W$ in a 
dilute polymer solution (compare I). We consider the special case 
$R \ll {\cal D}, {\cal R}_x$ and ${\cal D}, {\cal R}_x \ll l$.
Using Eq.\,(4.19) in I we obtain the corresponding effective 
free energy of interaction in three dimensions,
\begin{equation} \label{III70}
\Delta F_{depl}({\cal D}) \, = \, - \, 
n_{p\,} k_B T  A_{2,3} \, l R^2 \,
({\cal R}_x / R)^{1/\nu} \, 
[ 1 - M_M^{(W)}({\cal D} / {\cal R}_x) ] \, \, ,
\end{equation}
with the number density $n_p$ of the polymers in the 
bulk solution and the bulk normalized density profile 
$M_M^{(W)}(z / {\cal R}_x)$ of chain monomers in the 
half-space (without the cylinder)
as function of the distance $z$ from the wall $W$. 
This universal density profile can be determined
experimentally, e.g., by neutron reflectivity \cite{pai} 
(compare also Fig.\,5 in I). Note that Eq.\,(\ref{III70}), 
in which the universal amplitude $A_{2,3}$ enters
(see Eq.\,(\ref{III30})), is only valid for chains with EV 
interaction. Equation (\ref{III70}) gives rise to an 
{\em attractive\/} interaction between the rod and the wall.
The rhs of Eq.\,(\ref{III70}) is fixed by well-defined
quantities and is independent of nonuniversal model 
parameters \cite{kardar}.

%
\unitlength1cm
\begin{figure}[t]
\begin{picture}(10,7)
\put(0.5,0.5){
\setlength{\epsfysize}{6cm}
\epsfbox{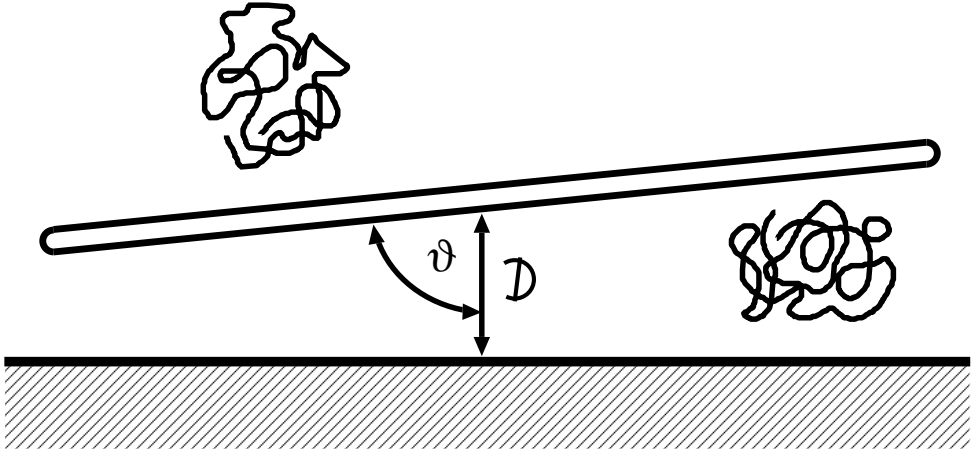}}
\end{picture}
\bigskip
\caption{Angular and positional coordinates $\vartheta$ and 
${\cal D}$ for a rod near a wall. The magnitudes 
$R \ll {\cal R}_x \ll l$ of length scales
for which Eq.\,(\ref{III100}) 
is expected to hold is shown schematically. As in 
Ref.\,[50] we assume that the rod has two caps with 
radius $R$ and that its cylindrical part has a length $l$.}
\label{fig_rod}
\end{figure}
%

Consider now a solution of rods whose number density is so low
that the interaction between the rods is negligible and
the rods behave like an ideal gas. In this case the density
of the rods near the wall is determined by the one-particle 
potential energy
\begin{equation} \label{III80}
U({\cal D}, \vartheta) \, = \, U_{W}({\cal D},\vartheta) \, + \, 
U_{depl}({\cal D},\vartheta) \, \, ,
\end{equation}
where $U_{W}$ equals infinity for rod configurations which overlap
with the wall $W$ and equals zero otherwise and $U_{depl}$
is the polymer-induced effective free energy of interaction. 
Both contributions depend on ${\cal D} = h - R$
with $h$ the distance between the center of the rod and the
wall and on the angle $\vartheta$ between the rod and the wall 
(compare Ref.\,\cite{sear} and Fig.\,\ref{fig_rod}). The density 
of rods $c({\cal D})$ with arbitrary orientations compatible 
with the presence of the wall is given by \cite{sear}
\begin{equation} \label{III90}
c({\cal D}) \, = \, c_{b} \, \frac{1}{2}
\int\limits_{-1}^{1} \, d (\cos \vartheta) \,
\exp[ - U({\cal D}, \vartheta) / k_B T ] \, \, ,
\end{equation}
where $c_{b}$ is the number density of the rods in the bulk solution.
Even for $R \ll {\cal R}_x \ll l$, in general one cannot
identify $U_{depl}$ on the rhs of Eq.\,(\ref{III80}) 
with $\Delta F_{depl}$ because Eq.\,(\ref{III70})  
holds only for a long, thin rod with 
$R \ll {\cal R}_{x}, {\cal D} \ll l$ and which is in 
addition {\em parallel\/} to the wall, i.e., 
$\vartheta = \pi / 2$. However, since the attractive
effective interaction $U_{depl}({\cal D},\vartheta)$ 
vanishes beyond a scale given by ${\cal R}_x$ one expects 
that $c({\cal D})$
will be largest in a region ${\cal D} \lesssim {\cal R}_x$
so that an approximation which is valid in this region
should be accurate in general \cite{sear}. 
Due to ${\cal R}_x \ll l$
the rods must be located almost parallel to the wall in 
this region so that it should be a good approximation 
to replace $U_{depl}({\cal D},\vartheta)$ by its 
value for $\vartheta = \pi / 2$, i.e., by 
$\Delta F_{depl}({\cal D})$ in Eq.\,(\ref{III70}). 
In this case the integration over $\vartheta$ in 
Eq.\,(\ref{III90}) can be carried out explicitly
\cite{sear} so that
\begin{equation} \label{III100}
c({\cal D}) \, \approx \, c_{0}({\cal D}) \,
\exp[ - \Delta F_{depl}({\cal D}) / k_B T ]
\end{equation}
with $c_{0}({\cal D}) = 2 c_{b\,} {\cal D} / l$ for
$0 \le {\cal D} \le l/2$ and $c_{0}({\cal D}) = c_{b}$ 
otherwise. Equation (\ref{III100}) is expected to be valid
provided $R \ll {\cal R}_{x}, {\cal D} \ll l$. Up to a 
certain degree of accuracy it is even possible to drop the 
restrictions on ${\cal D}$:  
since the prefactor $c_{0}({\cal D})$ on the rhs of 
Eq.\,(\ref{III100}) vanishes linearly with ${\cal D}$, the 
density of rods $c({\cal D})$ will be much smaller for 
${\cal D} \approx R$ than for the intermediate region
${\cal D} \approx {\cal R}_x$ where Eq.\,(\ref{III100}) applies.  
On the other hand, for ${\cal D} \gtrsim l$ the density of rods 
$c({\cal D})$ will be close to $c_{0}({\cal D})$ due to 
${\cal R}_x \ll l$, which is also consistent with 
Eq.\,(\ref{III100}). Therefore we expect that 
Eq.\,(\ref{III100}) is applicable even in the 
whole range of ${\cal D}$ provided 
$R \ll {\cal R}_{x} \ll l$. 
 

\subsection{Depletion interaction between spherical particles}
\label{secIIIB}

In Eqs.\,(\ref{I100}) - (\ref{I120})
the interaction between small spherical particles 
is expressed in terms of the universal
small sphere amplitude $A_{D,D}$ and the monomer density
correlation functions $C_m$ of a polymer chain in unbounded   
infinite space. Numerical values of the former 
for several spatial dimensions $D$ are summarized in 
Table\,\,\ref{table_amplitude}. For the latter we 
note the relations
\begin{equation} \label{IV10}
\int\limits_{{\mathbb R}^{D}}
d^{D} r_A \, C_2({\bf r}_A, {\bf r}_B) \, = \, 
{\cal R}_x^{\,2/\nu}
\end{equation}
and
\begin{equation} \label{IV20}
\int\limits_{{\mathbb R}^{D}}
d^{D} r_C \, C_3({\bf r}_A, {\bf r}_B,  {\bf r}_C) \, = \,
{\cal R}_x^{\,1/\nu} \, C_2({\bf r}_A, {\bf r}_B)
\end{equation}
which follow from the defining Eqs.\,(\ref{I70}) and (\ref{I110}).
Simple limiting behaviors arise if the relative distance 
$r_{AB} = |{\bf r}_A - {\bf r}_B|$ 
$-$ albeit being large on the microscopic scale $-$
is much smaller than other mesoscopic lengths.
For the pair correlation 
\cite{gennes2,cloizeaux,schafer,duplantier} this limiting
behavior takes the form
\begin{equation} \label{IV30}
C_2({\bf r}_A, {\bf r}_B) \, \to \, 
\sigma \, r_{AB}^{\,\,-(D - 1/\nu)} \, {\cal R}_x^{\,1/\nu}
\, \, , \quad r_{AB} \ll {\cal R}_x \, \, .
\end{equation}
For the triple correlation one finds
\begin{eqnarray}
C_3({\bf r}_A, {\bf r}_B,  {\bf r}_C) \, & \to & \,
\sigma \, r_{AB}^{\,\,-(D - 1/\nu)} \,
C_2\Big( \frac{{\bf r}_A + {\bf r}_B}{2}, {\bf r}_C \Big) 
\, \, , \label{IV40}\\[1mm]
& & r_{AB} \ll {\cal R}_{x\,}, 
|{\bf r}_C - ({\bf r}_A + {\bf r}_B)/ 2| \, \, . \nonumber
\end{eqnarray}
Here $\sigma$ is a universal bulk amplitude. 
For ideal chains $\sigma = \sigma^{(id)}$ is only 
defined for spatial dimensions $D > 2$ for which
\begin{equation} \label{IV50}
\sigma^{(id)} \, = \, \pi^{-D/2} \,
\Gamma\Big(\frac{D}{2} - 1 \Big) \, \, .
\end{equation}
For chains with EV interaction, however, $\sigma$ remains 
finite down to $D = 1$. Numerical values of $\sigma$ for 
several $D$ are summarized in Table\,\,\ref{table_amplitude}. 
In Appendix\,\ref{appC} we show how these values can be obtained.

\newpage

%
\begin{minipage}[t]{16cm}
\begin{table}[tbp]
\caption{Numerical values of the small sphere amplitude $A_{D,D}$
and of the short-distance amplitude $\sigma$ for chains with
EV interaction and for ideal chains.}
\label{table_amplitude}
\medskip
\begin{tabular}{ccccc} 
$D$ & $4$  &  $3$  &  $2$  &  $1$ \\ \hline
$A_{D,D}$       &  $19.739$  &  $9.82 \pm 0.3$  &  $3.810$  & (marginal) \\
$A_D^{\,(id)}$  &  $19.739$  &  $6.283$         &  $0$ \, (marginal)
& $-$ \\ \hline \hline
$\sigma$ &   $0.101$  &  $0.13$  &  $0.278$  & $1$ \\
$\sigma^{(id)}$ &   $0.101$  &  $0.318$ &  $\infty$ \, (marginal) & $-$
\end{tabular}
\end{table}
\end{minipage}\\[3mm]
%

For ideal chains the correlation functions $C_2$ and $C_3$
can be calculated in closed form. For the pair correlation 
one finds for arbitrary $D$
\begin{eqnarray}
& & C_2^{\,(id)}({\bf r}_A, {\bf r}_B) \, = \,
\pi^{-D/2} \, {\cal R}_x^{\,2} \, r_{AB}^{\,\,2 - D}
\label{IV60}\\[1mm]
& & \qquad \times \, 
\Big[ \Gamma\Big(\frac{D}{2}-1, \varrho^2 \Big)
- \varrho^{2\,} \Gamma\Big(\frac{D}{2}-2, \varrho^2
\Big) \Big] \nonumber
\end{eqnarray}
with $\Gamma$ the incomplete 
gamma-function \cite{abr} and 
$\varrho^2 = r_{AB}^{\,\,2} / (2 {\cal R}_x^{\,2}) = 
z_{AB}^{\,\,2} / 2$.
For $D = 3$ Eq.\,(\ref{IV60}) reduces to
\begin{equation} \label{IV70}
C_2^{\,(id)}({\bf r}_A, {\bf r}_B) \, = \,
\pi^{-3/2} \, ({\cal R}_x / z_{AB}) \,
S(z_{AB}^{\,\,2}/2) \, \, , \quad D = 3 \, ,
\end{equation}
where
\begin{equation} \label{IV80}
S(\varrho^2) \, = \, (1 + 2 \varrho^2) \sqrt{\pi} \,
\mbox{erfc}\,\varrho \, - \, 2 \varrho \exp(- \varrho^2)
\end{equation}
is the Fourier transform of the Debye scattering function
\cite{gennes2,cloizeaux,schafer}. 
For the triple correlation one finds in $D = 3$
\begin{eqnarray}
& & C_3^{\,(id)}({\bf r}_A, {\bf r}_B,  {\bf r}_C) \, = \,
\frac{1}{2 \pi^{5/2}} \, 
\Big[\, 
\frac{S( (z_{BA} + z_{AC})^2 / 2 )}
{z_{BA\,} z_{AC}} \, \label{IV90}\\[1mm]
& & \quad + \, 
\frac{S( (z_{AB} + z_{BC})^2 / 2 )}
{z_{AB\,} z_{BC}} \, + \,
\frac{S( (z_{AC} + z_{CB})^2 / 2 )}
{z_{AC\,} z_{CB}} \,
\Big] \, \, , \quad D = 3 \, . \nonumber
\end{eqnarray}
One can verify that the expressions (\ref{IV70}) and 
(\ref{IV90}) obey the short distance relations
(\ref{IV30}) and (\ref{IV40}) with $\sigma^{(id)} = \pi^{-1}$
from Eq.\,(\ref{IV50}). 

The limiting behavior (\ref{I100}) ceases to apply if the
mutual distance between the small spheres becomes comparable 
with the order of their radii. 
As an illustration we consider two spheres 
$A$ and $B$ with equal radii $R_A = R_B = R$. While for 
$R \ll r_{AB}, {\cal R}_x$ the reduced free energy of
interaction $n_p \, f_{A,B}^{\,\,(2)}$ is given by 
Eq.\,(\ref{I100a}) and, in particular, for
$R \ll r_{AB} \ll {\cal R}_x$ by 
\begin{equation} \label{IV100}
f_{A,B}^{\,\,(2)} \, \to \,
- \, (A_{D,D})^2 \, \sigma \, 
R^{2 (D - 1 / \nu)} \, {\cal R}_x^{\,1/\nu} \,
r_{AB}^{\, - (D - 1/\nu)} 
\end{equation}
due to Eq.\,(\ref{IV30}), one finds for 
$R$, $r_{AB} \ll {\cal R}_x$ with arbitrary $r_{AB} / R$ that
\begin{equation} \label{IV110}
f_{A,B}^{\,\,(2)} \, \to \, 
- \, (2 - {\cal M}) \, A_{D,D} \, R^{D - 1 / \nu} \,
{\cal R}_x^{\,1 / \nu} \, \, .
\end{equation}
Here 
\begin{equation} \label{IV120}
{\cal M} \, = \, {\cal M}({\cal D} / R)
\end{equation}
is independent of ${\cal R}_x$ and is a universal function of
${\cal D} / R$ with
\begin{equation} \label{IV130}
{\cal D} \, = \, r_{AB} - 2 R
\end{equation}
the distance of closest approach surface-to-surface 
between the two spheres $A$ and $B$. 
Equation (\ref{IV110}) holds because
on the large length scale set by ${\cal R}_x$
the `dumb-bell' composed of the two spheres with small $R$
and ${\cal D}$ can be considered in leading order 
\cite{dumb-bell} as a pointlike perturbation 
as in Eq.\,(\ref{I60}) in conjunction with the lower
part of Eq.\,(\ref{I61}), but with the amplitude 
$A_{D,D}$ replaced by an amplitude-function 
${\cal M} A_{D,D}$ which depends on ${\cal D} / R$
\cite{dumb-bell2}. Consistency of Eqs.\,(\ref{IV110})
and (\ref{IV100}) requires that 
\begin{equation} \label{IV140}
2 - {\cal M} \, \to \, A_{D,D} \, \sigma \, 
({\cal D} / R)^{- (D - 1/\nu)} \, \, , \quad 
{\cal D} / R \to \infty \, \, .
\end{equation}
In the opposite limit ${\cal D} / R \to 0$ of two 
{\em touching\/} spheres [19(b)] the function 
${\cal M}({\cal D} / R)$ approaches a constant
larger than one because the dumb-bell is a stronger 
perturbation than a single sphere. Similar to the small 
radius expansion (\ref{I60}) the small dumb-bell expansion has 
$-$ via the polymer magnet analogy $-$ 
its counterpart in the ${\cal N}$-component field theory.
An easy way to obtain the explicit expression for 
${\cal M} A_{D,D}$ is to calculate the energy density profile
$\langle - {\bf \Phi}^2({\bf r}) \rangle_{\text{db},\,\text{crit}}$ 
{\em at\/} the critical point of the field theory
in the presence of the two spheres with radius $R$ and Dirichlet 
boundary conditions which represent the dumb-bell (db) centered at 
the origin, and to compare the result with the corresponding 
result as derived from the small dumb-bell expansion in the form 
\begin{equation} \label{IV150}
{\cal M} A_{D,D} \, = \,
\lim\limits_{r \to \infty} \,
\frac{\langle - {\bf \Phi}^2({\bf r}) \rangle_{\text{db},\,\text{crit}}
\, - \, \langle - {\bf \Phi}^2 \rangle_{b, \, \text{crit}}}
{R^{D - 1/\nu} \, 
\langle \Psi(0) \, {\bf \Phi}^2({\bf r}) 
\rangle_{b, \, \text{crit}}} \, \, \, .
\end{equation}
Here $\Psi$ is the normalized energy density introduced in
Eq.\,(\ref{C70}) and the rhs of Eq.\,(\ref{IV150})
is taken in the limit ${\cal N} \searrow 0$. The evaluation 
of the numerator is simplified by means of a 
conformal transformation relating it to the 
corresponding quantity between two {\em concentric\/}
spheres [36(b)]. 

For ideal chains $-$ corresponding to a Gaussian field theory $-$
the latter quantity is known and leads to
\begin{mathletters} 
\label{IV160}
\begin{equation} \label{IV160a}
{\cal M} \, = \, 2 \, (\theta^{-1/2} - \theta^{1/2})^{D-2} \,
\sum\limits_{l = 0}^{\infty} \,
{D - 3 + l \choose l} \,
\left[ \theta^{- (l + (D-2)/2)} + 1 \right]^{-1} \, \, ,
\end{equation}
where $\theta$ is related to the dumb-bell parameter
${\cal D} / R$ via 
\begin{equation} \label{IV160b}
\frac{1}{2} \, (\theta + \theta^{-1}) \, = \,
1 \, + \, 2 \, \frac{\cal D}{R} \, + \, 
\frac{1}{2} \left( \frac{\cal D}{R} \right)^{2} \, \, .
\end{equation}
Equation (\ref{IV160}) provides an explicit expression for
$f_{A,B}^{\,\,(2)}$ in Eq.\,(\ref{IV110}). In particular
one can check Eq.\,(\ref{IV140}) by using the relations 
$A_{D,D}^{\,(id)} \, \sigma^{(id)} = 2$ and $\nu^{-1} = 2$ 
valid for ideal chains. In $D=3$ one finds
the leading behavior \cite{hantel}
\end{mathletters}
\begin{equation} \label{IV170}
\lim\limits_{{\cal R}_x \to \infty}
\frac{f_{A,B}^{\,\,(2)}}{2 f_{A}^{\,(1)}} \, \to \,
- \, (1 - \ln 2) \, + \, \frac{\cal D}{R} \, \frac{1}{6} \,
\Big( \ln 2 - \frac{1}{4} \Big) \, \, , 
\quad {\cal D} / R  \to 0 \, \, ,
\end{equation}
which determines not only the 
solvation free energy of but also 
the depletion force between two small touching spheres
in a dilute solution of ideal chains in $D=3$. 
Numerical evaluation of Eq.\,(\ref{IV160}) for
{\em arbitrary \/} ${\cal D} / R$ in $D = 3$ shows
that the crossover of $f_{A,B}^{\,\,(2)} / (2 f_{A}^{\,(1)})$
from the behavior given on the
rhs of Eq.\,(\ref{IV170}), valid for 
${\cal D} \ll R \ll {\cal R}_x$, to the behavior $- R / {\cal D}$, 
valid for $R \ll {\cal D} \ll {\cal R}_x$, is monotonic and 
without inflection point. Since this holds also for the 
crossover from $R \ll {\cal D} \ll {\cal R}_x$ to 
$R \ll {\cal R}_x \ll {\cal D}$ as implied by Eqs.\,(\ref{IV70}) 
and (\ref{I100a}), one finds that upon increasing the distance 
${\cal D}$ the reduced free energy of 
interaction $n_p f_{A,B}^{\,\,(2)}$ between two small spheres
is monotonically increasing and the attractive
force $\frac{\partial}{\partial {\cal D}} n_p f_{A,B}^{\,\,(2)}$ 
is monotonically decreasing in the whole range of ${\cal D}$.

This is different from the behavior of a particle with small 
radius interacting with a {\em planar wall\/} (compare 
Sec.\,\ref{secIIIA} and I). In this case the attractive force 
$\frac{\partial}{\partial {\cal D}} \Delta F_{depl}$ 
is not monotonically decreasing with increasing 
${\cal D}$ but exhibits a {\em maximum\/} at a 
distance ${\cal D}_{max}$ of the order ${\cal R}_x$ since 
the monomer density profile $M_M^{(W)}$ in Eq.\,(\ref{III70}) 
has a point of inflection. This qualitatively different 
feature applies not only for a thin cylinder but also for a 
small spherical particle near a wall \cite{B}. Another 
remarkable difference between the two cases is the behavior 
of the force in the limit $R, {\cal D} \ll {\cal R}_x$.
While in the case of two spheres 
$\frac{\partial}{\partial {\cal D}} n_p f_{A,B}^{\,\,(2)}$
increases as ${\cal R}_x^{\,1/\nu}$ for ${\cal R}_x \to \infty$, 
the force $\frac{\partial}{\partial {\cal D}} \Delta F_{depl}$
between the particle and the wall exhibits a {\em finite\/} 
limit for ${\cal R}_x \to \infty$.
This is plausible since the particle eventually moves into a 
region which is already depleted due to the presence of the wall.

It is interesting that for two touching
spheres in a solution of ideal chains
in $D=3$ the form of the normalized interaction free energy 
\begin{mathletters}
\label{eigen}
\begin{equation} \label{eigena}
f_{A,B}^{\,(2)} / (2 R)^3 \, = \,
- \, \frac{a}{2} \, ({\cal R}_x / R)^2 \, \, , \quad
R \ll {\cal R}_x \, \, ,
\end{equation}
for {\em small\/} radius as implied
by Eqs.\,(\ref{IV170}), (\ref{I20}), (\ref{I30}) and
$A_3^{\,(id)} = 2 \pi$ is very similar to its counterpart 
\begin{equation} \label{eigenb}
f_{A,B}^{\,(2)} / (2 R)^3 \, = \, 
- \, \frac{b}{2} \, ({\cal R}_x / R)^2 \, \, , \quad
R \gg {\cal R}_x \, \, ,
\end{equation}
for {\em large\/} radius following
from the Derjaguin approximation \cite{derjaguin}, which is
supposed to be exact in this limit. 
Both forms display the same power in the length ratio
${\cal R}_x / R$ and their amplitudes 
$a = \pi (1 - \ln 2) = 0.964$ and $b = (\pi/2) \ln 2 = 1.09$
are nearly the same.
Although we do not have an explicit
expression for the normalized interaction free energy
for ${\cal R}_x / R$ of order unity we expect that either of the 
two limiting forms (\ref{eigen}) provides a reasonable approximation
even in the intermediate regime. This is confirmed by the computer
simulation results of Ref.\,\cite{meijer} in which the chain is 
modelled as an $N$-step random walk on a simple cubic lattice with 
$N = 10$ or $100$ and the diameter $2 R$ of each of the two
touching spheres equals $10.5$ lattice constants. 
This corresponds to the values $0.06$ or $0.60$ of 
$\frac{1}{2} ({\cal R}_x / R)^2 = \frac{1}{6} N / (\frac{1}{2} 10.5)^2$
and each of our two forms leads to estimates which are fairly
close \cite{SSa} to the simulation results 
$0.04$ or $0.50$ displayed
in Fig.\,3 of Ref.\,\cite{meijer} for the
quantity $- \bar{\Omega}^{2b*} / \sigma_{\text{col}}^{\,3}$
there which is to be identified with 
$- f_{A,B}^{\,(2)} / (2 R)^3$ here.

\end{mathletters}

In order to be able to appreciate the results for the 
depletion interaction of particles with small radii $R$ as
obtained in this subsection it is instructive to compare them
with those of the PHS model \cite{asakura} extrapolated to the
case of small $R$ \cite{eintauch}. A force displaying a maximum 
at a distance ${\cal D}_{max}$ of order ${\cal R}_x$ for the 
effective interaction between a particle and a planar wall 
and a monotonical decrease of the force with 
increasing ${\cal D}$ for two particles of equal size
are also found within the PHS model when extrapolated to 
$R \ll {\cal R}_x$. However, the PHS model does not produce
the decrease of the absolute values of the free energy of 
interaction and of the force with decreasing $R$ but in the 
limit $R \to 0$ rather leads to finite quantities
which are independent of $R$. For example, in the case of 
a thin cylinder or a small sphere near a wall the maximum 
force in the PHS model is not proportional to 
$R^{d - 1 / \nu} \, {\cal R}_x^{\, 1 / \nu - 1}$
as for a flexible chain but rather to ${\cal R}_x^{\,d-1}$.
In the particle-wall case the PHS model also fails to predict 
that the force becomes independent of ${\cal R}_x$ for
$R, {\cal D} \ll {\cal R}_x$.

Even for the much studied case of a {\em large\/} sphere radius,
i.e., $R \gg {\cal R}_x$, for which the PHS
approximation is expected to work best, the deviation
of the PHS approximation from the Derjaguin result 
is considerable.
The PHS approximation implies $b = 1$ in Eq.\,(\ref{eigenb})
\cite{polykugel},
i.e., it leads to a free energy $f_{A,B}^{\,(2)}$
for two large touching spheres of equal size whose absolute value
is too small by about $10\,\%$. The same ratio $(\pi/2) \ln 2$ 
between the Derjaguin result and the PHS approximation appears 
in the case of a single large sphere touching a planar wall 
(compare footnote [28] in Ref.\,\cite{unirel}).


\section{Summary and concluding remarks}
\label{summary}

We have studied the interaction of mesoscopic particles 
(spheres, cylinders, and planar walls) with a dilute solution 
of long, flexible, free, and nonadsorbing
polymer chains which are depleted by the particles in good or
theta solvents. The properties for a single particle 
as well as the effective interaction between two or more 
particles have been considered.

One topic of main concern has been to investigate in a systematic
and quantitative way how the excluded volume (EV) interaction 
between the chain monomers modifies the ideal chain behavior.
Our results are in line with the plausible conjecture 
that {\em weaker\/} depletion effects arise from chains with
EV interaction than from ideal chains with the same Flory radius.
Another main topic has been the description of situations in which
the particle radius $R$ is small compared with the Flory radius 
${\cal R}_x$ so that the chain will coil around the particle
(compare Fig.\,\ref{fig1})
and in which the classic PHS treatment ignoring chain flexibility 
\cite{asakura} is clearly of no use. For example, consider
the limit $R / {\cal R}_x \to 0$ in which 
the spherical or cylindrical particle degenerates
to a point or a thin needle, respectively, on the scale of 
${\cal R}_x$: for flexible polymers both the
solvation free energy of the particle and its 
polymer mediated free energy of
interaction with other particles vanishes in this
limit whereas these two quantities remain finite for the
rigid polymers of the PHS model.

Our analysis is based on 
the polymer magnet analogy which maps 
the polymer problem with interactions within a single 
polymer chain and between a polymer chain and a particle
onto a Ginzburg-Landau $({\bf \Phi}^2)^2$ field
theory in the outer space of the particle with the order 
parameter field ${\bf \Phi}$ vanishing on the particle surface
(see Ref.\,\cite{eisen}, I, and Sec\,\ref{secIIvoran}). 
This allows us to resort to basic field-theoretical
tools such as the renormalization group and short-distance 
expansions which turn out to be extremely useful for the understanding 
of the polymer conformations in the presence of the particle(s).

In the following we summarize our main results starting with the
case of a single particle. The evaluation in I of the 
solvation free energy for immersing
the particle in a theta solvent (i.e., ideal chains)
has been generalized to the generic case of a good solvent (i.e.,
chains with EV interaction) by calculating the universal scaling
function $Y_{d,D}({\cal R}_x / R)$ (see Eq.\,(\ref{I20})).
For estimates based on a systematic perturbative approach it is
very useful to introduce the particle shape 
of a `generalized cylinder' (see Eq.\,(\ref{I10}))
which is characterized by the space 
dimension $D$ and an internal dimension $d$ encompassing cylinder, 
sphere and wall as special cases. The general results for
$Y_{d,D}(x)$ to first order in $\varepsilon = 4 - D$
are given by Eqs.\,(\ref{II20}) and (\ref{A250}).

(1) Our investigations in Sec.\,\ref{secIIA} of generalized 
cylinders with small curvature, i.e., $R \gg {\cal R}_x$,
provide strong evidence for the validity of the local and
analytic Helfrich-type expansion conjectured in Eq.\,(\ref{I25}).
With the help of Eq.\,(\ref{II160}) this expansion can be generalized
to arbitrary spatial dimensions $D$ so that we were able 
to obtain explicit expressions for the universal coefficients 
$\Delta \sigma$, $\Delta \kappa_1$, $\Delta \kappa_2$, 
and $\Delta \kappa_G$ appearing in the Helfrich Hamiltonian
to first order in $\varepsilon = 4 - D$.
While the results for the spontaneous curvature energy 
$\Delta \kappa_1$ in Eq.\,(\ref{II180}) and the mean and Gaussian
bending rigidities $\Delta \kappa_2$ and $\Delta \kappa_G$ in 
Eqs.\,(\ref{II190}) and (\ref{II200}) are new, the result 
in Eq.\,(\ref{II170}) for the surface tension $\Delta \sigma$ has
implicitly been noted before (see Eq.\,(4.7) in Ref.\,\cite{freed}).
All coefficients have absolute values smaller than those of 
their ideal chain counterparts. The latter are given by
the above expressions for $\varepsilon = 0$. The decrease of
the depletion effects due to the EV interaction can be traced back
to a corresponding behavior of the profile $M_E$ of the
end density (see Eqs.\,(\ref{II20}) and (\ref{A240})).
The simplest case is the surface tension $\Delta \sigma$ which
follows from the profile $M_E$ near a planar wall
and for which the decrease is consistent with a corresponding
decrease [19(a)] of the surface exponent $a_E$ in the behavior
$M_E \sim (z/{\cal R}_x)^{a_E}$ for distances $z$ from the wall much
smaller than ${\cal R}_x$.

(2) For small particle radius, i.e., $R \ll {\cal R}_x$,
our results for $Y_{d,D}(x)$ to first order in
$\varepsilon$ confirm the validity of the power law (\ref{I30})
within the region (\ref{I35})
and allow us to determine the $\varepsilon$-expansions 
of the universal amplitude $A_{d,D}$ (see Eq.\,(\ref{II471})).
The region (\ref{I35}) is shown shaded in Fig.\,\ref{fig2}
and includes the interior point
$(d,D) = (2,3)$ which represents a cylinder in three 
dimensions. This is different from the case of ideal chains 
in which Eqs.\,(\ref{I30}) and (\ref{I35})
are not valid below and on the line $d=2$.
Reliable estimates 
for the amplitudes $A_{3,3}$ and $A_{2,3}$ corresponding
to a sphere and a cylinder, respectively, 
for chains with EV interaction in three dimensions
have been obtained from the plausible assumption that the 
amplitude $A_{d,D}$ as function of $d$ and $D$ 
forms a {\em regular\/} surface over the base plane $(d,D)$
(see Fig.\,\ref{fig_shell}). 
The combination of the value of $A_{2,2}$ corresponding
to a disc in two dimensions (see Table \ref{table_amplitude})
with the $\varepsilon$-expansions of $A_{d,D}$ in Eq.\,(\ref{II471})
leads to the estimates in Eqs.\,(\ref{III20}) and (\ref{III30}) for 
$A_{3,3}$ and $A_{2,3}$. 
    
(3) Estimates of the full scaling functions $Y_{3,3}(x)$
and $Y_{2,3}(x)$ for the solvation free energy of a sphere and 
a cylinder in three dimensions are shown in Fig.\,\ref{plot} in terms 
of the functions $\Theta_{3,3}(x)$ and $\Theta_{2,3}(x)$ 
defined in Eq.\,(\ref{II500}). This shows the crossover from
the small curvature regime $x \ll 1$ with the coefficients 
$\Theta(0)$, $\Theta'(0)$, and $\Theta''(0)$ of their 
Taylor expansions about the regular point 
$x = 0$ being simply related to the surface tension
$\Delta \sigma$, the energy $\Delta \kappa_1$ of spontaneous curvature,
and the bending rigidities $\Delta \kappa_2$ and $\Delta \kappa_G$,
respectively (see Eqs.\,(\ref{II167}) and (\ref{II500})),
to the small radius regime $x \gg 1$ with the power law 
$\Theta_{d,3}(x \to \infty) \to A_{d,3\,} x^{1/\nu\, - 1}$.
As expected the curves in Fig.\,\ref{plot} for chains with
EV interaction are {\em below\/} the corresponding 
curves for ideal chains and imply a smaller solvation energy.
For chains with EV interaction the exponent 
$1/\nu - 1$ is not a positive integer and the expansion of
$\Theta_{d,3}(x)$ or $Y_{d,3}(x)$ about 
$x = 0$ cannot be a polynomial 
with a finite number of terms. This is in contrast with the
solvation free energy of a sphere in a solution of ideal
chains in which case $\Theta(x)$ is
a linear function of $x$ (see Ref.\,\cite{jansons} or I).

We continue by summarizing our results for the interaction 
between particles with a small radius. Since the values 
in Eqs.\,(\ref{III20}) and (\ref{III30}) of 
the amplitudes $A_{3,3}$ and $A_{2,3}$
completely determine the Boltzmann weight in Eq.\,(\ref{I60})
of a small sphere and a thin cylinder, the interactions of these
particles with other distant particles or walls are completely
determined, too \cite{burkhardt,diehl,amit,B,cardy}.  

(4) We have studied the interaction between a wall and a long thin 
cylindrical particle a distance ${\cal D}$ apart with radius $R$
and length $l$ for the case $R \ll {\cal R}_{x}, {\cal D} \ll l$ 
(compare Fig.\,\ref{fig_rod}). The dependence on ${\cal D}$
of the polymer mediated free energy of interaction is 
proportional to that of the monomer density $M_M^{\,(W)}$
of a dilute solution of chains in the half-space {\em without\/}
the particle (see Eq.\,(\ref{III70})).
The same applies for a small sphere near a wall (compare I).
Since $M_M^{\,(W)}$ has a point of inflection at 
${\cal D} \sim {\cal R}_x$ the attractive mean force between 
a wall and a thin cylinder or between a wall and a small sphere
somewhat surprisingly passes through a maximum as ${\cal D}$
increases. The increase $\sim {\cal D}^{1/\nu\, - 1}$ of the
force per unit length $l^{D-d}$ and unit bulk pressure
$n_{p\,} k_B T$ with the distance ${\cal D}$ in the region
$R \ll {\cal D} \ll {\cal R}_x$ is a consequence of its 
length dimension $d - 1$, its independence of ${\cal R}_x$,
and of the fact that the particle radius $R$ enters the force
only in the form of the power law $R^{d - 1/\nu}$ according to
Eq.\,(\ref{I60}). Our study of the situation of long chains is
complementary to that of short chains, i.e., ${\cal R}_x \ll R$,
considered in Ref.\,\cite{sear}. In the latter case the attractive
mean force of depletion is monotonically decreasing as
${\cal D}$ increases.

(5) The interaction between two small spherical particles 
$A$, $B$ of equal size and with a distance 
$r_{AB} = {\cal D} + 2 R$ between their centers
has been studied in Sec.\,\ref{secIIIB}
both for $R \ll {\cal D}, {\cal R}_x$ and
for $R, {\cal D} \ll {\cal R}_x$. In the former case we use
Eq.\,(\ref{I100a}) expressing the interaction in terms of
$A_{D,D}$, $R$, and the universal monomer density correlation
function $C_2$ of a single chain in unbounded space.  
In the latter case the `dumb-bell' composed of the two spheres 
is small on the scale of ${\cal R}_x$ and can in leading order
be considered as a pointlike object. This gives rise to an
expansion similar to Eq.\,(\ref{I60}) 
in conjunction with the lower part of Eq.\,(\ref{I61})
in which, however, the amplitude $A_{D,D}$ is replaced
by an amplitude {\em function\/} ${\cal M} A_{D,D}$
depending on $R$ and ${\cal D}$.
This is another type of a short-distance like operator   
expansion which can be used not only for the effective
free energy of interaction but also for other polymer 
properties $-$ such as the monomer density profile $-$ 
induced by the two spheres. 
Both cases overlap in the region $R \ll {\cal D} \ll {\cal R}_x$
in which the interaction free energy $f_{A,B}^{\,(2)}$ per unit bulk
pressure $n_{p\,} k_B T$ is given by Eq.\,(\ref{IV100}).
Numerical values of $A_{D,D}$ and the universal bulk amplitude
$\sigma$ in Eq.\,(\ref{IV100}) are summarized in 
Table\,\ref{table_amplitude} for various space dimensions 
and both for ideal chains and chains with EV interaction. 
The value for $\sigma$ in $D = 2$ derived in Appendix \ref{appC}
is a new result for a self-avoiding chain in the
unbounded plane.
For $D=3$ and ideal chains we explicitly calculate 
the two functions $C_2$ and ${\cal M} A_{D,D}$ 
(see Eqs.\,(\ref{IV60}) and (\ref{IV160})) and thus
present a complete and explicit expression for the free
energy of interaction between two small spherical particles
to leading order in the small quantity $R/{\cal R}_x$.
In contrast to the polymer mediated force between a small 
sphere and a wall, for two spheres of equal size the force 
is monotonically decreasing in the whole range of ${\cal D}$.
For the case of two touching spheres and 
arbitrary values of $R / {\cal R}_x$ 
we consider an approximative form of $f_{A,B}^{\,(2)}$
(compare Eq.\,(\ref{eigen})) and compare it with the results 
\cite{meijer} of simulations.

(6) As an illustration for the nonpairwise character of the
depletion interaction between particles we have evaluated
an explicit analytic expression for the three particle interaction
$f_{A,B,C}^{\,(3)}$ in the case of small spherical particles
and ideal chains in three dimensions. 
The expression follows by inserting the triple correlation
function in Eq.\,(\ref{IV90}) of the monomer density 
in the unbounded solution in Eq.\,(\ref{I100b}) 
and using that in this case $D - 1/\nu = 1$ and
$A_{D,D} = A_{3}^{\,(id)} = 2 \pi$.
The result is valid
in the region $R \ll r_{ij}, {\cal R}_x$ with $r_{ij}$ 
denoting the relative distances $r_{AB}$, $r_{AC}$, or $r_{BC}$
between the spheres and is complementary to the three-body
results presented in Ref.\,\cite{meijer} with $R$ of the order
of ${\cal R}_x$. In order to convey an idea of the relative 
importance of one-, two-, and three-particle contributions we 
summarize the results
\begin{eqnarray}
& & \Big( f^{(1)}, \, f_{A,B}^{\,(2)}, \, f_{A,B,C}^{\,(3)} \Big)
\label{S10} \\[1mm]
& & = \, 2 \pi R \, {\cal R}_x^{\,2} \,
\Big(1, \, - \, 2 \, \frac{R}{r_{AB}}\, , \,
2 R^2 \, \frac{r_{AB} + r_{BC} + r_{CA}}{r_{AB} \, r_{BC} \, r_{CA}}
\Big) \nonumber
\end{eqnarray}
for the special case $R \ll r_{ij} \ll {\cal R}_x$.
For three small spheres configurated on an equilateral
triangle with edge length $r$ the interaction 
$f_{A,B,C}^{\,(3)}$ is 
related to $f_{A,B}^{\,(2)}$ for two spheres at a distance
$2 r$ via
\begin{equation} \label{S20}
\Big( f_{A,B,C}^{\,(3)} \Big)_{r_{ij} \, = \, r} \, \Big/
\Big( - f_{A,B}^{\,(2)} \Big)_{r_{AB} \, = \, 2 r} 
\, = \, 6 R / r \, \, .
\end{equation}
This relation holds for an arbitrary ratio
$r / {\cal R}_x$ provided $R \ll r, {\cal R}_x$.

Another interesting type of three-body depletion interaction
arises for two spherical particles near a planar wall. If their
radii are small this situation can again be systematically 
investigated by means of Eq.\,(\ref{I60}) and the 
lower part of Eq.\,(\ref{I61}). In the same spirit
the investigations of three-body interactions could be supplemented
to cover cases in which the distance between two of the spheres
(or between one of the spheres and the wall) becomes of the order
of $R$ or smaller by means of the `small dumb-bell' expansion  
(or an expansion which applies to a sphere close to a planar wall
\cite{B}).

Finally, we summarize some of the field-theoretic developments 
on which our treatment of the particle-chain interaction is based.

(7) After a brief outline of the polymer magnet analogy
in Sec.\,\ref{secIIvoran} we relate
the density profile $M_E$ of chain ends to the local 
susceptibility in the corresponding magnetic system
for a generalized cylinder $K$
in a dilute polymer solution (see Eq.\,(\ref{xi})).
For such nonadsorbing chains the corresponding order parameter 
field $\bf \Phi$ vanishes at the surface of the particle.
With the Gaussian order parameter correlation function
outside $K$ as the unperturbed propagator we use renormalized 
perturbation theory with respect to a $({\bf \Phi}^2)^2$
interaction in order to obtain a systematic expansion in
the EV interaction of the polymer quantities below
the upper critical dimension $D_{uc} = 4$. The behavior 
of our one loop expressions
(see the function ${\cal C}_d(\tau)$ in Eq.\,(\ref{A260}))
in the limits corresponding to large $R$ and small $R$
is discussed in Appendix \ref{neuA}.

(8) We verify to first order in the EV interaction that
the {\em same\/} small radius amplitude appears for
{\em different\/} properties of a 
generalized cylinder with a small radius $R$.
In Appendix\,\ref{appB} we write Eq.\,(\ref{I60}) 
in terms of fluctuating densities (operators) in the
equivalent field theory.
The universal small radius amplitude $A_{d,D}$ for
polymers is obtained from a corresponding 
critical amplitude $\widehat{A}_{d,D}$ in the field theory
by multiplying with a universal noncritical bulk amplitude.
In the two-point correlation 
function with distances of the two points from the
generalized cylinder much larger than $R$ there appears the same 
amplitude $\widehat{A}_{d,D}$ at the critical point of the field theory
$-$ where the correlation length $\xi_+$ is infinitely large $-$ 
as in the behavior of the field-theoretic excess susceptibility
of the generalized cylinder for $\xi_+ / R \gg 1$. 
The latter is related to the power law behavior (\ref{I30}) 
of the function $Y_{d,D}(x)$ for $x = {\cal R}_x / R \gg 1$.
These considerations
are important to understand that the mechanism behind the 
small radius expansion is basically of the same type as that
behind the well-known short distance expansions in field
theories without boundaries \cite{amit,cardy}. 
Moreover, in case of a sphere our
result for $\widehat{A}_{D,D}$ to first order in
$\varepsilon$ confirms that this 
amplitude can be reduced to bulk and half-space amplitudes
as predicted from a conformal mapping 
\cite{BE85} (see the last but one paragraph in 
Appendix\,\ref{appB}).

(9) By studying the energy density profile 
in a Gaussian field theory with boundaries
we explicitly verify that not
only a single sphere but also a `dumb-bell' composed
of two spheres of equal size can be considered as a 
pointlike perturbation on sufficiently 
large length scales. At bulk criticality the profile for the 
dumb-bell can be obtained by means of a conformal 
transformation from the known profile between two
concentric spheres. For ideal polymer chains we thus
find the explicit form (see Eq.\,(\ref{IV160})) 
of the amplitude function ${\cal M} A_{D,D}$
addressed in paragraph (5) of this Summary. 

\bigskip

\section*{Acknowledgements}

We thank T. W. Burkhardt for helpful discussions.
The work of A. H. and S. D. has been supported by the 
German Science Foundation through
Sonderforschungsbereich 237 
{\em Unordnung und gro{\ss}e Fluktuationen\/}.


\newpage

\appendix

\section{The function ${\cal C}_{\lowercase{d}}(\tau)$}
\label{neuA}

The results of Sec.\,\ref{secII}
are based on the behavior of the function $Q_{d,D\,}(\eta)$ in
Eq.\,(\ref{A250}), in particular on the behavior of $C_{d}(\eta)$
in Eq.\,(\ref{A260}). The difficult part of the 
corresponding calculation consists 
in performing the sum over $n$ and the double integral over $q$ and 
$\psi$ in order to calculate ${\cal C}_d(\tau)$ according to
Eqs.\,(\ref{A80c}) and (\ref{A260b}). Here we derive
the asymptotic expansions of ${\cal C}_d(\tau)$ for 
large and small $\tau$, respectively, and give 
numerical values of ${\cal C}_d(\tau)$ for the 
crossover region $0 \lesssim \tau \lesssim 3$.


\subsection{${\cal C}_d(\tau)$ for $\tau \to \infty$}
\label{subsec_gr}

We calculate the coefficients
${\cal C}_{1}^{\,(d)}$ and ${\cal C}_{2}^{\,(d)}$ 
in Eq.\,(\ref{II40})
for $d = D$, $3$, and $2$ by expanding the rhs 
of Eq.\,(\ref{A260b}) for large $\tau$. To this end we
need the behavior of the integrand in Eq.\,(\ref{A260b})
for $R \mu \sqrt{t} = \sqrt{\tau}$ large and 
$(y_{\perp} - R)_{\,} \mu \sqrt{t} = (\psi-1) \sqrt{\tau} \equiv s$
arbitrary. This is consistent with the expectation that for the small
curvature expansion the important regime in terms of polymer variables
is $R / {\cal R}_x$ large and $(y_{\perp} - R) / {\cal R}_x$ arbitrary.

(a) $d = D$:
Since ${\cal C}_{1}^{\,(D)}$
and ${\cal C}_{2}^{\,(D)}$ belong to the one loop contribution
of $Q_{D,D\,}(\eta)$ we need to consider only ${\cal C}_{1}^{\,(4)}$
and ${\cal C}_{2}^{\,(4)}$
(compare the remarks below Eqs.\,(\ref{A110}) and (\ref{A260b})).
The central part of the calculation consists in expanding 
$g_s(\psi, \tau, \varepsilon = 0)$ in the integrand on the rhs
of Eq.\,(\ref{A260b}) for $\tau$ large and $s$ arbitrary.
Since $W_{n}^{(1)}(0) = (n+1)^2 / (2 \pi^2)$ for $\alpha = 1$
in Eq.\,(\ref{A80c}) the quantity $g_s(\psi, \tau, 0)$ is, apart 
from a factor $- \psi^2 / (2 \pi^2)$, given by 
\begin{equation} \label{II70}
\sum_{n=0}^{\infty} \, n^2 \,
\frac{ I_{n}(\sqrt{\tau}\,) }
     { K_{n}(\sqrt{\tau}\,) } \, \,
[ K_{n}(s + \sqrt{\tau}\,) ]^{2} \, \, .
\end{equation}
A first hint on how to evaluate the sum (\ref{II70}) for large
$\tau$ can be gained from recognizing that its leading behavior 
corresponding to a vanishing curvature must 
describe the half-space bounded by a planar wall. This is 
discussed after Eq.\,(\ref{A13}) and shows that the ratio 
$n/R$ has the meaning of the length of a wavevector parallel 
to the wall and that all values of $n$ are important for which 
$(n/R) / (\mu \sqrt{t}) = n / \sqrt{\tau} \equiv \omega$ or
$(n/R) (y_{\perp} - R) = n s / \sqrt{\tau}$ are of order 
unity. Thus for the general expansion for large
$\tau$ a large number of terms will contribute and the sum 
can be replaced by an integral plus corrections according to 
the Euler-MacLaurin formula \cite{bender}:
\begin{equation} \label{II80}
\sum_{n=0}^{\infty} \, F(n) \, = \,
\int\limits_{0}^{\infty} dn \, F(n) \, + \frac{1}{2} F(0) \, - \,
\frac{B_2}{2 !} \, F'(0) \, - \, \frac{B_4}{4 !} \, F'''(0) \, + \, 
\ldots \, \, .
\end{equation}
Here $B_k$ are Bernoulli numbers and the function $F(n)$ 
can be read off from the expression (\ref{II70}). 
For case (a) the analysis of this expression
shows that all contributions on 
the rhs of Eq.\,(\ref{II80}) apart from the
integral lead to orders of $\tau^{-1/2}$ higher than 
needed for the first three terms on the rhs of the
expansion (\ref{II40}) of ${\cal C}_4(\tau)$ (but
compare case (b) below). Upon introducing $\omega$ instead 
of $n$ as the integration variable the expression 
(\ref{II70}) turns into
\begin{equation} \label{II90}
\tau^{3/2} \, \int\limits_{0}^{\infty} d\omega \, \, \omega^2 \,
\frac{ I_{a}(a / \omega) } { K_{a}(a / \omega) } \,
\Big[ K_{a}
\Big( a \, \frac{s + \sqrt{\tau}}{\omega \, \sqrt{\tau}} \, \Big)
\Big]^{2}
\end{equation}
with $a = \omega \sqrt{\tau}$. 
For large $\tau$ the integral (\ref{II90}) can be simplified 
by employing the {\em uniform} asymptotic
expansion for large orders $a$ of the modified Bessel functions 
$I_a$ and $K_a$ which is provided, e.g., in 
the sections 9.7.7 and 9.7.8 of Ref.\,[40(a)].
In addition to the leading term (compare the discussion after 
Eq.\,(\ref{A13})) now also the correction terms containing the 
functions $u_0$, $u_1$, and $u_2$ given in section 9.3.9 
of the above reference have to be included. By inserting this
simplified integral into $g_s$ in Eq.\,(\ref{A260b}) 
one finds that the first three coefficients on the rhs 
of the expansion \,(\ref{II40}) of ${\cal C}_4(\tau)$ are 
determined by a number of double integrals over $s$ and $\omega$
which can all be calculated in closed form. This reproduces
the expression (\ref{II45}) for ${\cal C}_{0}$ 
$-$ and thus checks the assumption leading to it $-$ 
and yields the expressions for ${\cal C}_{1}^{\,(4)}$ and 
${\cal C}_{2}^{\,(4)}$ in Eq.\,(\ref{II100}).

(b) $d = 3$: Due to the additional integration over $q$ 
in Eq.\,(\ref{A80c}) the expression corresponding to
(\ref{II70}) now reads $\sum_{n=0}^{\infty} F(n + 1/2)$
where, using $W_{n}^{(1/2)}(0) = (n+1/2) / (2 \pi)$
and substituting $\kappa = q \, \tau^{- 1/2}$ in 
Eq.\,(\ref{A80c}), 
\begin{equation} \label{II110}
F(n) \, = \, \sqrt{\tau} \, n \, 
\int\limits_{0}^{\infty} d \kappa \,
\frac{ I_{n}(\sqrt{\tau} \, \sqrt{\kappa^2 + 1}\,) }
     { K_{n}(\sqrt{\tau} \, \sqrt{\kappa^2 + 1}\,) } \, \,
\Big[ K_{n} \Big( (s + \sqrt{\tau}) \, 
\sqrt{\kappa^2 + 1}\,\Big) \Big]^{2} \, \, .
\end{equation}
From the Euler-MacLaurin formula (\ref{II80}) one infers that,
in contrast to case (a), apart from the integral on the rhs
also the terms proportional to $F(1/2)$ and to $F'(1/2)$ have to be 
included in order to obtain the first three terms on the rhs of
the expansion (\ref{II40}) of ${\cal C}_3(\tau)$. 
Proceeding in the same way as in case (a)
one is led to consider modified Bessel functions $I_a$ and $K_a$
with order $a = \omega \sqrt{\tau} \sqrt{\kappa^2 + 1}$
and triple integrals over $s$, $\omega$, and $\kappa$.
One reproduces again the expression (\ref{II45}) for  
${\cal C}_{0}$ and finds, using $B_2 = 1/6$,
the expressions for ${\cal C}_{1}^{\,(3)}$ and 
${\cal C}_{2}^{\,(3)}$ in Eq.\,(\ref{II120}).

(c) $d = 2$: In this case the procedure is quite similar 
as in case (b). The expression corresponding to (\ref{II70}) 
now reads $F(0)/2 + \sum_{n=1}^{\infty} F(n)$ where
\begin{equation} \label{II130}
F(n) \, = \, \tau \, 
\int\limits_{0}^{\infty} d \kappa \, \kappa \,
\frac{ I_{n\,}(\sqrt{\tau} \, \sqrt{\kappa^2 + 1}\,) }
     { K_{n\,}(\sqrt{\tau} \, \sqrt{\kappa^2 + 1}\,) } \,
\Big[ K_{n} \Big( (s + \sqrt{\tau}) \, 
\sqrt{\kappa^2 + 1}\,\Big) \Big]^{2} \, \, .
\end{equation}
The analysis shows that only the integral on
the rhs of the Euler-MacLaurin formula (\ref{II80}) 
contributes to the first three terms on the rhs of 
the expansion \,(\ref{II40}) of ${\cal C}_2(\tau)$
(compare cases (a) and (b) above). One finds again the 
expression (\ref{II45}) for ${\cal C}_{0}$ and
in addition the expressions for ${\cal C}_{1}^{\,(2)}$ and 
${\cal C}_{2}^{\,(2)}$ in Eq.\,(\ref{II140}).


\subsection{${\cal C}_d(\tau)$ for $\tau \to 0$}
\label{subsec_kl}

The leading behavior 
of $C_{d}(\eta \to \infty)$ in Eq.\,(\ref{II300}) can be
inferred from the behavior for $\tau \to 0$ of the quantity
\begin{equation} \label{II320}
{\cal I}_{d}(\tau) \, = \, 
- \, \frac{\tau^2}{8 \pi^2} \, \, {\cal C}_{d}(\tau)
\end{equation}
with ${\cal C}_{d}(\tau)$ from Eq.\,(\ref{A260}). 
The behavior of ${\cal I}_{d}(\tau \to 0)$ exhibits
two types of leading terms. The first is the 
logarithmically divergent contribution 
$- \alpha / (4 \pi^2) \, \ln(1 / \sqrt{\tau}\,)$ 
which follows from the behavior 
$g_s(\psi, \tau, 0) \to - \alpha / (4 \pi^2)$ for 
$1 \ll \psi \ll 1 / \sqrt{\tau}$ as mentioned below 
Eq.\,(\ref{A82}). The second contribution is independent 
of $\tau$ and requires special care.
Its evaluation is facilitated by splitting 
${\cal I}_{d}(\tau)$ according to
\begin{mathletters} 
\label{II322}
\begin{equation} \label{II322a}
{\cal I}_{d}(\tau) \, = {\cal H}_d(\tau) + {\cal J}_d(\tau) \, \, ,
\end{equation}
where
\begin{eqnarray}
& & {\cal H}_{d}(\tau) \, = \, 
\int\limits_{1}^{\infty} \, d\psi \, 
\psi^{\, -1} \label{II322b}\\[1mm]
& & \times \, \left\{ 
\left[ 1 - \psi^{-\alpha\,}
\frac{K_{\alpha}(\psi \sqrt{\tau})}
{K_{\alpha}(\sqrt{\tau})} \right]^2 \,
g_s(\psi, \tau, 0) \, - \,
g_s^{\,(\text{as})}(\psi \sqrt{\tau}, 0) \right\}
\, \, , \nonumber
\end{eqnarray}
\begin{equation} \label{II322c}
{\cal J}_{d}(\tau) \, = \, 
\int\limits_{1}^{\infty} \, d\psi \, \psi^{\, -1} \,
g_s^{\,(\text{as})}(\psi \sqrt{\tau}, 0) \, \, .
\end{equation}
Here we have used Eqs.\,(\ref{A40}) and (\ref{A62}) and we
have added and subtracted the function 
$g_s^{\,(\text{as})}(\psi \sqrt{\tau}, 0)$
which is defined as in Eq.\,(\ref{A82}) and represents
the behavior of $g_s(\psi, \tau, 0)$ for 
$1 \ll \psi, \, \tau^{-1/2}$. In ${\cal H}_{d}$ one can
interchange the order of the integration over $\psi$ and 
the limit $\tau \to 0$ \cite{sieben} which results in the
finite limit
\end{mathletters}
\begin{equation} \label{II370}
{\cal H}_{d}(\tau \to 0) \, \to \, {\cal B}_{d} \, = \,
\int\limits_{1}^{\infty} \, d\psi \, \psi^{\, -1}
\Big\{ \left[ 1 - \psi^{- 2 \alpha} \right]^{2} 
\gamma_s(\psi, \varepsilon = 0)
\, + \, \frac{\alpha}{4 \pi^2} \Big\}
\end{equation}
where the function
\begin{equation} \label{II340}
\gamma_{s}(\psi, \varepsilon) \, = \, 
\, g_s(\psi, \tau = 0, \varepsilon)
\end{equation}
can be read off from Eq.\,(\ref{A80c}). The integral in
Eq.\,(\ref{II370}) is well-defined since $\gamma_s(\psi, 0)$
tends to $- \alpha  / (4 \pi^2)$ for large $\psi$
so that the logarithmic singularity is removed.
The integral in Eq.\,(\ref{II322c})
can be carried out explicitly and leads in conjunction with 
Eqs.\,(\ref{II322}) and (\ref{II370}) to 
\begin{equation} \label{II400}
{\cal I}_{d}(\tau \to 0) \, \to \, {\cal B}_{d} \, + \,
\frac{\alpha}{4 \pi^2} \,
\left[ \frac{\ln \tau}{2} - \ln 2 + 1 - \frac{\Psi(d/2)}{2}
+ \frac{C_E}{2} \right]
\end{equation}
where $\Psi$ is the psi-function and $C_E$ 
denotes Euler's constant. 
Inserting Eq.\,(\ref{II400}) in Eq.\,(\ref{II320})
and carrying out the inverse Laplace transform in 
Eq.\,(\ref{A260a}) leads to the result for 
$Y_{d,D}(x \to \infty)$ in Eq.\,(\ref{II440}).
We conclude this subsection by calculating the number
${\cal B}_d$ for $d = D$, $d = 3$, and $d \searrow 2$
(see Eq.\,(\ref{II411})).

(a) $d = D$: Since ${\cal I}_{D}$ belongs to the
one loop contribution of $Y_{D,D}$ we need to consider 
${\cal I}_{4}$ only (compare the remarks below 
Eqs.\,(\ref{A110}) and (\ref{A260b})). This amounts to inserting 
$\alpha = 1$ into Eq.\,(\ref{II370}) and the function $\gamma_{s}$ 
corresponding to a sphere in $D = 4$ which is given by 
$\gamma_{s}(\psi,0) = - (4 \pi^2)^{-1} [1 - \psi^{-2}]^{-2}$
\cite{zeitschrift} yielding ${\cal B}_4 \, = \, 0$.

(b) $d = 3$: For $2 < d < D$ the quantity ${\cal B}_d$ does not
vanish and can be evaluated numerically. For $d = 3$ this leads to
the value for ${\cal B}_3$ given in Eq.\,(\ref{II411}).

(c) $d \searrow 2$: In this limit ${\cal B}_d$ can be
calculated exactly. It is useful to substitute 
$\sigma = \psi^{2 \alpha}$ in Eq.\,(\ref{II370}) and to 
carry out the limit $\alpha = (d - 2)/2 \searrow 0$ for fixed 
$\sigma$ in the ensuing integrand. One finds that only 
the term for $n=0$ in Eq.\,(\ref{A80c}) survives this
limit with the result
\begin{equation} \label{II411cNEU}
{\cal B}_2 \, = \,
\frac{1}{8 \pi^2} \int\limits_{1}^{\infty} d \sigma \,
\sigma^{-1} \, \Big\{- \frac{\sigma - 1}{\sigma} + 1 \Big\} 
\, = \, \frac{1}{8 \pi^2} \, \, .
\end{equation}


\subsection{${\cal C}_d(\tau)$ in the crossover region 
$0 \lesssim \tau \lesssim 3$}
\label{subsec_mi}

For the convenience of the reader in Table\,\ref{table} we 
give some numerical values of ${\cal C}_d(\tau)$. From these
values an approximation for
the full function ${\cal C}_d(\tau)$ can be constructed 
by using its asymptotic behaviors for $\tau \to \infty$ and 
$\tau \to 0$ as derived in the above subsections and by
appropriate interpolation.

%
\begin{minipage}[t]{16cm}
\begin{table}[tbp]
\caption{Numerical values of $\ln {\cal C}_d(\tau)$
(see Eq.\,(\ref{A260b})).}
\label{table}
\medskip
\begin{tabular}{rrrrr} 
$\ln \tau$ & $d=2$ & $d=3$ & $d=4$ \\ \hline
 $ -10 $  &  $  18.816  $  &  $  21.308  $  &  $  22.223  $  \\
 $ - 9 $  &  $  16.795  $  &  $  19.175  $  &  $  20.108  $  \\
 $ - 8 $  &  $  14.773  $  &  $  17.027  $  &  $  17.980  $  \\
 $ - 7 $  &  $  12.755  $  &  $  14.865  $  &  $  15.835  $  \\
 $ - 6 $  &  $  10.744  $  &  $  12.692  $  &  $  13.672  $  \\
 $ - 5 $  &  $   8.748  $  &  $  10.511  $  &  $  11.488  $  \\
 $ - 4 $  &  $   6.773  $  &  $   8.328  $  &  $   9.282  $  \\
 $ - 3 $  &  $   4.830  $  &  $   6.160  $  &  $   7.057  $  \\
 $ - 2 $  &  $   2.931  $  &  $   4.024  $  &  $   4.836  $  \\
 $ - 1 $  &  $   1.085  $  &  $   1.946  $  &  $   2.640  $  \\
 $   0 $  &  $ - 0.700  $  &  $ - 0.054  $  &  $   0.504  $  \\
 $   1 $  &  $ - 2.422  $  &  $ - 1.966  $  &  $ - 1.540  $  \\
\end{tabular}
\end{table}
\end{minipage}


\section{Small radius expansion to one loop order}
\label{appB}

The relation (\ref{I60}) for polymers is
$-$ via the polymer magnet analogy $-$ 
closely related to a corresponding 
small radius expansion (SRE) in a $({\bf \Phi}^2)^2$
field theory with the Boltzmann weight 
$\exp( - \Delta {\cal H}_K \{ {\bf \Phi} \})$
which describes the presence of the generalized cylinder $K$ 
(compare Sec.\,\ref{secIIvoran} and 
Appendix \ref{appC}). Here we shall illustrate the 
SRE by considering the two-point correlation function {\em at\/} 
the critical point of the field theory in one loop order. 
This is particularly well suited to reveal the mechanism 
behind the SRE. Moreover it provides a significant check for
the operator character of the expansion because we shall find 
the {\em same\/} small radius amplitude $A_{d,D}$ as in 
Sec.\,\ref{secIIB}. 

Keeping $u$ and $\varepsilon = 4 - D$ as independent variables 
the SRE can be written in the form
\begin{equation} \label{B10}
\exp\left(- \Delta {\cal H}_K \right) \, \propto \,
1 \, - \, {\cal F}(\mu R, u; \varepsilon, d) \,
\mu^{2-d} \, Z_t \, \omega_K \, + \, \ldots
\end{equation}
with 
\begin{equation} \label{B15}
\omega_K \, = \, \left\{ \begin{array}{l@{\quad}l}
\int\limits_{{\mathbb R}^{\delta}}
d^{\, \delta} r_{\parallel} \, \,
{\bf \Phi}^2({\bf r}_{\perp}=0, {\bf r}_{\parallel})
\, \, ,    & d < D \, \, , \\[5mm]
\, \, {\bf \Phi}^2(0) \, \, ,  & d = D \, \, .
\end{array} \right.
\end{equation}
Here $\mu^{2-d} \, Z_t \, \omega_K$ is a renormalized and 
dimensionless operator and
\begin{equation} \label{B17}
{\cal F}(\mu R, u; \varepsilon, d) \, = \,
- \, {\cal A}_K^{\,(0)} \, (\mu R)^{d-2} \,
[1 + u F_1(\mu R; \varepsilon, d) \, + \, {\cal O}(u^2)]
\end{equation}
has an expansion in terms of $u$ with the coefficient  
$- {\cal A}_K^{\,(0)} = 2 \pi^{d/2} / \Gamma(\alpha) = \alpha
\Omega_d$ of the leading term corresponding to the Gaussian model 
(see Eq.\,(4.6) in I). The functions $F_i$ can be 
expanded in terms of $\varepsilon$ 
with coefficients which depend on $\mu R$ only via powers of 
$\ln(\mu R)$. In particular we shall find from the critical two-point
function that
\begin{equation} \label{B18}
F_1(\mu R; \varepsilon, d) \, = \, \frac{{\cal N} + 2}{3} \,
\Big[ \ln(\mu R) + f_1 + e_d + 
\frac{4 \pi^2}{\alpha} \, {\cal B}_d
\Big] \, \, + \, {\cal O}(\varepsilon)
\end{equation}
where
\begin{equation} \label{B150}
e_d \, = \, 1 + \frac{\ln \pi}{2} - \frac{\Psi(d/2)}{2} \, \, ;
\end{equation}
the quantity ${\cal B}_d$ has been introduced in Eq.\,(\ref{II370}).
The ellipses in Eq.\,(\ref{B10}) stand for contributions 
in which higher powers of $R$ are multiplied by powers of
$\ln R$. Standard renormalization group arguments imply that
for large $\mu R$ the function ${\cal F}$ is proportional
to $R^{d - 1/\nu}$ and that the rhs of Eq.\,(\ref{B10}) can
be written as $1 + {\cal A}_K R^{d - 1/\nu} \, \omega_K$ where
\begin{equation} \label{B19}
- \, {\cal A}_K \, = \, \mu^{2 - 1/\nu} \, Z_t \, D_L(u) \,
{\cal F}(1, u^{*}; \varepsilon, d)
\end{equation}
with $D_L$ from Eq.\,(\ref{A220}).
The universal polymer amplitude $A_{d,D}$ in Eq.\,(\ref{I60})
is related to ${\cal A}_K = {\cal A}_K({\cal N})$ via
\cite{AcalA}
\begin{equation} \label{B22}  
A_{d,D} \, = \, - \, {\cal A}_K(0) \, 2 \, \mu^{-2} Z_t^{-1\,} L \,
{\cal R}_x^{\,-1/\nu} \, \, .
\end{equation}
By using Eq.\,(\ref{A220}) one finds that the nonuniversal 
quantities $\mu, Z_t, D_L, f_1$ cancel and
\begin{equation} \label{B24}
A_{d,D} \, = \, \frac{2 \pi^{d/2}}{\Gamma(\alpha)} \,
\Big\{1 + \frac{\varepsilon}{4} 
\Big[ \frac{4 \pi^2}{\alpha} {\cal B}_d 
+ \frac{3}{2} - \frac{\ln 2}{2} - \frac{\Psi(d/2)}{2} \Big]
\, + \, {\cal O}(\varepsilon^2) \Big\} \, \, ,
\end{equation}
which indeed reproduces the first-order $\varepsilon$ results 
of $A_{d,D}$ in Eq.\,(\ref{II471}).

We now verify Eqs.\,(\ref{B10}) - (\ref{B150}).
Consider the two-point correlation function 
$\langle \Phi_j({\bf r}) \Phi_k({\bf r}\,') \rangle$ of the
field theory described by Eq.\,(\ref{A10}) at its critical 
point. For $u = 0$ the SRE follows from the explicit 
expressions in Eq.\,(\ref{A20}) for the Gaussian propagator
which by using Wick's theorem lead to
\begin{eqnarray} \label{BB90}
& & \langle \omega_{K\,} \Phi_j({\bf r}) \Phi_k({\bf r}\,') 
\rangle_{b,\,[0]} \label{b9}\\[1mm]
& & \, \, = \,
\frac{\delta_{jk}}{2 \pi^d} \,
(r_{\perp} r_{\perp}^{\,\,'})^{- \alpha}
\int\limits_{{\mathbb R}^{\delta}}
\frac{d^{\, \delta} P}{(2 \pi)^{\delta}}  \, \,
\exp [i \, {\bf P} \,
({\bf r}_{\parallel} - {\bf r}_{\parallel}^{\,\,' }) ] \,
(P/2)^{2 \alpha} \, K_{\alpha}(P r_{\perp}) K_{\alpha}(P r_{\perp}^{\,\,'}) 
\nonumber\\
& & \, \, = \, ({\cal A}_K^{\,(0)})^{-1} \, \lim_{R \to 0}
R^{- 2 \alpha} \Big\{
\langle \Phi_j({\bf r}) \Phi_k({\bf r}\,') \rangle_{[0]} - 
\langle \Phi_j({\bf r}) \Phi_k({\bf r}\,') \rangle_{b,\,[0]} 
\Big\} \, \, .
\nonumber
\end{eqnarray}
Here $\langle \, \, \rangle$ is a cumulant average with the subscript 
$[0]$ indicating $u = 0$ and with $b$ denoting the 
unbounded bulk space in absence of $K$. Obviously Eq.\,(\ref{BB90})
verifies the SRE for the Gaussian model.

Consider now the first order in $u$ contribution:
\begin{equation} \label{BB100}
\langle \Phi_j({\bf r}) \Phi_k({\bf r}\,') \rangle_{[1]} \, = \,
- \, \delta_{jk} \, \frac{{\cal N} + 2}{3} \, 8 \pi^2 f \mu^{\varepsilon}
u \, R^{2 \alpha} \, J({\bf r}, {\bf r}\,')
\end{equation}
where
\begin{equation} \label{BB110}
J({\bf r}, {\bf r}\,') \, = \, 
\int\limits_{y_{\perp} > R} \, d^D y \, \, 
G({\bf r}, {\bf y}; R) \, G({\bf r}\,' , {\bf y}; R) \, 
I(y_{\perp}, R) \, \, ,
\end{equation}
\begin{equation} \label{BB120}
I(y_{\perp}, R) \, = \, R^{-2 \alpha} \, G({\bf y}, {\bf y}; R) \, = \,
\frac{R^{-2 \alpha}}{\cal N} \, \Big\{
\langle {\bf \Phi}^2({\bf y}) \rangle_{[0]} - 
\langle {\bf \Phi}^2({\bf y}) \rangle_{b,\,[0]} 
\Big\} \, \, .
\end{equation}
The first order expression given in Eq.\,(\ref{BB100})
has the same structure as the one in Eq.\,(\ref{A60}) and we have used
Eq.\,(\ref{A54c}). Note that in the present dimensional regularization
scheme and at $t_0 = 0$ the bulk quantity 
$G_b({\bf y}, {\bf y}) = 
\langle {\bf \Phi}^2({\bf y}) \rangle_{b,\,[0]} / {\cal N}$ vanishes. 
We have exploited this in order to write the 
last expression in Eq.\,(\ref{BB120}) 
in such a form which allows us to make contact with Eq.\,(\ref{BB90}) 
and which implies
\begin{equation} \label{BB130}
I(y_{\perp}, 0) \, = \, \frac{{\cal A}_K^{\,(0)}}{\cal N} \,
\langle \omega_{K\,} {\bf \Phi}^2({\bf y}) \rangle_{b,\,[0]} \, = \,
- \, y_{\perp}^{\,- d + \varepsilon} \, \frac{\alpha}{4 \pi^2} \,
\Big\{ 1 + \varepsilon \, e_d \, + \, {\cal O}(\varepsilon^2) \Big\} \, \, .
\end{equation}
The function $I(y_{\perp}, R)$ is related to
$\gamma_s(\psi,\varepsilon)$ in Eq.\,(\ref{II340}) by
\begin{equation} \label{BB140}
y_{\perp}^{\,d - \varepsilon} \, I(y_{\perp}, R) \, = \,
\gamma_s(y_{\perp}/R, \varepsilon)
\end{equation}
and Eq.\,(\ref{BB130}) is consistent with
$\gamma_s(\infty, 0) = - \alpha / (4 \pi^2)$
as mentioned below Eq.\,(\ref{II340}). In order to 
verify Eqs.\,(\ref{B10}) - (\ref{B150}) we decompose 
$J({\bf r}, {\bf r}\,')$ according to 
\begin{equation} \label{BB150}
J \, = \, J_{(i)} \, + \, J_{(ii)} \, + \, J_{(iii)} 
\end{equation}
with
\begin{mathletters}
\label{BB160}
\begin{equation} \label{BB160a}
J_{(i)}({\bf r}, {\bf r}\,') \, = \,
\int\limits_{{\mathbb R}^{D}} d^D y \, \, 
G_b({\bf r}, {\bf y}) \, G_b({\bf r}\,', {\bf y}) \, 
I(y_{\perp}, R=0) \, \, ,
\end{equation}
\begin{equation} \label{BB160b}
J_{(ii)}({\bf r}, {\bf r}\,') \, = \,
- \int\limits_{0 < y_{\perp} < R} d^D y \, \, 
G_b({\bf r}, {\bf y}) \, G_b({\bf r}\,', {\bf y}) \, 
I(y_{\perp}, R=0) \, \, ,
\end{equation}
\begin{eqnarray}
J_{(iii)}({\bf r}, {\bf r}\,') \, & = & \,
\int\limits_{y_{\perp} > R} d^D y \, \, \Big\{ 
G({\bf r}, {\bf y};R) \, G({\bf r}\,', {\bf y};R) \, 
I(y_{\perp}, R) \label{BB160c}\\
& & \qquad \quad \quad - \, 
G_b({\bf r}, {\bf y}) \, G_b({\bf r}\,', {\bf y}) \, 
I(y_{\perp}, R=0) \Big\} \nonumber \, \, .
\end{eqnarray}
In the following we analyze the behavior of the rhs of 
Eq.\,(\ref{BB100}) for $R \to 0$ which arises from each of 
these contributions $J_{(i)}$, $J_{(ii)}$, and $J_{(iii)}$.

\end{mathletters}

Rewriting $I(y_{\perp}, 0)$ in the integrand on the rhs of 
Eq.\,(\ref{BB160a}) by means of the first equality in Eq.\,(\ref{BB130}) 
one finds (see Fig.\,\ref{graph})
\begin{equation} \label{BB170}
\langle \Phi_j({\bf r}) \Phi_k({\bf r}\,') 
\rangle_{[1], \, (i)} \, \to \, {\cal A}_K^{\,(0)} \, R^{2 \alpha} \,
\langle \omega_{K\,} \Phi_j({\bf r}) \Phi_k({\bf r}\,')
\rangle_{b, \, [1]} \, \, .
\end{equation}

In order to evaluate the leading contribution of $J_{(ii)}$ 
for small $R$ one can set ${\bf y}_{\perp} = 0$ in the two bulk
propagators $G_b({\bf r}, {\bf y})$ and $G_b({\bf r}\,', {\bf y})$
in the integrand on the rhs of Eq.\,(\ref{BB160b}).
Its remaining dependence on $y_{\perp}$ as given by 
the last expression in Eq.\,(\ref{BB130}) leads
to a pole in $\varepsilon$.
This results in (see Fig.\,\ref{graph})
\begin{eqnarray}
& & \langle \Phi_j({\bf r}) \Phi_k({\bf r}\,') 
\rangle_{[1], \, (ii)} \, \to \, {\cal A}_K^{\,(0)} \, R^{2 \alpha} \,
\langle \omega_{K\,} \Phi_j({\bf r}) \Phi_k({\bf r}\,')
\rangle_{b, \, [0]} \label{BB180}\\[1mm]
& & \qquad \times \, \Big\{ (Z_t)_{[1]} \, + \, 
\frac{{\cal N} + 2}{3} \, u \,
\Big[ \ln(\mu R) + f_1 + e_d \Big] \Big\} \nonumber
\end{eqnarray}
where $(Z_t)_{[1]} = ({\cal N} + 2) u / (3 \, \varepsilon)$ 
is the contribution to $Z_t$ 
of first order in $u$ (see Eq.\,(\ref{A54a})).

\newpage

%
\unitlength1cm
\begin{figure}[t]
\begin{picture}(10,9)
\put(1.5,0.5){
\setlength{\epsfysize}{8cm}
\epsfbox{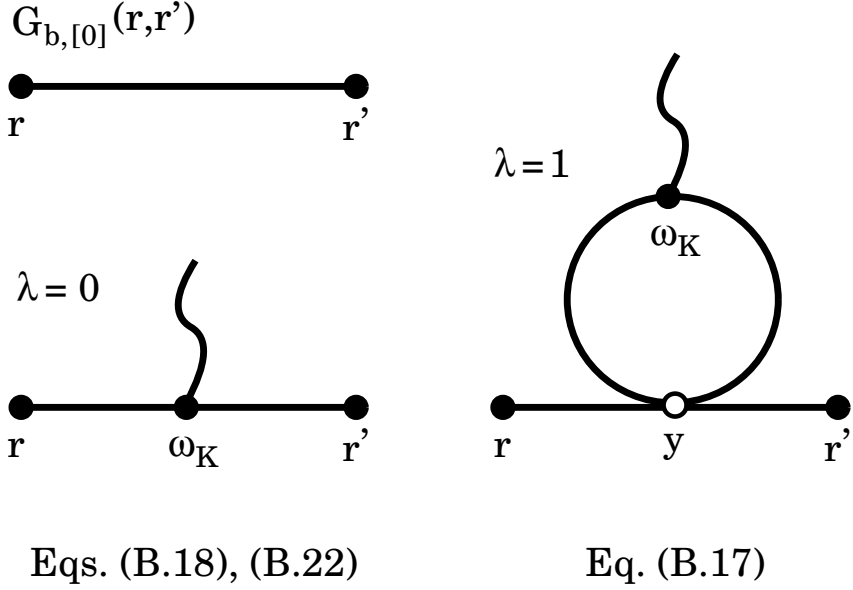}}
\end{picture}
\medskip
\caption{Diagrammatic representation of
$\langle \omega_{K\,} \Phi_1({\bf r}) \Phi_1({\bf r}\,')
\rangle_{b, \, [\lambda]}$ appearing on the rhs of 
Eqs.\,(\ref{BB170}), (\ref{BB180}), and (\ref{BB210}). 
The solid lines correspond to the bulk Gaussian propagator 
$G_{b, \, [0]} \equiv G_b$ and $\lambda = 0, 1$ denotes the loop order.
The wiggly lines indicate the insertion of the operator 
$\omega_K$ located at the `axis' of the generalized 
cylinder $K$.}
\label{graph}
\end{figure}
%

\medskip

For the leading contribution of $J_{(iii)}$ for $R \to 0$
it is sufficient to confine the integration over ${\bf y}_{\perp}$ 
on the rhs of Eq.\,(\ref{BB160c}) to the restricted region
$R < y_{\perp} < \sqrt{R \, r_{\perp}^{(<)}}$ with   
$r_{\perp}^{(<)} = \min(r_{\perp}, r_{\perp}^{\,\,'})$.
The reason is that in the remaining integration region
not only the ratios $R / r_{\perp}$ and $R / r_{\perp}^{\,\,'}$
but also $R / y_{\perp}$ are small so that in leading order 
one can insert $R = 0$ into the first term in curly brackets, 
which is then cancelled by the second term.
In particular, in the restricted region $y_{\perp}$ is smaller
than $r_{\perp}$ and $r_{\perp}^{\,\,'}$. By inserting the
representation (\ref{A20}) for the external legs 
$G$ and $G_b$ in Eq.\,(\ref{BB160c}) one finds that only
the terms for $n = 0$ contribute to the leading behavior 
of $J_{(iii)}$ for which one obtains
\begin{equation} \label{BB190}
J_{(iii)}({\bf r}, {\bf r}\,') \, \to \,
\frac{\Omega_d}{2} R^{\varepsilon} \, 
\langle \omega_{K\,} \Phi_1({\bf r}) \Phi_1({\bf r}\,')
\rangle_{b, \, [0]} \, \, \beta_d \, \, , 
\end{equation}
\begin{equation} \label{BB200} 
\beta_d \, = \, 
\int\limits_{1}^{\infty} \, d\psi \, \psi^{\, - 1 + \varepsilon}
\Big\{
\left[ 1 - \psi^{- 2 \alpha} \right]^{2} \gamma_s(\psi, \varepsilon)
\, - \gamma_s(\infty, \varepsilon) \Big\}
\end{equation}
with $\gamma_s$ from Eq.\,(\ref{BB140}).
The quantity $\beta_d$ arises as the limit for $R \to 0$ of the 
expression \cite{B11}
\begin{eqnarray}
& & \int\limits_{1}^{\sqrt{r_{\perp}^{(<)}/R}} d\psi \, 
\psi^{-1+\varepsilon} \, [\Gamma(\alpha+1)]^2 \, 
\Big(\frac{R P \, \psi}{2} \Big)^{- 2 \alpha} \label{BB205}\\
& & \qquad \, \times \, \Big\{ \Big[I_{\alpha}(R P \, \psi) - 
\frac{I_{\alpha}(R P)}{K_{\alpha}(R P)} K_{\alpha}(R P \, \psi) \Big]^2 
\gamma_s(\psi, \varepsilon) - 
[I_{\alpha}(R P \, \psi)]^2 \, \gamma_s(\infty, \varepsilon) \Big\} \, \, .
\nonumber
\end{eqnarray}
Of course, 
$\beta_d \to {\cal B}_d$ for $\varepsilon \to 0$ and the
present procedure for $R / r_{\perp}^{(<)} \to 0$ at the
critical point of the field theory leading to Eq.\,(\ref{BB200}) 
should be compared with the procedure for $R / \xi_+ \to 0$ 
leading to ${\cal B}_d$ in Eq.\,(\ref{II370}). 
Equation (\ref{BB190}) implies (see Fig.\,\ref{graph})
\begin{eqnarray}
\langle \Phi_j({\bf r}) \Phi_k({\bf r}\,') 
\rangle_{[1], \, (iii)} \, & \to & \, {\cal A}_K^{\,(0)} \, R^{2 \alpha} \,
\langle \omega_{K\,} \Phi_j({\bf r}) \Phi_k({\bf r}\,')
\rangle_{b, \, [0]} \label{BB210}\\[1mm]
& & \times \, \frac{{\cal N}+2}{3} \, u \, 
\frac{4 \pi^2}{\alpha} \, {\cal B}_d \, \, \, + \, 
{\cal O}(u \, \varepsilon) \, \, . \nonumber
\end{eqnarray}
Equations (\ref{BB170}), (\ref{BB180}), and (\ref{BB210}) 
corroborate the SRE in Eqs.\,(\ref{B10}) - (\ref{B150})
to first order in $u$ in the case of the critical correlation 
function. Note the recurrent character of 
the SRE which is typical for operator product expansions [39(b)].
A graphical representation of the bulk correlation
function with insertion of the operator ${\omega}_K$ is 
shown in Fig.\,\ref{graph}. 

Apart from particles in a polymer solution there are other  
physical systems the SRE can be applied to such as
spherical or cylindrical particles in liquid $^4\text{He}$ near the
$\lambda$ point and nonmagnetic inclusions in a ferromagnet of 
Ising or Heisenberg type near the Curie point. For these systems 
the parameter ${\cal N}$ takes the values $2$, $1$, and $3$,
respectively. A useful characterization of the small sphere or
the thin cylinder in these cases is provided by the 
universal amplitude
\begin{equation} \label{BB220}
\widehat{A}_{d,D}({\cal N}) \, = \, - \, {\cal A}_K({\cal N}) \, \,
\sqrt{ \frac{B_{\Phi^2}({\cal N})}{\cal N} }
\end{equation}
with the amplitude $B_{\Phi^2}$ of the bulk correlation function 
$\langle {\bf \Phi}^2({\bf r}) {\bf \Phi}^2(0) \rangle_b \, = \,
B_{\Phi^2\,} r^{-2(D - \frac{1}{\nu})}$ at the critical point. 
For example the change in free energy per unit `length' $l^{\delta}$ 
which arises upon immersing the generalized cylinder $K$ into the 
bulk system displays a singular dependence on $t \sim (T - T_c) / T_c$ 
given by
\begin{eqnarray}
& & \Big[ - \frac{k_B T}{l^{\delta}} \,  
\ln \langle e^{- \Delta {\cal H}_K} \rangle_{b,\,t\,} \Big]_{sing}
\label{BB230}\\[1mm] 
& & \, \, = \, - \, k_B T_c \, R^{d - 1/\nu} {\cal A}_K({\cal N}) \, 
[\langle {\bf \Phi}^2 \rangle_{b,\,t\,} ]_{sing} \nonumber\\[2mm]
& & \, \, = \, k_B T_c \, R^{d - 1/\nu} \, \xi^{- (D - 1/\nu)} \,
\widehat{A}_{d,D}({\cal N}) \, \widehat{E}({\cal N}) \nonumber
\end{eqnarray}
with the universal bulk amplitude
\begin{equation} \label{BB235}
\widehat{E}({\cal N}) \, = \, 
[\langle {\bf \Phi}^2 \rangle_{b,\,t\,} ]_{sing} \, \xi^{D - 1/\nu}
(B_{\Phi^2} / {\cal N})^{-1/2}
\end{equation}
which characterizes the temperature dependence of the bulk energy 
density. From Eqs.\,(\ref{B10}) - (\ref{B19}) and from the 
dependence of $B_{\Phi^2}({\cal N})$ on $\varepsilon$ 
one obtains the following explicit expressions:
\begin{mathletters} \label{BB240}

\begin{equation} \label{BB240a}
\widehat{A}_{D,D}({\cal N}) \, = \, \frac{1}{\sqrt{2}}
\, + \, {\cal O}(\varepsilon^2) \, \, ,
\end{equation}
\begin{eqnarray}
& & \widehat{A}_{3,D}({\cal N}) \, = \, \frac{1}{\sqrt{2} \, \pi} \,
\Big\{ 1 \, + \, \frac{\varepsilon}{2}
\, \Big[C_E + \ln \pi \label{BB240b}\\[1mm]
& & \qquad + \, \frac{{\cal N}+2}{{\cal N}+8}
\Big( 16 \pi^2 {\cal B}_3 + 2 \ln 2 - 1 \Big) \Big] \Big\} 
\, + \, {\cal O}(\varepsilon^2) \, \, , \nonumber
\end{eqnarray}
and
\begin{equation} \label{BB240c}
\widehat{A}_{2,D}({\cal N}) \, = \, 
\varepsilon \, \frac{{\cal N}+2}{{\cal N}+8} \,
2^{3/2} \pi {\cal B}_2 \, \, + \, {\cal O}(\varepsilon^2) \, \, .
\end{equation}

\end{mathletters}

The first-order $\varepsilon$ result (\ref{BB240a}), which we have 
obtained by carrying out the calculation 
{\em directly\/} in the outer space of a 
sphere, confirms the prediction
\cite{burkhardt}
\begin{equation} \label{BB250}
\widehat{A}_{D,D}({\cal N}) \, = \, 
\sqrt{ (A_O^{\,\Phi^2})^2 / ({\cal N} B_{\Phi^2}) }
\end{equation}
which follows from relating the half-space (hs) profile 
$\langle {\bf \Phi}^2(z) \rangle_{\text{hs}} = 
A_O^{\,\Phi^2} (2 z)^{- (D - \frac{1}{\nu})}$
with the distance $z$ from a planar wall with Dirichlet 
boundary conditions $O$ at the bulk critical point
to the profile $\langle {\bf \Phi}^2(r) \rangle$ in the outer 
space of the sphere by means of a conformal transformation 
\cite{BE85} (for ${\cal N} = 1$ compare the explicit result 
in the first Eq.\,(20) of Ref.\,[36(a)]). 
The consistency of the above results is expected but remarkable
since the finite conformal map changes the geometry under 
consideration. 

It is helpful to summarize the relationship between 
the universal amplitude $A_{d,D}$ for polymers and 
the universal amplitude $\widehat{A}_{d,D}$ for the
field theory in terms of the symbolic equation
\begin{equation} \label{B28}
A_{d,D} \, \Psi \, = \, \widehat{A}_{d,D} \,
\sqrt{ {\cal N} / B_{\Phi^2} } \, \, {\bf \Phi}^2
\end{equation}
which applies inside averages or correlation functions
for ${\cal N} \searrow 0$ with
$\Psi$ defined in Eq.\,(\ref{C70}).
Since $\Psi = \sqrt{b_{\Psi} / B_{\Phi^2} } \, {\bf \Phi}^2$
with $b_{\Psi} = ({\cal R}_x^{\,1/\nu} / (2 L_0))^2 B_{\Phi^2}$
from Eq.\,(\ref{Czehn}) one has
\begin{equation} \label{B29}
A_{d,D} \, = \, \Big( \widehat{A}_{d,D} \, 
\sqrt{ {\cal N} / b_{\Psi} } \, \Big)_{{\cal N} \searrow 0} \, \, \, .
\end{equation}
For a spherical particle, in particular, the polymer
amplitude $A_{D,D}$ can be expressed in terms of the 
critical universal amplitude ratio on the rhs of
Eq.\,(\ref{BB250}) and the noncritical
universal bulk amplitude $b_{\Psi}$. In $D=2$ both are 
explicitly known [19(b)] and lead to the value 
$A_{2,2} = 3.81$ in Table \ref{table_amplitude}.


\section{Short distance amplitude for polymer density correlations}
\label{appC}

Here we calculate the universal amplitude $\sigma$
in Eqs.\,(\ref{IV30}) and (\ref{IV40}). While for 
$D = 4$ it coincides with the corresponding ideal chain 
value $\sigma^{(id)} = \pi^{-2}$ from Eq.\,(\ref{IV50}),
in $D = 3$, $2$, and $1$ the amplitude $\sigma$ is different
from $\sigma^{(id)}$.

(a) $D = 3$: In this case results are available 
\cite{schafer,duplantier} for the normalized scattering form 
factor $H(Q)$ which is defined for general $D$ by
\begin{equation} \label{C10}
C_2({\bf r}, 0) \, = \, {\cal R}_x^{\,2/\nu} \,
\int\limits_{{\mathbb R}^D}
\frac{d^{D}Q}{(2 \pi)^D}
\, \exp(i \, {\bf Q} \, {\bf r}) \, H(Q) \, \, ,
\end{equation}
where $H(0) = 1$ as implied by Eq.\,(\ref{IV10}).  
The amplitude $h_{\infty}$ in the power law
\begin{equation} \label{C20}
H(Q \to \infty) \, \to \, h_{\infty} \, 
(Q^2 {\cal R}_x^{\,2} / 2)^{-1/(2 \nu)}
\end{equation}
is related to $\sigma$ by 
\begin{equation} \label{C30}
\sigma \, = \, h_{\infty} \, 2^{-1/(2 \nu)} \, \pi^{-D/2} \, \,
\Gamma\Big( \frac{D - 1/\nu}{2} \Big) \Big/
\Gamma\Big( \frac{1}{2 \nu} \Big) \, \, .
\end{equation}
From the accepted \cite{schafer,duplantier} approximate value
$h_{\infty} \approx 1.1$ in $D = 3$ one infers via Eq.\,(\ref{C30})
the value $\sigma \approx 0.13$ (see Table\,\,\ref{table_amplitude}).

(b) $D = 2$: In this case one can obtain a fairly
accurate estimate for $\sigma$ by combining a numerical 
estimate for a ratio of gyration radii 
of ring- and open-chain-polymers with conformal 
invariance and Bethe ansatz results for the $O({\cal N})$ 
vector model by invoking the polymer magnet analogy (PMA)
\cite{gennes2,cloizeaux}.
By using the language of the Ginzburg-Landau field theory
(compare Sec.\,\ref{secIIvoran}) the necessary relations 
of the PMA can be written in a way which makes the 
generalization to $D = 2$ obvious. 
The polymer average
\begin{eqnarray}
& & \left\{ \rho({\bf r}_A) \, \rho({\bf r}_B) \, 
\rho({\bf r}_C) \right\}_{\bf y} \label{C40}\\[2mm] 
& & \quad = \, \frac{ {\cal L} \left\langle 
\Psi({\bf r}_A) \, \Psi({\bf r}_B) \, \Psi({\bf r}_C) \,
\Phi_1({\bf y}) \int d^{D} y' \, \Phi_1({\bf y}') \right\rangle } 
{ {\cal L} \left\langle 
\Phi_1({\bf y}) \int d^{D} y' \, \Phi_1({\bf y}') \right\rangle } \nonumber
\end{eqnarray}
is expressed in terms of cumulant averages 
$\langle \, \, \rangle$ of the field theory.
Here ${\cal L} = {\cal L}(t_0 \to L_0)$
denotes an inverse Laplace transform defined as in
Eq.\,(\ref{E30}) and relates the strength $t_0$
in the thermal perturbation
\begin{equation} \label{C50}
{\cal H}_{th} \, = \, \int\limits_{{\mathbb R}^{D}}
d^{D} r \, {\cal T}({\bf r}) \, \, ,
\end{equation}
\begin{equation} \label{C60}  
{\cal T}({\bf r}) \, = \,
\frac{t_0}{2} \, {\bf \Phi}^2({\bf r}) \, \, , 
\end{equation}
of the Hamiltonian {\em at} the critical point of 
the field theory to the bare `chain length' $L_0$ which 
$-$ apart from a nonuniversal proportionality factor $-$   
equals the number of monomers in the polymer chain.
The scaling dimension of the quantity
\begin{equation} \label{C70}
\Psi({\bf r}) \, = \, {\cal R}_x^{\,1/\nu} \, \frac{1}{E} \,
{\cal T}({\bf r}) 
\end{equation}
equals its inverse length dimension $D - 1 / \nu$. 
Here $E = t_0 L_0$
is the exponent which appears in ${\cal L}$ 
(compare Eq.\,(\ref{E30})). 
The rhs of Eq.\,(\ref{C40}) has the normalization 
property that by integrating the numerator 
over, say, ${\bf r}_A$ one can replace
$\int d^{D} r_A \, \Psi({\bf r}_A)$
by ${\cal R}_x^{\,1/\nu}$ (compare the discussion related to
Eq.\,(18) in Ref.\,\cite{unirel}). This is consistent 
with the corresponding normalization property 
$\int d^{D} r_A \, \rho({\bf r}_A) = 
{\cal R}_{x}^{\,1/\nu}$ for the lhs of Eq.\,(\ref{C40})
as implied by Eq.\,(\ref{I70}).
  
Short distance properties such as those in 
Eqs.\,(\ref{IV30}) and (\ref{IV40})
follow from the operator product expansion (OPE)
\begin{equation} \label{C80}
\Psi({\bf r}_A) \, \Psi({\bf r}_B) \, \to \,
\sigma \, r_{AB}^{\, \, - (D - 1 / \nu)} \,
\Psi\left( ({\bf r}_A + {\bf r}_B) / 2 \right) \, \, ,
\end{equation}
which is 
equivalent to the well-known OPE of energy density operators
[39(b),64]. The amplitude $\sigma$ is expressed as
\begin{equation} \label{C90}
\sigma \, = \, \zeta \, \sqrt{b_{\Psi}} \, \, .
\end{equation}
Here $b_{\Psi}$ is the universal bulk amplitude in
\begin{equation} \label{Czehn}
\langle \Psi({\bf r}) \Psi(0) \rangle_{\text{crit}} \, = \,
b_{\Psi} \, r^{- 2 (D - 1/\nu)} 
\end{equation}
with $\langle \, \, \rangle_{\text{crit}}$
denoting the average at the critical point of the field theory 
and $\zeta$ is the amplitude which replaces $\sigma$ in
the corresponding OPE for the normalized energy density
$\widetilde{\Psi} = \Psi / \sqrt{b_{\Psi}}$.
By using results of Refs.\,\cite{dotsfat,beg88} one finds that
in $D = 2$ for ${\cal N} \searrow 0$ 
\begin{equation} \label{C100}
{\cal N}^{1/2} \, \zeta \, \to \,
(216 \, \pi)^{1/2} \, 
\left( \frac{\Gamma(2/3)}{\Gamma(1/3)} \right)^{9/2}
\, \approx \, 1.21
\end{equation}
(compare Eq.\,(7.161) in Ref.\,[19(a)] where $\zeta$ 
is denoted by $c_2$).
The amplitude $b_{\Psi}$ has been 
calculated in the Appendix of Ref.\,[19(b)] 
by using results of Ref.\,\cite{cardymuss} so
that in $D = 2$ for ${\cal N} \searrow 0$ one has
\begin{equation} \label{C110}
{\cal N}^{-1/2} \, \sqrt{b_{\Psi}} \, \to \, \frac{\kappa}{\pi} \,
\left( \frac{5}{6} \, 
{\cal R}_x^{\,2} \Big/ {\cal R}_{\text{ring}}^{\,2} \right)^{2/3}
\end{equation}
where $\kappa = 0.226630$ and 
${\cal R}_x^{\,2} / {\cal R}_{\text{ring}}^{\,2} \approx 6.85$
\cite{cardyguttm} is the ratio of ${\cal R}_x^{\,2}$ 
of an open polymer chain and the mean square radius of gyration
${\cal R}_{\text{ring}}^{\,2} = {\cal R}_{x, \, \text{ring}}^{\,2} + 
{\cal R}_{y, \, \text{ring}}^{\,2}$ of a {\em ring polymer}
with the same number of monomers.
Equations (\ref{C90}) - (\ref{C110}) lead to the value
$\sigma \approx 0.278$ in Table\,\,\ref{table_amplitude}. 

(c) $D = 1$: In this case 
the behavior of a chain with excluded 
volume interaction is that of a rigid rod of length ${\cal R}_x$.
Thus $\nu = 1$ and $C_2({\bf r},0)$ equals 
${\cal R}_x - |{\bf r}|$ for $|{\bf r}| \le {\cal R}_x$
while it vanishes for $|{\bf r}| > {\cal R}_x$.   
The assumption that Eq.\,(\ref{IV30}) still holds for $D = 1$
leads to the value $\sigma = 1$ in 
Table\,\,\ref{table_amplitude}.



\begin{thebibliography}{99}
\bibitem{confusion}We use the phrase `excluded
volume interaction' exclusively for the repulsive interaction between
chain-monomers and not for the interaction between a
colloidal particle and a chain-monomer.
\bibitem{walk}Equation (\ref{eq117}) also applies for long walks
on a lattice which must not visit the region occupied by the
mesoscopic particle \cite{meijer} provided one considers the
variation of the partition function on length scales much larger 
than the lattice constant.
\bibitem{ideal}These authors also discuss the crossover from 
rigid rods to flexible polymers by considering chains composed
of two rigid segments joined by a hinge.  
Ideal chains inside a generalized cylinder $K$ 
can be readily discussed by using methods as those 
described in Ref.\,[60] in I.
\bibitem{medium}Throughout this work we treat the 
actual common solvent of the particles and the polymers
as an inactive background medium so that
without polymers the dilute particles behave as an ideal gas. 
In the context of Eq.\,(\ref{I50}) the `solvent' of the particles 
is formed by the polymers which are separated from the
ideal gas phase of the particles, e.g., by a semipermeable 
membrane.
\bibitem{lengths}Examples for such lengths are 
the distances from the particle of a fixed end of the chain,
of specific points of the local monomer density distribution 
of the chain, or of another boundary which is impenetrable for the chain.
\bibitem{geometry}We briefly outline the argument: Following
Ref.\,\cite{david} the curvature contributions should depend on 
how the surface is embedded in the space ${\mathbb R}^{D}$, i.e., 
they should be derivable from the local extrinsic curvature tensor 
${\bf K}_{ij} = K_{ij\,} {\bf n}$
where ${\bf n}$ is the local unit vector normal to the surface 
(we only consider orientable surfaces). The principal local radii of
curvature $R_i$ are the inverse of the $D - 1$ eigenvalues of the
matrix $(K_{ij})$. Therefore $K_m$, $K_m^{\,2}$,
and $K_G$ as defined in Eqs.\,(\ref{II160})
are the only independent scalar quantities 
to first and second order in $1 / R_i$ 
which can be deduced from $K_{ij}$ and which are invariant under
permutations of the indices $i = 1, \, \ldots \, , D - 1$.
\bibitem{shape}The dashed line in 
Fig.\,\ref{fig2} is known rather accurately by means
of the $\varepsilon$-expansion of $\nu(D)$
in conjunction with the exact value of $\nu(D=2)$ \cite{amit}.
\bibitem{log}For example, for $(d,D) = (2,4)$ and ideal chains
the reduced free energy increase $n_{p\,} f_K^{\,(1)}$ 
vanishes for $R \to 0$ as
$\sim {\cal R}_{x}^{\,2} / \ln({\cal R}_{x} / R)$ \cite{joanny2} 
involving a logarithm
which should be compared with the power law decay 
$\sim {\cal R}_{x}^{\,1 / \nu} R^{2 - 1/\nu}$ with 
$1 / \nu \approx 1.7$ for chains with EV interaction in $(d,D) = (2,3)$.
\bibitem{methods}Methods based on conformal invariance allow one to reduce
$A_{D,D}$ for arbitrary $D$ to a half-space amplitude \cite{BE85}
and, for the particular dimension $D = 2$, even to a bulk amplitude
\cite{beg88}. The latter can be evaluated by using Bethe ansatz 
results \cite{cardymuss} and the ratio 
${\cal R}_x / {\cal R}_{\text{ring}}$ considered in Eq.\,(\ref{C110}).
\bibitem{dumb-bell2}This small dumb-bell operator expansion can be used
for a calculation not only of the solvation free energy of the 
dumb-bell but also of other observables such as the monomer density
profile at distances from the dumb-bell which are much larger than
$R$ and ${\cal D}$.
\bibitem{hantel}Here one may replace the sum in 
Eq.\,(\ref{IV160a}) by an integral according to the 
Euler-MacLaurin formula in Eq.\,(\ref{II80}) where 
the terms proportional to $F(0)$ and $F'(0)$ have 
to be included.
\bibitem{derjaguin}The Derjaguin expression of the free energy
of interaction between two large spheres with the same radius $R$
follows from the corresponding free energy between a sphere
and a planar wall in Eq.\,(14) in Ref.\,\cite{unirel}
by replacing the prefactor $R$ by $R/2$.
The reason is a corresponding replacement in the
local distance $\widetilde{D}(r_{\parallel})$
in Ref.\,\cite{unirel}.
\bibitem{SSa}For $N = 100$ our crude estimate is off by only
$20$\,\%. The reason for the larger deviation in case of $N = 10$
could be that $N = 10$ is still too small for a comparison
with the universal asymptotics.
\bibitem{eintauch}A discussion of the validity of the
PHS model in the case
of the solvation free energy for a single particle is given in I.
\bibitem{polykugel}The effective radius $\widetilde{\cal R}$
of the polymer `spheres' may be adjusted to ${\cal R}_x$
by comparing the free energy change on immersing a single
large sphere in the unbounded solution. This leads
to $\widetilde{\cal R} = \sqrt{2/\pi} \, {\cal R}_x$
(compare Eq.\,(3.10) in I).
\bibitem{sieben}The function 
$g_s^{\,(\text{as})}(\psi \sqrt{\tau}, 0)$
has the property that the integral in Eq.\,(\ref{II322b}) 
is dominated by contributions from the 
$\psi$-interval $1 < \psi < \varphi$ with 
$1 \ll \varphi \ll \tau^{-1/2}$.
Alternatively, we observe that the integrand in Eq.\,(\ref{II322b})
approaches its limit for $\tau \to 0$ uniformly with respect to 
$\psi$ (compare section 6.2 in Ref.\,\cite{bender}).
\bibitem{AcalA}Compare Eq.\,(4.8) in I where $A_{d,D}$ is 
denoted by $- \widetilde{{\cal A}}_K D^{\frac{1}{2\nu} - 1}$ and
where the bare `chain length' $L_0 = \mu^{-2} Z_t^{-1} L$ is
denoted by $L$. See also the argument in Ref.\,[19(b)] 
according to which $A_{d,D}$ can be read off by rewriting the rhs
of Eq.\,(\ref{B10}) for large $\mu R$ as 
$1 - A_{d,D\,} R^{d - 1/\nu\,}\widetilde{\omega}_K$ where
$\widetilde{\omega}_K$ follows from $\omega_K$ 
in Eq.\,(\ref{B15}) upon replacing
${\bf \Phi}^2$ by $\Psi$ as defined in Eq.\,(\ref{C70}).
\end{thebibliography}
\end{document}